CONSCIOUSNESS AND THE PHYSICAL WORLD

EDITED PROCEEDINGS OF AN INTERDISCIPLINARY SYMPOSIUM ON
CONSCIOUSNESS HELD AT THE UNIVERSITY OF CAMBRIDGE IN JANUARY 1978

EDITED BY

B. D. JOSEPHSON
Cavendish Laboratory, Cambridge

and

V. S. RAMACHANDRAN
California Institute of Technology, U.S.A.

Contents

Foreword

F. J. DYSON

This book stands in opposition to the scientific orthodoxy of our day. The orthodox dogma is stated by the biologist Jacques Monod in his book Chance and Necessity with characteristically French sharpness: "The cornerstone of the scientific method is the postulate that nature is objective. In other words, the systematic denial that true knowledge can be got at by interpreting phenomena in terms of final causes — that is to say, of purpose." Monod labels those who disagree with him "animists". The arch-animist is Teilhard de Chardin, for whom Monod reserves his deepest scorn: "The biological philosophy of Teilhard de Chardin would not merit attention but for the startling success it has encountered even in scientific circles. . . . There is no inert matter, and therefore no essential distinction between matter and life. . . . For my part I am most of all struck by the intellectual spinelessness of this philosophy. In it I see more than anything else a systematic truckling, a willingness to conciliate at any price, to come to any compromise. Perhaps, after all, Teilhard was not for nothing a member of that order which, three centuries earlier, Pascal assailed for its theological laxness.' '

The authors of this book are not followers of de Chardin. They represent a variety of scientific disciplines and a variety of philosophical viewpoints. But they are all, according to Monod's definition, animists. That is to say, they are not willing to exclude a priori the possibility that mind and consciousness may have an equal status with matter and energy in the design of the universe. They are trying to extend the boundaries of scientific discourse so that- the subjective concepts of personal identity and purpose may come within its scope. They are all to some extent exposing themselves to the charges of ideological laxity with which Monod lambasted de Chardin. They are accepting a certain risk that their orthodox colleagues will consider them a little soft-headed.

I am delighted to see that the contributors to this book include more biologists than physicists. In recent years biologists have usually been more inhibited than physicists in stepping outside the accepted norms of scientific respectability. Monod was, after all, a biologist. In dealing with the problems of consciousness, physicists have had courage but no competence, biologists have had competence but no courage. In this book we see some examples of competence combined with courage.

Why have the biologists during the last century been so inhibited? I believe they are still suffering from the after-effects of the great nineteenth-century battle between the evolutionists led by Darwin and Huxley, and the churchmen led by Bishop Wilberforce. The high point of the battle was the great debate in Oxford in 1860 during which Bishop Wilberforce asked Huxley whether he was descended from a monkey on his grandfather's or on his grandmother's side. Huxley won the debate, but the biologists are still fighting the ghost of Bishop Wilberforce. In the bitterness of their victory over the forces of religious orthodoxy, they have made the meaninglessness of the universe into a new dogma. "Any mingling of knowledge with values is unlawful, forbidden", says Monod.

The authors of this book have defied Monod's anathema. They have wandered freely over the borderland between science and philosophy, where knowledge and values are inextricably mixed. I believe they have brought back some insights which will be illuminating not only to scientists but also to anybody with a philosophical turn of mind who enjoys pondering over the mysteries of mind and consciousness.

Introduction

V. S. RAMACHANDRAN
Trinity College, Cambridge

This book is about consciousness and is based on a symposium on that subject held at the University of Cambridge on 9-10 January 1978. Usually, books on scientific or philosophical subjects are edited by experts on the subject matter of the book itself. I make no apology for the fact that this particular book is an exception to that rule — since there can really be no such thing as an "expert" on a subject as nebulous as consciousness.

Although scientists often have their own private views on consciousness they are usually reluctant to talk about these views. There are two reasons for this. Firstly, scientists are generally unwilling to venture into realms outside their legitimate scope or to speculate on questions for which there can be no precise empirically demonstrable answers. Secondly, there is a widely prevalent superstition among them that interest in such "fringe areas" is a sign of woolly thinking and declining intellectual vigour. Perhaps this explains their curious silence and their unwillingness to publish philosophical speculations.

The purpose of the Cambridge conference was to encourage distinguished scientists to express their views on the relationship of conscious experience to the physical world.* To add a sense of proportion we also invited a professional philosopher (G. Vesey) and a person claiming psychokinetic powers (Suzanne Padfield). By doing this we have tried to represent as wide a spectrum of views on consciousness as possible.

And as the reader will notice, the spectrum is very wide indeed — ranging from Barlow's materialistic account (that consciousness is nature' 3 ' 'trick' '

'We are grateful to Research Corporation of New York for a grant out of which this conference was supported.

to chain us to our herd) to Josephson's view that minds may even have oertain attributes of their own (e. g. "creativity" or "intelligence' ') to help channel the activity of physical brains towards specific goals. Yet in spite of these wide-ranging views, some of them flatly contradicting each other, a surprising degree of communication was achieved between the various speakers. What emerged was this book, whose contents I shall attempt to summarize in this Introduction. *

The publication of an interesting book by Popper and Eccles,' The Self and its Brain, coincided with the conference, and since the ideas in that book are rather similar in spirit to some of those which were discussed at the conference, it may be relevant to begin our survey with some of Popper's ideas. Popper calls himself a "dualist" and "interactionist' ', and believes in what he calls World 1 (the material universe, including physical brains), World 2 (individual human minds) and World 3 (language, culture, science and other products of World 2). He suggests that although World 3 originally emerged as a product of World 2, it seems to have acquired a life of its own and is no longer chained to individual minds. He speaks of World 3 "objects" like numbers, ideas, numerical concepts, etc., which are in some respects analogous to the physical objects of World 1. Calling ideas and numbers objects may sound like an elaborate joke to some readers, but in defence of his thesis Popper points out that:

(a) World 3 has a quasi-independent status and would exist even if individual men died.

(b) Many World 3 attributes are unplanned consequences of collective culture (e. g. Goldbach's conjecture and other hidden properties of number systems that are discovered by mathematicians just as an archaeologist discovers a World 1 object).

(c) World 3 properties are often novel and "emergent", i.e. irreducible to the properties of individual minds — just as brains may have properties which are irreducible to single neurons.

((1) Finally, one can imagine chains of causation in World 3 that are logically independent of (though necessarily accompanied by) physical causation in World 1. For instance, two computers that are grossly

*The speakers were encouraged to correspond with each other after the conference and this additional discussion is also included in the book.

different physically can nevertheless operate according to the same
" standards of logic' ' (which are World 3 entities).

Popper also emphasizes that Worlds 2 and 3 are symbiotic since culture can "feed back" to enrich and expand individual minds. "Matter", he argues, "can thus transcend itself by producing mind, purpose and a world of the products of the human mind. One of the first of these products is language. In fact I conjecture that it was the very first of these products, and that the human brain and the human mind evolved in interaction with language." Elsewhere: ". . . As selves, as human beings, we are all products of World 3 which, in its turn, is a product of countless human minds."

It is important not to evade the chicken-or-egg aspects of this theory. Fortunately both authors (Eccles and Popper) give some thought to the apparently insuperable problem of how a closed system like the physical universe can "interact" with minds. Eccles begins by making the deliberately outrageous suggestion that the physical world is in fact not a closed system and that World 2 can directly influence the activity of brains.* The self-conscious mind, according to him, may act on certain "open" elements in the nervous system (such as synaptic clefts), which are so minute that even Heisenbergian uncertainty can influence their behaviour. The activity of these structures could then become magnified to account for brain events corresponding to human "choice" or "creativity".

Not everyone would find this view very satisfactory. If a combination of sub—atomic uncertainty (World 1) and the constraints of rational thought (World 3) can account for human freedom and creative enterprise, then what need is there for World 2? There is, after all, nothing logically impossible about World 1 objects (brains) creating World 3 without the intervention of World 2; so Eccles's own argument seems to suggest that minds are redundant by—products of evolution!
In spite of these difficulties The Self and its Brain contains some bold and powerful arguments for dualism and is sure to provide a valuable stimulus

'The authors seem to rely largely on introspection for arriving at some of these conclusions. For instance, the fact that people can reverse Necker cubes or engage in adventurous mountain climbing (Popper, p. 146) is cited as evidence for the view that the conscious self has "taken over" the activities of brains!

to new enquiry. If the Cambridge Symposium (embodied in this book) provides a similar stimulus, it will have achieved its purpose. It begins, appropriately, with a scholarly chapter by G. Vesey which contrasts sharply with some of the more light-hearted chapters in the book. The other contributors include three psychologists (R. L. Gregory, N. K. Humphrey and M. J. Morgan), three physical scientists (B. D. Josephson, H. C. Longuet-Higgins and D. M. MacKay), two physiologists (H. B. Barlow and myself) and a psychiatrist (M. Roth).

THE SOCIAL DIMENSIONS OF CONSCIOUSNESS

Chapters 4 and 5 form the core of the book and deal with speculations on the possible evolutionary significance of consciousness. Barlow's suggestion (Chapter 5) is novel and surprisingly simple. He begins by rejecting "parallelism" (i.e. the view that consciousness simply parallels any complex neural event such as the activity of MacKay's "supervisory" system, described in Chapter 6) on the grounds that if consciousness merely parallels complex neural events there is no reason why only a tiny fraction of such events should emerge into awareness. He suggests, instead, that consciousness may have emerged as an evolutionary novelty among social animals to permit gregariousness and communication. Thus consciousness, according to him, is "interaction and not a property". We feel pain only in order to communicate it, and if the need to communicate it had not arisen (e.g. in non-social animals like frogs or lizards) there would only be reflex withdrawal unaccompanied by the subjective sensation of pain. Perhaps the fact that people generally shout when jabbed with a needle supports Barlow's argument, but then why is the pain often felt after the shout?

Barlow also suggests that archetypes of other people are modelled into our brains by natural selection, and that consciousness consist either of real conversations with other individuals or of imaginary conversations with those archetypes (psychologists would call this "internal rehearsal"). Consciousness in his view is synonymous with communication. It would be biologically useless to communicate certain brain events (like the pupillary light reflex and reflex arcs regulating visceral functions, etc.) and therefore these events never emerge into consciousness.

Note that Barlow is not merely saying that communication adds an extra dimension to consciousness (a point that is already implicit in Popper's ideas), but that communication is consciousness. What he claims to have found is a correlation between certain kinds of neural events and consciousness — namely those neural events which are involved exclusively in communicating with other brains. Of course Robinson Crusoe was also conscious, but that is because his brain was engaged in imaginary dialogues with archetypes of other people.

Humphrey (Chapter 4) also emphasizes social aspects of consciousness but in a sense of his argument is the exact converse of Barlow's. He points out that a person who has never felt (say) pain cannot meaningfully understand or interpret the behaviour of another person being exposed to painful stimuli and would consequently be unable to communicate* effectively with him. From this example, he argues that the biological function of the sensation of pain lies in its usefulness for social interaction. Thus we feel pain in order better to understand the pain felt by others. He argues further that such subjective sensations evolved primarily to permit an animal to attribute reasons for its own behaviour and consequently to make sense out of the behaviour of other members of the social group.

Although at first sight Humphrey's argument seems flatly to contradict Barlow's, there is really no fundamental inconsistency, since both authors emphasize the importance of social factors and suggest that consciousness may have an evolutionary function. Thus, while Humphrey suggests that introspection is necessary for modelling archetypes of other people, Barlow regards conversations with archetypes as almost synonymous with introspection. Barlow speaks of communication with people "enriching" our conscious experience whereas Humphrey points to people who seek out new subjective experiences in order to enrich communication with others! A biologically inclined philosopher might support Barlow, but Humphrey's more introspective account seems closer to common sense.

To use Popperian terminology, Barlow is suggesting that World 2 (mind) is compulsorily parasitic on World 3 (which includes languages and culture). This is a bold departure from Popper's own interactionist view that Worlds 1, 2 and 3 exist independently while interacting to enrich each other. Humphrey, on the other hand, sticks to the Popperian tradition, and his view would be consistent with the suggestion that World 3 (as well as simple communication with others which is a necessary antecedent of World 3) would not have arisen if World 2 had not crept into physical brains at some stage in evolution, i.e. in his account World 2 would necessarily precede World 3. However, it is not clear whether either author would want to argue that the survival value of World 3 actually exerted selection pressure for the emergence of World 2. This, it seems to me, is the crux of the whole debate.

---

*Here, and elsewhere, I use the word communication in its widest sense (and interchangeably with social interaction). The word should not be taken to mean verbal communication alone.

These considerations must lead us to a synthetic View of the evolution of mind. Perhaps at some stage in phylogeny, consciousness emerged as an incidental by—product of certain complex neural events. This new property was unplanned for, but once it emerged it made communication possible, since animals could begin to "introspect" and (by analogy) make sense out of one another's behaviour. Since communication has survival value, natural selection seized upon these neural events which were associated with consciousness and this in turn led to a mutually reinforcing interaction between collective culture and individual minds.

Such an account would be wholly consistent with Humphrey and Popper but would also leave several questions unanswered. Implicit in all the views presented so far is the assumption that consciousness is causally important for communication. For if it were not causally important then natural selection could not have favoured its emergence and its absence would have made no difference to the course of evolution. On the other hand, if its presence does make a difference we would have to assume that minds can actually exert an influence (however indirect) on the course of events in the physical world — particularly on a small portion of the physical world consisting of communicating brains. The implication of this would be that (a) the physical world is not a "closed system" and that (b) minds cause communication and do not merely accompany it.

This gets us into logical difficulties. Can consciousness really cause neural events? Stimulating the cortex can lead to mental events (e. g. phosphenes), but the converse would be hard to demonstrate empirically. Josephson accepts "mind acting on brain" as being almost axiomatic but is there any evidence to justify such a view? Unfortunately we are not even sure of what cause—and—effect means when talking about brain events and mentation. Gregory (Chapter 2) points out that our common—sense notions about causation are hopelessly muddled, and he illustrates this with the example of night following day. Obviously day does not cause night; nor are they both caused by some third agent. Instead we see the night-day sequence as part of our conceptual model of the solar system. Similarly, the nature of brain—mind causation may become clearer when we start seeing it as part of a larger (hitherto undiscovered) conceptual scheme. Perhaps the causal links between brain events and mentation belong to a logical category that is quite distinct from, and are of a much more subtle nature than, the causation we talk about in the context of objects and forces. (Though, heaven knows, these words beg enough questions themselves!)

And nowhere is the problem of disentangling cause and effect more difficult than in the World 2 .2' World 3 interactionism proposed by Popper and Eccles. Could World 3 have arisen at all in the absence of at least a rudimentary World 2, and if so could the survival value of World 3 have exerted any selection pressure for the emergence of World 2? Did the dim introspective abilities of Proconsul necessarily antedate his ability to com-municate, and if so did the culture which emerged from such communication propel him onwards to become Homo erectus? The theories of Barlow and Humphrey (as well as Popper's interactionism) may well contain partial answers to these important questions.

WHAT IS CONSCIOUSNESS?

While engaging in philosophical discussions of this kind there is always the tendency to forget that problems of consciousness are not merely of academic interest. To a patient in a hospital, experiencing intense pain or anguish, what we have said so far in this chapter, and any talk about consciousness being a "ghost in the machine", would seem curiously irrelevant or even perverse (Gregory, p. 31). Fortunately, this deficiency is remedied by Roth (Chapter 8), who surveys the phenomenology of consciousness from a clinical point of view, and by Longuet—Higgins (Chapter 3), who examines the validity of common-sense criteria which people generally use for deciding whether someone is conscious or not. Relying largely on common sense, Longuet—Higgins argues that the encodability of events into memory seems to be an invariant correlate of conscious experience — i.e. if a person remembers something, he must have been conscious of it in the first instance. This seems to be generally true, but it is not difficult to think of possible exceptions. For instance, we often remember dreams vividly and attribute consciousness to dreams while recalling them later, but does it necessarily follow that we were conscious during the dream?

Josephson's approach to consciousness (Chapter 7) differs radically from those of the other contributors. Most scientists start with the brain and ask themselves why certain brain events seem to be associated with consciousness. Josephson's point of departure, on the other hand, is in consciousness itself, which he suggests can be empirically studied by introspection.

He begins with consciousness as a "given thing" and points out that our minds seem to have certain obvious attributes like creativity or intelligence. He regards these attributes as being almost axiomatic since we know them to be there from our own personal experience. Might we then not start with these almost axiomatic observations on consciousness and then try to arrive at more general "laws" of behaviour? Josephson points out that there already exists an extensive introspective–phenomenological account of consciousness to be found in the Eastern philosophical literature? He uses ideas from this literature and tries to construct a theory of consciousness based on concepts borrowed from systems engineering.

Professional psychologists frown on introspection largely because other professional psychologists would frown on them if they did not. There is, after all, no a priori reason for starting with brains and working up towards consciousness instead of vice versa. In fact, to a person untrammelled by conventional scientific training, Josephson's approach might seem much more simple and straightforward. Galileo and Newton began with observations about the physical world and went on to construct laws (such as the laws of motion) of steadily increasing explanatory power. Why sneer on the same approach being used for studying our own conscious experience?

Until now we have considered the evolutionary origins of consciousness and tried to answer the question "What is consciousness?" We must now turn to more ancient philosophical issues — like free will and personal identity. In my own contribution (Chapter 9) I have tried to point out that there are really two kinds of personal identity which I have dubbed "empirical identity" and "ontological identity". The empirical identity question is philosophically trivial and has the form "What criteria do people generally use when trying to identify an agent A' as being the same as an agent A whom they have seen in the past?". The ontological identity question (i.e. what criteria should be used when trying to decide whether A' is existentially the same as A who lived in the past) is much more important and can be stated in the form of a series of "thought experiments". I have argued that nothing more can be said about personal identity than what is contained in these thought experiments.

## FREE WILL — AN EVOLUTIONARY APPROACH

Any theory of consciousness must eventually contend with the problem of free will and determinism. If every event in the universe (including brain events) is the inevitable outcome of preceding events, then in what sense are our actions really free? Of course, if a person were completely free his behaviour would be chaotic. Freedom of behaviour (and consequently the will) is necessarily limited by environmental constraints, and hence the question of freedom arises only at what might be called "choice-points", where an agent is called upon to choose between alternate courses of action.

The situation is analogous to a donkey located exactly between two haystacks. Obviously the donkey would not starve to death. He would eventually move towards one haystack or the other, and one would be tempted to describe his choice as being random. A human being in a similar situation might claim that he was exercising the privilege of free will.

If there were no special reason for favouring one haystack, the donkey's choice would either (a) depend on a hidden, thermodynamic bias in the immediately preceding state of the animal's nervous system, or (b) be truly random. Such randomness could arise from a magnification of Heisenbergian uncertainty (see Eccles).3 The sequence of events would be identical in a man but an illusion of free will would accompany the events. Two questions arise. Firstly, why are events at choice-points accompanied by the subjective feeling of free will? And secondly, is there any sense in which an agent's behaviour may be said to be truly free?

Why human behaviour at choice-points is accompanied by the subjective sensation of ''willing'' is difficult to answer. I do not get this feeling (even at choice-points) if my behaviour is triggered off by (say) an epileptic fit. So the presence of intervening variables, as opposed to a straightforward S—R sequence, and knowledge (or belief) that I could have acted otherwise are both necessary conditions for claiming to have chosen freely. Further I must be aware of the outcome of my action and must intend that outcome. (For there can be irrelevant consequences of my action which I am aware of but do not intend — see Kenny[4])

The criteria specified above are mainly self-testimonial. Further, they would be possible only in a nervous system that was capable of projecting itself into the future to anticipate consequences of different kinds of simulated behaviour. (Hence our donkey could not have acted freely.) The system could then use feedback from such anticipations to make what one could call a decision — based on certain goal criteria. If the anticipated consequences are the same for either of two kinds of behaviour then an element of randomness may be deliberately introduced to break the deadlock.

Thus free will seems logically possible only in situations where the outcomes of two anticipated courses of action are equally desirable (e.g. choosing between two identical peanuts — where all of Kenny's criteria would be satisfied). Yet, oddly enough, it is precisely in situations like this that a person often declines having chosen freely and says: "My choice was not based on any particular reason —it was random. . . ." One is almost tempted to conclude that free will exists only among philosophers!

We experience willing even in situations where one choice is clearly preferable to the other. The fact that rational considerations lead to one choice and not the other does not seem to be incompatible with feeling free (i.e. feeling that we could have acted otherwise). The sense of choosing freely seems to parallel closely the activity of the system in the brain that is involved in assessing priorities of action in the light of certain goal criteria.
Actions uncoupled from this system (e.g. automatisms) are not "willed". Why the activity of this system should be accompanied by a feeling of conscious choice is a mystery, but we can speculate on its biological origins.

Perhaps belief in free will provides the drive or incentive to explore various strategies of action by turning and tossing over ideas in one's mind (just as hunger provides the drive for exploring one's physical environment). A drive of this kind would discourage passive acceptance of environmental constraints — and would therefore have obvious survival value. What demarcates Jean—Paul Sartre from Homo habilis may be free will rather than language or consciousness!

This analogy between hunger and free will may not be as superficial as it sounds. Consider a hypothetical organism living in an environment where food is always available in plenty. Such an organism would eat and excrete in a continuous and uninterrupted cycle and would never need to feel hungry. Hunger must have evolved as part of a control system to regulate the state of nutrition of the animal, when food supply became scarce and intermittent. A fall in blood sugar generates hunger and this in turn goads the animal on to look for food. Consistent with this argument is the fact that carnivores probably experience more intense hunger than herbivores, and plants and trees do not feel hungry at all.

Now in my view, just as the conscious sensation of hunger leads us to explore the environment around us, the inner feeling of freedom goads us on to explore strategies of action in an imaginary world which we construct in our minds. We then see ourselves as active agents striving to do things in this imaginary world; and this is possible only because we feel free.

Consider a fatalist who feels a sense of inevitability about his own future. To him all actions would seem futile and pointless. In extreme cases, such individuals are often profoundly depressed since they feel they have "lost control" over themselves. Conscious beings need to feel free in order to justify planning for the future and even to justify their very existence. As Sartre would put it, we need to believe in the permanent possibility of consciousness ". . . effecting a rupture with its own past, of wrenching itself away from its past . . .". So, if consciousness is "nature's joke" to chain us to each other (Barlow, Chapter 5), free will may be nature's joke to permit human beings to plan their own future without feeling like puppets in a Laplacian world.

An animal will work only for a tangible reward that lies well within his reach. What characterizes all human actions, on the other hand, seems to be the willingness to participate in what Bronowski[5] has called

"unbounded plans". Instead of going through a specific sequence of steps leading to a reward, we often adopt global strategies of action directed towards more general aims which we call values or ideals. This ceaseless striving towards abstract and sometimes even unattainable goals (such as "truth" or "perfection") may also depend crucially on our belief in our freedom. Free will may therefore turn out to be a biologically useful delusion that has been built into our brains by natural selection, i.e. those who believed in their ability to will survived and those who did not died out.

This delusion certainly exists in World 2 —and it may also partly exist in World 3 (e. g. French Existentialist literature). It is a sort of imaginary carrot that keeps the donkey in us running all the time — and maybe that is what Sartre really meant when he said, ''It is not enough to will; it is necessary to will to will."

"Free will" may even have a specific anatomical locus—the frontal lobes. Evidence for this view comes from the fact that frontal lobotomy patients are often impulsive and unwilling to deliberate. They also report losing all initiative (or "drive") and have no interest in planning for the future; although they remain conscious, alert and intelligent in every other respect. Since the illusion of willing has survival value it may have become incorporated into the circuitry of the frontal lobes as these structures became progressively larger during the transition from Proconsul to H. habilis. It could be argued that once this feeling emerged men began to experience events as being done by them rather than happening to them; and this in turn may have given rise to other socially useful feelings such as "conscience" and moral responsibility. Theologians would no doubt find this theory rather distasteful and I myself find it intuitively unappealing; but that, of course, is additional evidence in its favour. In fact it is probably our innate sense of freedom that makes us feel a bit odd when someone points out that all our actions are determined exclusively by preceding brain events.

Of course, it is interesting to ask why this whole debate over freedom v. determinism arose in the first place. The physicist's assertion—that even what we usually call a choice is determined exclusively by preceding brain states— seems to conflict with our inner experience (i.e. our feeling that we could have acted differently). So we jump to the conclusion that there is some kind of paradox here to be explained. But is it really legitimate to contrast feelings with factual assertions about brain states? Why not consider the possibility that what our feelings assert may simply be wrong? Perhaps what we really have here is a pseudo-paradox that is based on the unwarranted assumption that because we feel free, we must also in some sense be free.

Let me illustrate this with a more familiar example. Each of us has a feeling of what might be called "self-importance" or selfishness built into him. And a person will continue to feel selfish even if a biologist assures him that objectively speaking his existence has no more value than anyone else's. The brain is biologically programmed to value itself, for if it did not value itself there would be no motivation for a person to preserve his safety or to plan his future. In fact, it could even be suggested that an error or perversion in this mechanism is what sometimes leads to feelings of uselessness and futility ("depression") culminating in suicide.

Maybe the free will illusion has similar biological origins. One of its functions, as we have seen, may be to provide motivation for exploring novel strategies of action and for "non-conformist" behaviour. It is also important to note that practically all our notions about legal, moral and ethical issues are parasitic on the assumption of human freedom — i.e. the assumption of a distinction between responsible and irresponsible actions. So perhaps it is these World 3 entities that exerted the selection pressure for the emergence of free will as a useful superstition in our minds. For without this superstition we could make no sense out of even such commonly used words as "deceit", "cunning", "kind", "fickle", "impulsive", "determined", "deliberate", and so on.*

Note the similarity between some of these ideas and Humphrey's theory on the evolution of mind. While Humphrey would probably argue that we feel free in order to make sense out of each other's actions, we could go a step further and suggest that the need to attribute freedom (or lack of it) to someone's choices arises only in the specifically social context of law, ethics and morality — i.e. all those institutions which seem to provide a cohesive force for the orderly organization of society. (For instance, punishing "irresponsible" behaviour would reinforce "responsible" behaviour and encourage people to deliberate more and to refrain from acting impulsively.)

Until now we have been considering what might be called the phenomenology of free will. But is there any other strictly logical sense in which human actions may be said to be free?

Consider yourself faced with a difficult choice—say between A and B, which are equally desirable. Your choice, we have argued, 1S

'Note that this argument deliberately evades the "mind-body problem" in its classical form -i.e. why there should be any feelings at all associated with neural events. But it is perhaps just as meaningful to ask why the feeling 'of freedom (and associated neuracircuitry) evolved as it is to ask why (say) hunger or pain evolved.

determined either by a hidden bias or (in the absence of such a bias) by random neural events caused by magnifications of Heisenbergian uncertainty. If this is true then how can you justify your claim to freedom other than by pointing out that you feel free?

## LOGICAL INDETERMINACY

Imagine that a super-scientist is watching you from behind a tree and that he has access to complete information about your brain state and about your local environment. He then tries to make an accurate prediction of the detailed future state of your brain including your choice (A or B) and writes this down on a piece of paper. After you make your "free" choice he triumphantly shows you the slip of paper to prove that he was right. He could repeat this a hundred times and he would be right each time . . . until you begin seriously to start doubting your free will.

But you would be wrong to do so — or so at least MacKay would argue (Chapter 6). For you could challenge the scientist to state whether his prediction is one that you would be correct to accept as inevitable in every detail before you make the choice. The fact is that if the prediction were embodied in your brain it could no longer be valid in full detail. MacKay argues that deterministic predictions even in a Laplacian world are valid in full detail only for a detached external observer (like our super-scientist) and are not valid for you since they would have no unconditional claim to your assent. His prediction would be rendered invalid in detail the moment it was embodied in your brain, since the state of your nervous system would change. Even if he could take these anticipated changes into account while writing his prediction he cannot claim that you would be in error to disbelieve it, since if you disbelieved it, it would then be incorrect in detail. These and other versions of MacKay's arguments are too well known to require repetition here. For a more comprehensive review of his ideas see his chapter in Cerebral Correlates of Conscious Experience (ed. P. A. Buser and A. Rouguel-Buser, Elsevier/ North-Holland, 1978).

Finally, let us consider another related "thought experiment". When facing a choice between A and B supposing you were to decide "I shall deliberately do the opposite of whatever prediction the determinist makes". You can now challenge your determinist friend confidently with this self-fulfilling prophecy—since whatever prediction he now makes would ipso facto be rendered false! There is no way in which your decision to contradict his prediction can be embodied in the final prediction which is shown to you. He, in turn, may point out to you that although your behaviour is now no longer predictable by him it is still determined by his prediction.

I shall conclude this chapter with a quotation from Russell, which, I hope, conveys the essence of what we have tried to achieve in this book:

> Philosophy is to be studied, not for the sake of any definite answers to its questions, . . . but rather for the sake of the questions themselves; because these questions enlarge our conception of what is possible, enrich our intellectual imagination, and diminish the dogmatic assurance that Closes the mind to speculation; but above all because, through the greatness of the Universe which philosophy contemplates, the mind also is rendered great, and becomes capable of that union with the Universe which constitutes its highest good.

PART I

General

CHAPTER 1

What Defines Privacy?

G. VESEY
The Open University, Milton Keynes

The purpose of this conference is stated as being "to make a scientific study of subjective experience and to explore the relationships between subjective experience and the objective world". Examples are given of subjects which might be discussed. One of them is "What defines the privacy and personal nature of a person's conscious experience?" Perhaps the hope is that a definition can be found which does not put subjective experience beyond the pale for the scientist. Whether I can fulfil that hope, expressed in these terms, I am not sure. But at least I can show willing, by giving my paper the title "What defines privacy?' '.

One other preliminary remark. In the statement of the purpose of the conference there is the phrase "to make a scientific study". To what is "scientific" opposed, in this phrase? If it is opposed to "philosophical" then perhaps I have stumbled into the wrong conference. But probably the term "scientific" was meant in opposition to "unsystematic" or otherwise disreputable. After all, scientists, quite properly, reflect on the concepts with which they operate, and sometimes make what may be called "conceptual innovations". I am not sure that most scientific revolutions are not in large measure conceptual in character; for example, new ways of thinking about the relationship of space and time. Perhaps the required definition of privacy will be a new way of thinking about the relation of what a person says about himself and what others, observing him, can say about him.

But before talking about new ways, perhaps I had better say something about the old ones. At the risk of telling you what you know very well, I shall briefly survey the relevant history of the concept of mind; that is, the history of answers, not to the question "What defines privacy?", but to the question "What defines mentality?". I will concentrate on just three philosophers: René Descartes, whose Second Meditation is sub-titled "The Nature of the Human Mind: it is better known than the Body";¹ Franz Brentano, author of Psychology from an Empirical Standpoint;² and G. E. Moore, who once read a paper to the Aristotelian Society with the title "The Subject-matter of Psychology' ".³

Descartes defined mentality in terms of thinking and extension, meaning by "thinking"— doubting, understanding, asserting, denying, willing, and so on;⁴ and by "extension"— being spread out in space. Minds think but are not extended, he said; matter is extended but does not think. In other words, mind and matter are two distinct substances, even though a mind and a portion of matter are providentially united in a person in this earthly existence.⁵ This dualism of thinking non-extended mind and non-thinking extended matter gave rise to the intractable problem of how the two substances interact, a problem which a number of philosophers tried to solve by invoking God. I shall not go into that, but will skip across the centuries, two and a third centuries to be exact, to Franz Brentano.

Brentano was more concerned with the inadequacy of Descartes's definition of mentality than with the problems to which it gave rise. The negative characteristic, of not being extended in space, he thought, does not serve to mark off what is mental from what is not mental. "A large number of not unimportant psychologists teach that the phenomena of some or even all of our senses originally appear apart from all extension and spatial location. In particular, this view is very generally held with respect to sounds and olfactory phenomena."⁶ Certainly one does not talk of the shape and size of a sound or smell as one does of, say, a colour patch. The argument is: If Descartes was right then sounds and smells, not being extended, should be mental; but obviously they are not mental — they are not the sort of things that think; so Descartes was wrong.

But Brentano was not content with showing Descartes to be wrong about the negative characteristic. He held that, in stating the positive characteristic to be thinking, Descartes had done no more than state the problem. What is thinking? What is it that is common to doubting, understanding, asserting, denying, willing, and so on, and that does not characterize any physical phenomenon?

Brentano's answer was as follows:

Every mental phenomenon is characterized by what the Scholastics of the Middle Ages called the intentional (or mental) inexistence of an object, and what we might call, though not wholly unambiguously, reference to a content, direction toward an object (which is not to be understood here as meaning a thing), or immanent objectivity. Every mental phenomenon includes something as object within itself, although they do not all do so in the same way. In presentation something is presented, in judgement something is affirmed or denied, in love loved, in hate hated, in desire desired and so on.

This intentional in»existence is characteristic exclusively of mental phenomena. No physical phenomenon exhibits anything like it. We can, therefore, define mental phenomena by saying that they are those phenomena which contain an object intentionally within themselves.7

With the qualification "though not wholly unambiguously" Brentano recognized that the expressions "reference to a content", "direction toward an object" and "immanent objectivity" stand in need of further elucidation. He later attempted such elucidation, as did other philosophers, such as his one-time pupil Edmund Husserl. I think that G. E. Moore may well have read Brentano's Psychol0gy—he certainly read his book on ethics — and that Moore's paper on the subject-matter of psychology was an attempt at a non—technical version of Brentano's thesis.

But before I come on to Moore let me just say one other thing about Brentano. In one respect he was still very much in the Cartesian tradition. The respect is that of knowledge. What sort of knowledge do people have of their own doubts, understandings, beliefs and so on? Is it like the knowledge they have of objective material things? We feel drawn to say that it is not. We feel drawn to distinguish two ways of knowing things, an inward way and an outward way. Thus John Locke distinguishes between "reflection", the mind's turning inward upon itself, and "sensation", the source of our ideas of external objectsf'

Apart from one being inward and the other outward, Locke regards reflection and sensation as being very similar. He evidently regards sensation as being the more familiar mode of observation, for he explains reflection in terms of it. "Though it be not sense, as having nothing to do with external objects, yet it is very like it, and might properly enough he called internal sense."9 He accepts unquestioningly that both sensation and reflection are modes of observation: he refers to "our observation, employed either about external sensible objects, or about the internal operations of our minds perceived and reflected on by ourselves". 10

Brentano draws our attention to a difference which seems to have escaped Locke when he was writing the above. Outer observation is inherently fallible: one may not actually be perceiving what one thinks one is perceiving. This means that, strictly speaking, so-called outer perception is not really perception at all. One does not perceive external objects; one infers their existence from one's ideas of them. So, far from it being appropriate to assimilate inner perception to outer perception, as Locke does, we must acknowledge that mental phenomena are "the only phenomena of which perception in the strict sense of the word is possible' '." In this, Brentano is more consistently Cartesian than Locke. Descartes had said that the mind is better known than the body.

G. E. Moore, in his paper on the subject-matter of psychology, sets out to answer the questions "What kinds of 'entities' are 'mental' or 'psychical' entities? And how are those which are 'mental' entities distinguished from those which are not?"'2 He says that certain kinds of entities seem to him to be undoubtedly mental. They are the acts of consciousness named by the words "seeing , remembering", "imagining", "dream—ing' ', "thinking", "believing", "resolving" and so on. Whenever a person performs such an act, Moore says, he is always "conscious of" something or other. 13 But when one sees a colour and when one remembers it, one is conscious of it in very different senses. Apart from both being acts of consciousness they may have nothing in common. Moore does not know how to explain what he means by "consciousness" except by saying that each of the acts he has named is an act of consciousness. The sense in which to be a mental entity is to be an act of consciousness is, he thinks, the most fundamental sense of the word "mental".

But being an act of consciousness, he says, is not the only characteristic of mentality. There is a characteristic which cannot be said to be a "meaning" of the term "mental", but which may be proposed as a criterion of what is mental. The characteristic is that of being directly known by one mind only." In one word: privacy. Moore is doubtful whether privacy is a characteristic which belongs to all mental acts. He has in mind the abnormal phenomena of co—consciousness in a case of split personality. Other philosophers do not share his doubts. Brentano had said that since mental phenomena are the objects of inner perception, "it is obvious that no mental phenomenon is perceived by more than one individual".15 And John Wisdom says: "The peculiarity of the soul is not that it is visible to none but that it is visible only to one.""'

I said that I would briefly survey the relevant history of answers to the question "What defines mentality?" at the risk of telling you what you already know very well. I have now done so and we have seen that the two chief contenders, once Descartes is out of the way, are, first, what Brentano calls "intentionality" and Moore calls "consciousness of"; and, secondly, privacy — defined in terms of direct knowledge, inner perception, or visibility only to one. Incidentally, the thesis that mentality is characterized by privacy, so defined, is known by critics of the thesis as "the doctrine of privileged access".

I have little doubt that the organizers of this conference are familiar with the traditional "privileged access" definition of privacy. Moore and Wisdom are both Trinity College, Cambridge, philosophers, and philosophical ideas have a way of percolating through into other disciplines. And yet they suggest as an example of the sort of subject we should discuss "What defines the privacy and personal nature of a person's conscious experience?". I can only suppose that they find the "privileged access" definition unsatisfactory for some reason, and are looking for a new definition. This supposition leads me, naturally enough, to the question "What might their reason be for being dissatisfied with the traditional conception of privacy?".

I can only speak for myself. My reason for being dissatisfied with the inner-observation account is this. We think of people doubting, understanding, believing, remembering, imagining, hoping, knowing, fearing, expecting, and so on. How do we know the difference between these things —between, say, remembering and imagining, or believing and knowing? How is a person able to say which he is doing? According to the privileged access definition of privacy, the inner observation account, the answer is as follows. We know about these things in the same sort of way as we know about the difference between, say, cats and dogs, except that the perception is an inward perception and the phenomenon perceived is a private phenomenon. An act of remembering is phenomenally different from an act of imagining, or expecting, or understanding; and we know which we are doing because we can recognize the act in question.

I am dissatisfied with the inner-observation account because that answer, to put it bluntly, seems all wrong, a complete fabrication. Let me explain what I mean, by means of a partial analogy. Consider the question "What is the difference between a pain in one's foot and a pain in one's stomach?". "the most natural and immediate answer", says William

James, is that the difference is one '' of place pure and simple". But James rejected that answer, for reasons we need not go into," in favour of the so-called "Local Sign" theory of Hermann Lotze. According to this theory a person can say where a pain is because of something qualitative about the pain which is for him a sign of the location of the cause. He has learnt to associate this quality with a definite part of the body. Wilhelm Wundt agreed with Lotze, and described the local sign as "a peculiar qualitative colouring, which is independent of the quality of the external impression". But a later psychologist, Oswald Kulpe, put the theory down to what he called "metaphysical prepossession". He wrote, with reference to the local sign theory of Lotze and Wundt: "The thought upon which this whole theory is based is that the impressions must all be of a conscious nature. And here we see the influence of metaphysical prepossession. It was difficult to conceive that an unequivocal relation obtaining between tactual impressions and visual ideas, or other factors subserving localization, could have arisen without conscious direction, by way of purely physiological connection. But there is no justification for the assumption of these conscious intermediaries in the facts of consciousness itself."

In short, Lotze fabricated local signs, signs of location, in order to provide a specious answer to the question "How can a person say whether he has a foot-ache or a stomach-ache?". Similarly, I suggest, advocates of the privileged-access doctrine fabricate phenomenally different acts of consciousness in order to provide a specious answer to the question "How can a person say whether he is remembering, or imagining, or expecting, or understanding?". The analogy is partial only in that there is nothing in the case of remembering, imagining, and so on, to correspond to the physiological explanation of how a person can say where he feels pain.

But I have said enough about the old way of defining privacy. It is high time to go on to a new way.

The new way I want to outline is the way proposed by a third Cambridge philosopher, Ludwig Wittgenstein, in the last lectures he gave before he resigned from his chair here. They were the lectures on Philosophical Psychology given in 1946-7, attended by people like Peter Geach, now Professor of Philosophy at Leeds, and Norman Malcolm, Professor at Cornell. Geach took notes. Unfortunately they have not been published and I do not intend to quote from them. There are passages in Wittgenstein's published works in which he makes the same or similar points.

The first and possibly the most important point for our purposes is this. Psychological verbs—that is, verbs like "believe", "expect", "hope", "imagine", "intend", "know", "remember"— are characterized, Wittgenstein says, "by the fact that the third person of the present is to be verified by observation, the first person not".'8 In other words there is an asymmetry between the third person singular present tense use of a psychological verb — for example, "He expects ...' — and the first person singular present tense use —"I expect". That someone else expects something IS something I find out about by observation; that I expect something is not something I find out about by observation. It is neither something I 'ind out about by outer observation, nor something I find out about by inner observation. It is not something I find out about, and a fortiori not something I find out about in this way or that. If I say to someone "I expect he'll be here in a moment" and they say "How do you know?", they are asking how I know, or why I think, he will be here in a moment. It would be perverse in the extreme to take their question as meaning: how_do I know I am expecting, as opposed to believing, hoping, imagining, intending, remembering? I do not need any internal evidence to use the word "expect". If I say to someone ''I hope you'll come to the dance", and he, having two left feet so far as dancing goes, says "Are you sure?", he is questioning my sincerity—do I really want him to come? —not my powers of recognizing the psychological state I am in. I do not need to have Something to go on, to say "I hope you'll come' '. Norman Malcolm has. a rather nice phrase for this— "the autonomous status of self-testimony".'9

YOU may be thinking: What has the autonomous status of self—testimony got to do with privacy? Well, private is opposed to public. Consider the two utterances "I'm longing for a cigarette" and "I'm over eleven stone in weight ._About the latter someone may say: "You're not; your scales need adjusting . The fact that it is my weight that is in question does not mean that what I say goes. But with my longings and such like, it is different. What I say goes. And it is a large part of treating a person as a person that we accept this; that is, that we treat people as having the first and last word about certain things. It is the reluctance of behaviourists like B.2F. Skinner to accept this that makes us think he treats people as things. 0 The privacy that matters is not that of having access to private Objects of some sort: it is that of being treated as a person as opposed to being treated as an object among other objects.

There is one big question that I have left unanswered. According to the privileged-access doctrine, we know about the difference between believing, expecting, hoping, imagining, intending, knowing, remembering, and so on, through some sort of inner perception. They are phenomenally different acts of consciousness, and we have simply to attend to the "mental phenomena", as Brentano calls them. If the privileged—access doctrine is a false doctrine, if it is a myth that the words "believe", "expect", "hope", "know", "remember", etc., have meaning by being names of introspectible mental processes," then how do they have meaning? How do we distinguish between them?

At this point there is an awful temptation to take an easy way out. It is very tempting to fall back on the distinction between inner observation and outer observation, and to say that if we do not know about these things by inner observation then we must know about them by outer observation. That is, they must be words for various kinds of behaviour. We get the concept of expecting, for instance, from observing expectant behaviour. Along with this answer goes a causal interpretation of intentionality. To say that someone's longing is a longing for a cigarette is to predict that a cigarette will satisfy it, at least temporarily.

Wittgenstein rejected this answer as vehemently as he rejected the inner observation answer. I have heard it said that Wittgenstein was a behaviourist. Nonsense! To say that he was a behaviourist is to ignore all that he says about grammar, about what he calls "language games", and about rules of language. In his 1946-7 lectures be explicitly contrasted any attempt to elucidate psychological concepts in terms of phenomena, whether inner or outer, with an elucidation in terms of the grammar of linguistic utterances. Unfortunately I have not left myself with time even to begin a summary of his teaching on these matters. In particular I cannot hope to convey anything of what he says about intentionality ("It is in language that it's all done"21). What I will do is to leave you with five quotations, in four of which the term "language-game" occurs, and a short story.

The quotations are these:

. . . a concept is in its element within the language—game."

The question is not one of explaining a language-game by means of our experiences, but of noting a language—game."

Look on the language—game as the primary thing. And look on the feelings, etc., as you look on a way of regarding the language-game, as interpretation."

. . . the term "1anguage-game" is meant to bring into prominence the fact that the speaking of language is part of an activity."

Words are deeds."

And the story is as follows. Once upon a time a Martian, with telepathic powers, landed on earth in his flying saucer. He put on his invisibility shield and set off to investigate the possibility of communicating with the natives. He came across children engaged in some activity with pieces of card. They talked quite a lot, but our Martian soon noticed that one word, "snap", was evidently regarded as particularly important. He decided that if he was to communicate with these strange beings he must learn what this word stood for. He had never heard of Wittgenstein, but back on Mars he had come under the influence of the great Martian philosopher, Lok Jon, who taught that words have meaning by standing for ideas, and that ideas are either ideas of sensation or ideas of reflection. Using his telepathic powers our Martian soon discovered that when one of the children used the word "snap" he often had an experience like the experience he, the Martian, had back home when he managed to jump clear across one of the canals —an experience of triumphant excitement. Perhaps the child was reporting this experience. And yet when the second child said "snap", a second after the first, the experience was more like the one he had when he fell short and landed in the liquid nitrogen. Perhaps, then, the word stood for some feature of the cards. He had noticed that the cards had marks on them, and it was not long before he realized that the word "snap" was used when the marks were the same. And yet that could not be it, for he noticed that if one child said "snap" very quickly, the second child instead of agreeing would look as if he did not agree at all with the description. So, in despair, our Martian left as silently as he had come, and went back to tell Lok Jon that the earthlings made language—like noises but that he could not make out what on earth the noises were meant to refer to.

Well, that is the story — a very simple one. The moral of the story is that not all words and expressions are used to refer to something. Consider the expressions "I know", "I hope", "I remember", "I mean" and "I understand". Are they used to refer to something? We have a metaphysical prepossession to say that they are~— that they are used to refer to various mental processes." So long as we are in the grip of that prepossession we shall continue to be worried about "the relationships between subjective experience and the objective world".

24. Ibid., §656.

25. Ibid., §23.

26. Philosophical Grammar, Pt. I, §131.

27. Zettel, §211.

Discussion

MACKAY:

It is not too strong to say (p. 25)that with my subjective experience, "What I say goes' '? This may be true as a matter of social convention; but ontologically we must surely recognize the possibility of lying—as in malingering, for example, or when a mischievous student subject fools an experimental psychologist. I would agree that the privacy that matters is not that of having access to private objects, but it is, I think, that of having private experiences about which we can (in principle) lie. If we knew more about the physiological correlates of specific experiences (e.g. of seeing red rather than seeing blue, or seeing one line as longer than another) we might even hope to check objectively the probable truthfulness of such reports. What is at issue is a question of fact: did he, or did he not, have the experience he reports? To "treat him as a person" is doubtless a necessary condition for gaining evidence on this question, but it is not of itself sufficient to resolve it.

VESEY:

I may say "I expect he'll be here soon" to someone, and then later say to someone else "I only said that to buoy her up; actually I had no idea whether he was coming or not". And I may say ''I hope you'll come to the party" to someone I hope will not come, simply to be pleasant to him. I do not wish to deny the possibility of my uttering sentences beginning "I expect . . ." or "I hope ..." and of what I say being in some sense wrong. But in what sense? That is the question. Is what I said wrong in a sense which supports the thesis that someone who says ''I expect ...", or "I hope .. .", meaning what he says, is right about something? Suppose someone says "I hope you'll come", not just to be pleasant.
What he says is not wrong, in the relevant sense. Is its not being wrong a matter of his not being wrong about something? Specifically, is it a matter of his being right—about his "subjective experience"?

My answer to this last question is "No". If I say "The carpet is blue" there is the question of whether I correctly perceived it. There is no similar question with "I hope you'll come". That is what I meant when I said, apropos of such utterances, "What I say goes". Such utterances are not reports, correct or incorrect, of something being experienced. The notion of private mental processes is the product of the idea that such utterances must be reports. And this, in turn, is one aspect of the idea that all linguistic utterances, except questions and commands, have the same linguistic role, that of stating what is the case.

JOSEPHSON:

What the organizers had in mind as a subject for the conference was the possibility of being able to produce a general theoretical framework from which could be derived details of the phenomenon of privacy, and not just a dictionary definition of that term.

I do not find your discussion of such words as "believe", "expect" and "hope" very helpful in regard to the problem of privacy of personal experience, as it seems to me that there is a straightforward way to deal with the problems you raise. Just as a hot body gives rise to thermal radiation whose colour gives an indication of its temperature, we can postulate that the nervous system, as a result of its prior exposure to verbal and non-verbal sensory input, generates under certain conditions verbal output which is a reflection of, and in a certain sense a description of, its internal state. The fact that I may not be able to observe the state of the nervous system directly (as opposed to being aware of the descriptions generated) is not of fundamental importance: I may similarly be able to tell the temperature of a hot body by its colour without being in a situation to feel the heat directly. With this idea in mind, I postulate that when sentences containing words such as "believe", "expect" and "hope" are spoken, there is a corresponding property of the nervous system, caused by its response to the given situation. The connection Wittgenstein proposed between non-observability and privacy seems to me to be fortuitous, and in my experience, when belief, expectation, hope are sufficiently intense they are observable, in the form of emotions.

There is a reason, I believe, why psychological states are often not observable. Unlike ordinary information, which may have to be analysed intellectually before being used, emotions (more visibly with emotions such as fear or hate) produce their effects on behaviour directly, without the mediating influence of the intellect. Therefore we tend just not to bother to attend to such happenings and are not aware of them.

# CHAPTER 2

## Regarding Consciousness

RICHARD L. GREGORY
Brain and Perception Laboratory, Medical School, University of Bristol

"Problems of consciousness" are often regarded as of philosophical but of no other interest. For neurologists problems of consciousness can be practical— requiring urgent decisions on disturbingly metaphysical grounds. A boy, whom I knew with affection, fell on his head with grievous injury and remained in coma for many months. Although unable to speak, or respond to those around him, he did have facial expressions. Sometimes he appeared distressed: perhaps he was in pain. Was he conscious, suffering? Our usual means for deciding were absent as he could not communicate: yet perhaps he was in dire need of pain-relieving measures. In such a case as this the extreme behaviourist creed that statements of "mental events" can be reduced to accounts of behaviour — that consciousness and all its works should be exorcised as the absurd "ghost in the machine" —seem irrelevant, even actively evil. They do not at all help to deal with normal life situations; less still such extreme human issues demanding deep understanding and urgent solution.

It is remarkable that anaesthetics are given to remove pain, and indeed all consciousness, while we have no understanding of what they do; or how they or consciousness are related to brain function. Possibly anaesthetics and analgesics will become significant research tools for discovering just which features of brain structure and function are especially associated with consciousness; and hopefully how they are related. The question, What is this relation?, is the classical Problem of Consciousness.

## HOW IS CONSCIOUSNESS RELATED TO BRAIN FUNCTION?

Is consciousness causally important? Does consciousness affect brain processes? We know that brain activity is affected by normal or artificial physical stimuli; this is clear from electrical brain recordings. We know also that some physical stimuli produce sensations. So the question may be put: "Is consciousness a kind of stimulus affecting the nervous system and behaviour?"

Are conscious states mental stimuli? This violates the accepted technical meaning of "stimuli", which are regarded in physiology and psychology as physical events affecting the nervous system. The notion of non-physical stimuli looks like a move from a different game; not like anything accepted in physiology or physics. Is consciousness, conceived in this way, so odd that it will never be accepted by any future physics? Or is consciousness odd in the way that magnetism is odd, fitting, if with some difficulty, into accepted physics? This issue forces us, I think, to consider what we can accept as science—and so what "paranormal" should be taken to mean. I shall discuss this now, for use later  in a rather different context.

## IS CONSCIOUSNESS PARANORMAL?

Telepathy and telekinesis are alleged phenomena which—if accepted as genuine phenomena and not conjuring tricks — may be accepted in one of two ways: (a) they could be explained by processes, or whatever, which though so far unrecognized would be acceptable to science; or (b) they are due to processes, or whatever, unacceptable to science. This second alternative subdivides into two kinds of unacceptable: (i) unacceptable now, in current science, (ii) unacceptable for ever, in any possible future science. To justify this last possibility (which is the strong meaning of " paranormal"), it would be necessary to show that something going on is in principle unacceptable, in any possible or conceivable natural science. Are there any examples? There are many cases of the weaker sense-phenomena moving from "unexplained" to "explained" as science changes. Examples are: lightning; the lode stone, or compass; movement of muscles; and very many other formerly mysterious but clearly established phenomena. These were not, however, generally considered as paranormal before they were explained in terms acceptable to natural science. What is odd about telepathy, telekinesis and pre-cognition is that they are "paranormal" in the sense that no such explanation is expected. Relegation to paranormal status seems to mean, not only that they cannot now be explained; but that they cannot in principle be explained within, or brought into, any conceivable acceptable science. Does consciousness fall into this category; though its "existence" (if this is the correct word) is not in serious doubt? If so, consciousness is a paranormal phenomenon.

No doubt this would be strengthened by establishing other phenomena as paranormal. Meanwhile, it seems wise to attack the claim that consciousness is a paranormal phenomenon; for this is to say that we cannot now (weak sense) or ever in the future (strong sense) explain consciousness in terms acceptable to science.

IS CONSCIOUSNESS BIOLOGICALLY SIGNIFICANT?

If consciousness affects matter (especially brain states) then is it causal —in the sense that, for example, hitting a nail with a hammer affects the nail? If it has no such effect (no effect on brain or behaviour) why should consciousness have developed biologically? If we suppose that sticks and stones are conscious this would not be a special problem; but given that only organisms are conscious, we must suppose that consciousness has developed in organic evolution — and so it should have survival value. But how can it have survival value if it has no causal effects on behaviour? Any account which supposes that consciousness has biological significance must surely suppose that it causally affects some matter: which implies either that it can (like magnetism) be incorporated into physics though it is odd; or that it is (in the strong sense) paranormal. Personally, I do not see how we can predict future science (or lay down defining criteria for what is for ever to be "acceptable" science) to preclude "paranormal" from being "normalized", by incorporation into some future science. But if consciousness has no causal roles to play then science would have to accept non—causal entities or processes; and neo—Darwinians would have to accept features developed by natural selection which, though not causal, have survival value.

A way out from this is to suppose that consciousness is a biological fluke. But it seems to have developed through many species over a long biological time span; so this is unlikely. A further way out may be to say that consciousness is an accidental property (or, similarly, an emergent property) of complex neural tissue or function— or of whatever increases physically at the top end of evolution. To deny causal significance to consciousness is likely to be dull indeed. If consciousness is a logically necessary characteristic of high complexity, or some such, this would be formally like: "This x is an extended object, so x must be coloured", though we may not know the colour, and the colour does not matter. This would parallel a logically necessary (rather than a contingent) status to consciousness, which would be hardly more interesting, at least for biologists, than supposing it to be an inconsequential ineffective one—off fluke having no effects. All this is, however, to assume that something without effects is pointless. Do we understand cause sufficiently clearly to make this claim?

THE CONCEPT OF CAUSE

As David Hume pointed out, the notion cause is not plain. It is more than sequence, or correlation; for causes have, or at least are supposed to have, one-way direction in time. Thus if event A causes event B, A must have occurred before B. There may, however, be a common ancestral causal event, C. If C causes A and B, then A and B may occur simultaneously, or in either order of precedence. So causes cannot be simply inferred from observed sequences: though particular causal hypotheses may be ruled out by observed sequences.

There can be sequences or regularities for which no cause is attributable. A classical case is, day following night. Day does not cause night, nor night cause day, yet they follow in sequence (so far) without exception. We understand this from our conceptual model of the solar system - the earth rotating and so on. Within this model it is absurd to say that day causes night, or that night causes day. They are not causally related; in the way that a nail is driven by a hammer, or the hammer becomes worn or dented by hitting nails. Neither is there a simple ancestral common cause. We might say that day is "caused" by the sun shining on a given region of the earth, and that the sequence day—night is given by the earth's rotation with respect to the sun. We must appreciate the complex solar system model to ascribe any such causal relation, linking day with night. This inspires the thought that perhaps we cannot link brain states and consciousness causally except by some general model, which we do not as yet appreciate. This would be an adequate account of psycho-physics. This model might look very odd indeed, and yet be acceptable to science, if we may judge from weird current accounts of particle physics invoking extremely counter-intuitive concepts and linking relations which indeed are hardly causal in any traditional sense. Though extremely odd they are (though I do not entirely understand why) acceptable science. This is so for the more familiar statistical regularities of matter from random (and perhaps uncaused) quantal jumps of sub-atomic particles. Indeed, randomness is itself a tricky issue, if only because it can be attributed to (1) lack of evidence of cause; or (2) very many causal influences; or (3) no causal influences. Quantal jumps of sub-atornic particles are generally accepted as individually uncaused; but, when averaged in large populations, they give macroscopic stability and predictable lawfulness to objects. A candle flame is a beautiful example of statistics of quantal jumps of energy producing the appearance of an object, with sharp outlines (the flame), from a continuous temperature gradient.

Interestingly, the apparent size of the flame is set by the spectral sensitivity characteristic of our eyes. The same considerations apply to the apparent size of the sun as photographed at different wavelengths. I consider that such examples go to show that what seem objects of the physical world—as well as what seems causal and caused—is no simple matter; but are given by matching characteristics of data and observer. It is this relation which determines our consciousness of reality.

The importance for object-perception of this kind of matching applies to time as well as space. This is clear from the dramatic change of appearance of speeded—up or slowed—down film (say of the movements of clouds) which shows that what are taken as objects is very different when the time-scale is changed, with respect to the temporal characteristics of our visual sensory channel. This, in turn, affects how we attribute cause — for cause is applied to what are accepted as separate, but interacting, objects. This has relevance to the classical consciousness problem: Does consciousness interact with brain states: or are they ultimately the same? There is, of course, far more to how we classify patterned stimuli as representing more or less separate objects: this includes stimulus pattern or shape characteristics, and it involves the whole of perceptual learning. This issue has conceptual relevance to the classical consciousness problems: Does consciousness, or do states of consciousness, interact with brain states, or are they ultimately the same? It is relevant also to how conscious states are structured; and to how we should think of one state affecting another which, following Hume, is a central question in controversies on the nature of self-identity.

To say that consciousness and certain brain states are the same (and so have no causal relations) is some kind of identity theory. To say that consciousness and brain states are separate implies that there is some kind of relation between them —which may or may not be causal. If causal, it might be two-way interaction or one-way control, by brain or by mind. So we have a number of possibilities to examine, and for each we should consider whether there is or could be empirical evidence. I shall not go exhaustively (and exhaustingly) through all these; but will extend a physical analogy suggested by Leibniz, which may be of some help at least as an aide-mémoire.

## KINDS OF MIND—BRAIN PARALLELISM COMPARED WITH SYNCHRONIZED CLOCKS

Consider two clocks which keep the same time. They thus run 'in parallel", in time. It is time-parallelism which is suggested for mind—brain (or psycho—physical) parallelism. Brain states and consciousness may appear to be quite different — so we may imagine our clocks as very different though they run parallel in time. There are several ways of achieving parallel-time clocks. Each suggests a relation to be considered for mind-brain parallelism. The relations are between Masters, Slaves and Repeaters — by analogy with electrically synchronized clocks. It is worth noting that all these clock systems do exist. These are the three kinds of clock:

1. Master clocks: which are autonomous timekeepers.

2. Slave clocks: which are autonomous timekeepers, but receive frequent corrections from the Master.

3. Repeater dials: which follow or count pulses (including domestic electric clocks run off the mains). They depend on uninterrupted Master signals to keep going, but may have ''catch-up" mechanisms to restart them correctly.

Regarding Consciousness 37

Suppose that the brain and mind are separate entities, with physical and mental events running parallel in time, like a pair of our clocks. Suppose also that there may be some kind of causal link (analogous to electrical clock-pulses) to synchronize them. We can look at clock systems of these kinds and ask which (perhaps by assuming horological technology, or more general knowledge of interacting systems) would work best.

1. Master—Master: a pair of Master clocks keeping the same time, though without causal links of any kind. This is the most unlikely horologically, as in practice clocks always differ in their rates.

2. Master—Slave: the autonomous Master provides occasional corrections to the Slave. This is an autonomous clock, which is generally set to run either slightly fast or slow, so that the correcting signals are of only one sign. If the correction signals are lost, the Slave will continue running, though slow or fast.

3. Master—Repeater: a Master providing a steady flow of signals to a Repeater dial. If the signals are interrupted, when restarted the Repeater dial will always be slow by the lost time, until it is reset. Resetting may be automatic.

4. Slave—Slave.' there might be a pair of (or better very many) interacting Slaves, each correcting the others so that there is a pooled average time. There would be small varying discrepancies but on average they would agree with each other in this democracy.

The remaining logical possibilities remain empty, for neither Slave nor Repeater can drive a Repeater or a Master Clock, and one Master driving another denies the "Master" status. The Master—Repeater relation is most used, as it is simple and cheap. It has the grave disadvantage that a momentary loss of Master signal gives a permanent error, which must be corrected. The Master—Slave is interesting, in that the two clocks are not continually but only on average synchronized.

The least likely alternative is (l) the Master——Master relation, in which neither affects the other. It is virtually impossible to get a pair of independent clocks to run in step, so we should not on this analogy expect independent mind—brain parallelism. (Of course one might use this for fanciful theories: for example that there is loss of synchronism with ageing—and that this explains increasing absent-mindedness and lack of contact with reality in old people.)

All these analogy—examples adopt the underlying assumption of a representational account of perception; and that Direct Realism is false. A Direct Realist horological example would be sundials. They keep time by direct "contact" with an aspect of the physical world, which is accepted as time. Since adopting atomic time standards, and rejecting the earth's rotation and so the sun's shadow as time, sundials now can —and do—run fast or slow (apart from the equation of time correction for mean solar time). So the realist account would now be given by atomic clocks rather than by sundials. On this analogy with sundials or atomic clocks consciousness is selections of physical reality: features of physical reality are held to be conscious. This gives the brain no role beyond that of a passive filter. I assume here that this doctrine is false: that the brain does have some special importance for consciousness and that other objects are not conscious, and do not confer consciousness.

This might be an empirical issue for experiment. Indeed I think one can give strong empirical evidence against Direct Realist theories of perception; though I shall not give the evidence here.

EVIDENCE FOR CAUSAL EFFECTS OF CONSCIOUSNESS

We seldom doubt that physical events, especially stimuli of various kinds, affect consciousness. The sensation of a pin stuck in the finger is sufficient example. It is not, however, at all so clear that this sensation, or any other, has any effect. Indeed, the sensation of pain generally, and perhaps always, comes too late to serve as a cause of action—such as withdrawing the hand from hurt. We start to feel the pain after the (reflex) withdrawal of the hand. Is there any evidence for consciousness affecting behaviour?

William James discusses mind affecting brain in The Principles of Psychology (1890), under the heading "The Intimate Nature of the Attentive Process". On page 434 he gives as examples (his italics):

1. The accommodation or adjustment of the sensory organs

and

2. The anticipatory preparation from within of the ideational centres concerned with the object of which the attention is paid.

The point he makes here is that the eyes are adjusted prior to (in anticipation of) what visual signals will be needed; so their movements are not always controlled by physical stimuli. This may occur in darkness — in the absence of any visual stimuli — according to purely internal processes of mental images.

The question is: Do we know that it is mental events which are moving the eyes; or is it internal physical events? This question must be asked, for it is clear that there are physical (physiological) processes capable of moving the eyes from within. (If everything inside the skull could be said to be mental, the situation would be much simpler!)

William James goes on to consider various perceptual examples of what he thinks might be mind controlling brain. He cites an interesting observation of Helmholtz's, which I am not sure has been investigated since. Helmholtz found that simple stereograms could be made to fuse by will, when presented as after-images from the flash of a single spark, so that eye movements are ineffective. Helmholtz says: "If I chance to gain a lively mental image of the represented solid form (a thing that often occurs by lucky chance) I then move my two eyes with perfect certainty over the figure without the picture separating again." This must be a "central" effect because the after-images are stuck to the eyes. -

William James, however, is careful to point out that it could be other physical brain processes which are affecting the fusion of the images — in which case such effects would be no evidence for mind affecting brain. I shall quote James in full here from Principles (p. 447):

When, a few pages back, I symbolized the ldeational preparation element in attention by a brain—cell played upon from within, I added "by other brain-cells, or by some spiritual force' ' without deciding which. The question ' 'which? ' ' is one of those central psychological mysteries which part the schools. When we reflect that the turnings of our attention form the nucleus of our inner self; when we see that volition is nothing but attention; when we believe that our autonomy in the midst of nature depends on our not being pure effect, but a cause — Principium quoddam quod fait foedera rumpat, Ex infinito ne causam causa sequatur — we must admit that the question whether attention involves such a principle of spiritual activity or not is metaphysical as well as psychological, and is well worthy of all the pains we can bestow on its solution. It is in fact the pivotal question of metaphysics, the very hinge on which our picture of the world shall swing from materialism, fatalism, monism, towards spiritualism, freedom, pluralism, — or else the other way.

James himself inclines to thinking that attention and will do have significant effects — especially (p. 453): "It would deepen and prolong the stay in consciousness of innumerable ideas which else would fade more quickly away."

James ends this discussion by objecting to the materialist analogy with the sense of will occurring during difficulty as being merely physical (like the turbulence of rivers in constricted regions) (p. 454). He waxes eloquent:

Meanwhile, in view of the strange arrogance with which the wildest materialistic speculations persist in calling themselves "science", it is well to recall just what the reasoning is, by which the effect—theory of attention is confirmed. It is an argument from analogy, drawn from rivers, reflex actions and other material phenomena where no consciousness appears to exist at all, and extended to cases where consciousness seems the phenomenon's essential feature. The consciousness doesn't count, these reasoners say; it doesn't exist for science, it is nil; you mustn't think about it at all.
The intensely reckless character of all this needs no comment. . . . For the sake of that theory we make inductions from phenomena to others that are startlingly unlike them; and we assume that a complication which Nature has introduced (the presence of feeling and of effort, namely) is not worthy of scientific recognition at all. Such conduct may conceivably be wise, though I doubt it; but scientific, as contrasted with metaphysical, it cannot seriously he called.

Whatever the reader makes of William James's impassioned prose, in favour of mind controlling brain, it may be agreed that he states very clearly this traditional view: that consciousness has significant if small effects on perceiving, thinking and behaviour.

This view has recently been defended by Sir Karl Popper and Sir Jack Eccles, in The Self and its Brain (1977). They take two cases of what they regard as evidence for mind affecting brain:

1. The reports of people having their brains electrically stimulated while undergoing brain surgery, who experience streams of memories of other mental images, and at the same time are aware that they are in the operating theatre.

2. Speeding or slowing down of Necker cube reversals by act of will.

These are taken as evidence — but without explicit reference to the alternative which James considers, though does not like — that other physical brain processes produce or inhibit changes in the Necker cube reversals, or whatever; and that the two experiences of the brain—operation patient are given by two sets of physical brain processes. Actually it is not at all clear to me why this particular example is supposed by Popper and Eccles to have special power to persuade us; I can read a book, listen to the radio and feel hungry at the same time: which of these is supposed mental, which physical? They should have considered not only two, but three or more simultaneous experiences or activities; do we have a mind (or brain) for each?

For Popper and Eccles mind and brain are separate, with weak and slow-acting control of brain by mind. Thus, Popper writing on page 514:

. . . there are two kinds of illusions — illusions delivered to us or imposed upon us by the brain; and illusions which have a mental origin, let us say. wish-fulfilment. It is apparently built into our organism and into the whole "mechanism of interaction" between the brain and the mind that the mind should be in many respects dependent on the brain, in order not to fall too easily into that kind of illusion which we experience in fantasy.

I would say that this whole field can be used to show at the same time a kind of gulf and also a kind of dependence between the self-conscious mind and the brain.

Do illusions give us the right to make such statements? Granted we can perceive one thing and know at the same time that this perception is false —but surely it by no means follows that one of these, perception or knowledge, is mental and the other physical. Indeed computers could not be given check procedures involving recognizing discrepancies if this were so. All we can infer is that the brain can process more than one thing at a time, and that discrepancies can be noted—as when we recognize an illusion. Of course this could be rewritten "the mind can process more than one thing at a time . . ." but this again gives no reason for saying that there is evidence here of mind and brain as separate, interacting entities. I conclude that James's alternative (which he disliked) is in no way discounted by such evidence. In every case. it seems, one can argue that it is some other brain process which intervenes —not mind or con-sciousness.

WHEN ARE WE MOST AWARE?

It is worth considering under which conditions we are most conscious. or aware —and relate these to our ability at skills and how we behave. I think I am most aware at surprise, and when things go wrong. Thus when driving a car I am scarcely conscious of the situation until something surprising happens. Consciousness seems to be associated with mismatches between predictions and events as signalled. If this is correct, and if consciousness does have causal effects, I suppose it might be supposed to select fresh predictive models— "internal models" as Kenneth Craik called them — or fresh hypotheses of reality or fiction. But again it is not at all clear that this observed association is evidence of a causal relation; or if it is causal, in which direction the cause goes. There is no reason to suppose that awareness causes changes of predictive models or perceptual hypotheses even if we are most aware in these situations.

Should we look beyond normal brain function for evidence of mind affecting matter—to alleged paranormal phenomena? I think there are severe logical problems in such an undertaking, at least before we are clearer what we should mean by "paranormal". This we shall now briefly discuss.

## EVIDENCE OF CAUSAL CONSCIOUSNESS FROM PARANORMAL PHENOMENA

If it could be shown that mind affects matter other than via the nervous system, would we have evidence for causal mind?

There is not, it seems, any clear evidence from neurology that behaviour is controlled by a separate interacting mind or consciousness: so why should breaking the neural link —as for telekinesis where distant objects unconnected by nerve fibres, or the body, are supposed to move by mental cause — give better evidence?

Returning to our discussion of the meaning of "paranormal" (p. 32) we distinguished between a weak and a strong sense. The weak sense concerns phenomena which are odd and difficult to explain, but which might receive explanation within present or conceivable future science. The strong sense of "paranormal" concerns phenomena which are still odder and more difficult to explain, for it is being claimed that explanation is not possible within present or any future science. This at once raises the question: can we make the claim with certainty that phenomena cannot be explained by any future science? I fail to understand how such a claim could be justified. It is tantamount to saying that only limited paradigm shifts of science are possible: but how can this be shown? After all, there have been quite remarkable paradigm shifts during this present century.

What the strong sense amounts to, I think, is that we cannot at present see how such alleged phenomena can be reconciled within current or predictable future paradigms of science. This claim may depend purely upon limited imagination. Even if telepathy, telekinesis, or whatever were shown convincingly to occur (and at present I am not convinced), the phenomena could not, without highly questionable assumptions, be used to justify mind affecting brain. This is more than a question of whether this is the best hypothesis (which holds for interpretations of all observations and empirical claims) for a sufficiently drastic paradigm change might change "paranormal" to "normal". If I am right that claims of paranormal can only be made within fixed paradigm systems, I do not see how any inferences from what are claimed as paranormal can be made: for the (strong and weak senses of) "paranormal" preclude inferences within accepted data-bases and inference structures.

## IDENTITY THEORIES

Although the notion of mind as being different and essentially separate from, though somehow causally linked to, the brain is the traditional view through the recorded history of philosophy and religions, there is remarkably little if indeed any evidence for it, and there are severe conceptual difficulties. It would be much neater to suppose that mind and brain are essentially one—different aspects of the same thing. To show how this could be so is the aim of Identity Theories.

I shall introduce the notion with the words not of a philosopher, but of a nineteenth-century naturalist. who was a personal friend of Darwin, and who concerned himself with the evolution of mind—George John Romanes (1848-94). In his Mind and Motion (1885), Romanes writes (see Body and Mind (1964), ed. G. Vesey, p. 183):

We have only to suppose that the antithesis between mind and motion — subject and object—is itself phenomenal or apparent: not absolute or real. We have only to suppose that the seeming quality is relative to our modes of apprehension; and, therefore, that any change taking place in the mind, and any corresponding change taking place in the brain, are really not two changes, but one change.

This is a remarkably clear statement of the mind—brain identity notion. Romanes continues:

When a violin is played upon we hear a musical sound and at the same time we see a vibration of the strings. Relatively to our consciousness, therefore, we have here two sets of changes, which appear to be very different in kind: and yet we know that in an absolute sense they are one and the same: we know that the diversity in consciousness is created only by the difference in our modes of perceiving the same event — whether we see or whether we hear the vibration of the strings. Similarly we may suppose that a vibration of nerve-strings and a process of thought is really one and the same event, which is dual or diverse to our modes of perceiving it.

The great advantage of this theory is that it supposes only one stream of causation, in which both mind and motion are simultaneously concerned.

Romanes regards this identity as an hypothesis— for which there is no evidence but which is for him "the only one which is logically possible, and at the same time competent to satisfy all the facts alike of the outer and the inner world' ".

This is in many ways a highly attractive account. It does, however, raise questions which are far from resolved. In the first place, much clearly hinges here on what we mean by "identity". It cannot be taken to mean that everything we say of brain states we can say of consciousness. The problem indeed is to find anything in common! So this is no "surface" identity. What criteria for identity should, then, be satisfied for brain states to be accepted as identical with conscious states? If we have to accept criteria for "identity" which would be accepted in other cases (such as electron flow and lightning, or electro-magnetic radiation and light) just how like such cases does the mind—consciousness relation have to be to be accepted as an "identity"?

If it is a unique case — and this is the trouble about consciousness — it is helpful to apply criteria taken by such analogies? At least two criteria for "identity" between two things (A and B) would normally be demanded. First, that there are precisely related time relations between the occurrences, or changes, in A and in B. Secondly, that A and B are coincident in space. Whether there are exact time relations between brain states and consciousness seems to be an empirical question which might be answered by experiment. Spatial coincidence of brain states and consciousness poses a deeper problem; for it seems misleading to say that consciousness occupies space. How then can brain states and consciousness be identical, if one occupies space and the other does not! Could this, though, be a case where we take over criteria of "identity" from common physical object examples which are inappropriate for this peculiar case?

Since the brain exists in space but experiences do not, for the identity theory to be accepted, should we relax the usual spatial requirement for identity? We do often allow that two things can be in some ways identical though they have spatial differences. For example, a brick and a pail of water may have the same weight. Their weights are identical though their shapes, etc., are quite different. Can mind—brain identity be such a non-spatial identity? This would require that some non-spatial characteristic of the brain is identical with consciousness. For the brick and bucket of water we found a common non-spatial property (weight) and many others could be suggested —so non-spatial identity of consciousness and brain would not be a unique kind of identity. What, then, should we suppose is identical?

## DO WE LIVE IN QUOTATION MARKS?

It does seem clear that brain states symbolize events and concepts; so they are like words in a book. A kind of identity which I consider should be explored here is between symbol and symbolized. It is not identity in a wide strict sense: it is rather a stands for, or an equivalent to, relation. The sentence: "there are six beans in this box" stands for, and for some purposes is equivalent to, the box with six beans in it, provided that the sentence can be read. Perhaps consciousness is reading the world. The nearest I can come to an understanding is to say that: the brain puts reality into quotation marks. We seem to live inside our brain's quotations.

## MACHINE CONSCIOUSNESS

Mind seems more mysterious than matter; but if we ask "What is matter?" do we get an answer? If we ask "What are the ultimate particles of matter made of?" we get no more of an answer than we do when we ask "What is mind?". We can, however, say a great deal about relational aspects of matter: about generalizations and laws which make predictions possible; and especially conceptual causal models, which give unique intellectual satisfaction. It is this which is absent from accounts of mind and consciousness. Most accounts of mind are analogies with accounts of physical substances and the sometimes surprising "emergent" properties of, especially, chemical combinations of atoms into other substances having (at least on inadequate accounts) surprising properties which seem to pop up mind-like: as ideas, inventions, and indeed perceptions seem to emerge from situations. Ideas are sometimes seen as embedded in inexperienced "mind substance". Physics is, however, more concerned with structures than substance. In physics, surely, the term "substance'' is as mysterious and probably as meaningless as the term "mind'' conceived as an underlying primeval glue, sticking bits of consciousness and behaviour together to give the self-identity of a person. Hume was surely right to reject this, but until we have functional models approaching the adequacy of accounts in engineering, mind must appear unexplained. The best hope for developing precise and detailed models of mind seems now to be the procedures for problem solving developed for "artificial intelligence" of robot machines. As they begin to solve problems we find difficult (and even impossible), and as they begin to learn, and recognize objects with their television-camera eyes, so we must ask, "Will they become conscious?"

Philosophers might be persuaded that they are, if the machines spend time speculating on whether we are conscious. For the common man, it may depend, rather, on whether they share jokes and opinions which inspire our loyalty. Animals and humans which do not share these are doubtfully conscious.

Suppose, though, that A.I. machines prove never to be highly successful. Would this be attributed to their lack of consciousness? This could be evidence that consciousness is causally important in us. This might be shown if A.I. reaches a ceiling too low, and unexplainable by limited computing power and concepts of intelligence. If, on the other hand, A.I. machines do come to rival us, and we hold that they are not conscious, then it will be hard to hold that consciousness has causal effects for us—at least for anything for which machine performance rivals ours. This would, say, rule out consciousness as important for problem-solving, though it might allow consciousness the role of setting up aesthetic preferences and goals and action. Such conclusions could follow empirically from successes or failures of machine replications of human capacities to decide, solve and do.

Meanwhile, I see no reason to suppose that consciousness is a separate entity, affecting brain and behaviour. Some kind of identity with brain function seems a better bet; but if so, it is a limited identity and hard to specify. One might guess that it is something to do with symbols and how they symbolize. If this is so (and the notion is vague) we may have to suspect that future machines, capable of rivalling us by the power of symbols, will be our conscious brothers.

Discussion

VESEY:

I have two related comments, both about the problem of "the gap between matter and mind". The first is about one of the supposed solutions to the problem, the so-called "Identity Theory". The second is much more general.

You said something about "the status of verification criteria" for the identity theory. As you know, the people who hold the theory say that the identity in question is an empirical or contingent one, like the identity of a flash of lightning and an electrical discharge (J. J. C. Smart, Philosophical Review, LXVIII, 141-56, 1959). Evidently you don't think that answers your question about the status of the theory. Why not? More specifically, would you agree with the following criticism of the assimilation of mind-brain identity to lightning—electricity identity? The statement "Sensations are identical with brain processes" is about a whole philosophical category of things. It isn't just about, say, sensations of being tickled. But the statement "Flashes of lightning are identical with electrical discharges" is not about a whole, philosophical category of things. It isn't about the whole class of things in the ordinarily accepted world of experience—as opposed to that of things in the world of the physical sciences. It is not even about the whole class of visible phenomena. In short, the two statements are not on a par. A closer parallel to "Sensations are identical with brain processes" would be "Things in the ordinarily accepted world of everyday experience are identical with things in the world of the physical sciences". But to say that is to let the cat—meaning the status question—well and truly out of the bag. If "Sensations are identical with brain processes" is like "Things in the ordinarily accepted world, etc.", is it a scientific hypothesis? (If so, what would falsify it?)
Or is it a methodological postulate? (If so, why would not a statement of isomorphism, or what used to be called "psycho-physical parallelism", serve as well?) And so on.

My second, much more general, comment is intended to undercut the whole endeavour in which identity theorists and other are engaged. It occurs to me that if I were to ask you a question about the concept of perception — for example, "Should we think of perception in stimulus-response terms?" —- you would not be at a loss for an answer. It is only when a question like "What is the relation of mind and matter?" is asked that people don't know what to say. This makes me wonder whether the fault is not in the formulation of the question — implying, as it does, that we know something the word "mind" stands for, and know something else the word "matter" stands for, but somehow can't penetrate to how the two things are connected. Do you share my feeling that if we have an adequate conceptual understanding of things that distinguish people from lumps of matter —I mean things like their being able to perceive things, and to do things for which they can be held morally responsible, and, more than anything else, to enter into conversation with us — we know all there is to know about the concept of mind? Do you, like me, think that the idea that there is some sort of higher-order truth about the relationship of two substances, mind and matter, is a myth left over from our Cartesian past?

# CHAPTER 3

# Is Consciousness a Phenomenon?

H.C. LONGUET—HIGGINS

University of Sussex

In this short contribution to our discussion of "Consciousness and the Physical World" I do not propose to offer solutions to any scientific problems about consciousness, but merely to make some observations on how we use the word "conscious", and on whether consciousness can legitimately be regarded as a "phenomenon" in the same sense as gravity or morphogenesis, to be explained in ordinary scientific terms.

I completely agree with Godfrey Vesey, and Wittgenstein before him, that many intellectual headaches are due to negligence about the use of words, and can be dispelled by proper attention to the everyday use of language. The word "consciousness" occurs most naturally in contexts such as "I lost consciousness" or "He regained consciousness", where the state of consciousness is clearly being contrasted with more passive states such as sleep, coma or trance. When we visit a seriously injured person, and cannot tell whether he is conscious or not, what is the nature of our concern? We are wondering, surely, whether he is aware of what is going on around him—whether he is having experiences, pleasant or painful, which he might subsequently be able to recall. If, on a later occasion, he can accurately report events in which he was involved at a particular time, then we have no doubt that he was conscious at that time. So although the ability to commit experience to memory may not suffice to define the conscious state, it does seem to be a peculiarly characteristic property of that state.

If consciousness is hard to define, self-consciousness is even harder. But the commonplace sentiment "I was acutely self-conscious" points the way to some relevant considerations. It indicates that the speaker was observing himself and his actions in a way similar to that in which other people might be observing him. Most people feel sure that monkeys, for example, must be conscious, and possibly even self-conscious. But any satisfactory definition of consciousness, or of self—consciousness, should in principle be applicable to any system, biological or other, which was capable of processing information. If someone were to design an apparently intelligent robot, the definition ought to enable us to decide whether or not the robot was conscious, by studying in detail the programme which controlled it. At present we have only the haziest notions as to what criteria might be relevant, but presumably they would have to be couched in logical or psychological terms, rather than in the language of electronics or solid-state physics. Presumably the system would have to possess a memory, both of its experiences and of its own responses to those experiences; presumably also. its representation of the world would have to include a representation of the robot itself, for it to meet the criteria of self-consciousness. But there would be formidable problems of principle, to do with the appropriateness of any such psychological account of the physical processes taking place inside it; and the implementation of such proposals is a task altogether beyond the scope of present achievements in "artificial intelligence".

A quite different approach to the concept of consciousness takes as its starting-point the theory of observation, as often propounded in connection with the interpretation of modern physics, especially quantum mechanics. The orthodox view is that the complete "Laplacian" account of physical reality is a myth, and that all we can hope for is statistical laws which correlate descriptions of the world at different times. These descriptions must be couched in terms of "observations" (of complete sets of commuting observables). What intervenes between two states so specified is amenable to mathematical calculation but not to observation; the only "phenomena" admitted by the theory are the observations themselves, and the concept of an observation seems to depend crucially on the concept of an observer. So if consciousness is that state of being which enables the observer to observe, it must belong to a different ontological category from anything that he observes, and cannot be classified as a physical phenomenon, in the strict sense of that term.

The latter part of this argument is my own gloss on the orthodox theory of observation, but comes close to the view expounded by Heisenberg in his book The Philosophy of Physics. But whether or not it stands up to critical examination, it does suggest that attempts to construct a scientific account of consciousness may be doomed to failure from the start. We may succeed in understanding, in evolutionary terms, how creatures have evolved which can evidently commit their experiences to memory and thereby profit from their failures and successes, but that is an altogether different enterprise from trying to describe a subjective state in objective terms.

When this meeting was first being planned, its provisional title was "Possible Effects of Consciousness on the Physical World". So let me devote the rest of this paper to some issues which that title suggests.

First, it is undeniable that human beings affect the world all the time, not only in accidental ways, such as exhaling carbon dioxide, but also by design — through conscious decisions, translated into action. Unless one is a Cartesian dualist, perplexed as to how the mind can affect the body, there need be no mystery, in principle, about our ability to do things on purpose. Have we ourselves not designed machines which, under the control of computer programmes, can respond in quite complex ways to stimuli from their environments; and as for our own bodies, do we not possess brains which, beyond doubt, carry out the logical processes which we describe as our thoughts? It is unnecessary, and solves no problems, to postulate the existence of a ' 'homunculus' ' sitting at the controls of the brain (possibly somewhere near the pineal gland) and transforming the aspirations of the soul into physical stimuli acting on the brain; the required transformations would be just as problematical as the mind-body interaction hypothesis itself.

One need not suppose, then, that the microscopic cerebral events which mediate consciousness are any different physically from those which have been studied experimentally in similar systems. But those events are of no particular interest in themselves, except to a neurophysiologist or neurochemist. What concerns us as human beings is their collective outcome, which we can only interpret in terms of concepts such as motive, intention, decision and action. An action, as Vesey reminds us, is much more than a complex train of physical events, it is something that a person does, and something we may or may not hold him responsible for.

The exercise of the will, which we normally regard as a manifestation of consciousness, presents the psychologist — and the philosopher — with a number of difficult problems. Certain sorts of behaviour for which people used to be held responsible are now seen as unconscious or uncontrollable; other kinds of activity, conventionally classified as ' 'autonomic' ' , are now known to be accessible to voluntary control. There is much evidence that people can be trained to control their pulse rates and skin responses, and to weep spontaneously, though at the present time the physiological mechanisms are quite obscure.

But I suspect that more may have been in the minds of the organizers: perhaps such putative phenomena as levitation, psychokinesis, telepathy and clairvoyance. It would, of course, be unhelpful merely to dismiss such claims as founded on delusion, deceit or experimental incompetence. But the fact remains that a mere set of observations does not constitute a natural phenomenon. To establish a new phenomenon which contravenes accepted laws demands that the relevant observations be exhibited as instances of a clearly stated generalization, and that a convincing reason be given for accepting this generalization in the face of the evidence for the laws in question. And finally, the authentication of one or more "psychic" phenomena would not put an end to the matter: we should then be hard put to it to understand what need human beings have for hands, feet, eyes, ears and voices.

Discussion

VESEY:

The first sense of the word "consciousness" you mentioned was that in which the state of consciousness is opposed to states such as sleep, coma or trance. I suppose there may be borderline cases but, for the most part, I think, we know where we stand with that sense of "consciousness". That is, we can usually tell whether someone is conscious or not; and the question "What is the use of being conscious?" has the obvious answer "Well, if everyone were always asleep, or in a coma or trance, our days would be numbered". Now, two of our fellow symposiasts, Nick Humphrey and Horace Barlow, evidently regard the usefulness question as requiring some other, less obvious, answer. Presumably they are not using the term "consciousness" in the sense in which it is opposed to sleep, etc. So, in what sense are they using it'? How is their use of the term related to the one in which consciousness is opposed to sleep? What are the criteria of application of the term in their sense? Is one of the senses more basic than the other?

JOSEPHSON:

You say it solves no problems to postulate a "homunculus" sitting at the controls of the brain transforming the aspirations of the soul into physical stimuli acting on the brain. But if we are trying to understand a complex system, it is always helpful to try to subdivide it into components having particular roles, and the hypothesis you refer to may be an extremely useful one in the long run, even if it is far from leading to a complete solution of the problem when standing by itself. In biological systems we do this subdivision all the time (with conceptual components such as the circulatory system and the immune system), and even in artificial intelligence we find such divisions of function in the most advanced systems, such as Sussman's Conceptual Model of Skill Acquisition (HACKER). Separation of the total system into the knower, what he knows, and the consequences of that knowledge, may be the most important step we can take towards the understanding of the nature of intelligence.

PART II

Consciousness and Behaviour

# CHAPTER 4

## Nature's Psychologists*

N. K. HUMPHREY
University of Cambridge

On the temple at Delphi was written the stern message "Know thyself". Did the oracle realize she was uttering an evolutionary imperative? I shall argue presently that self-knowledge, and through it the possibility of "intuitive" knowledge of others, has made an essential contribution to the biological fitness of man and certain other social animals. The means to self-knowledge have consequently been promoted and perfected by selection. Within this argument lies a theory of the evolution of consciousness; within it, too, lie some humbler ideas about the evolution of overt behaviour.

In The Nature of Explanation' Kenneth Craik outlined an "Hypothesis on the nature of thought", proposing that "the nervous system is .. . a calculating machine capable of modelling or paralleling external events. ... If the organism carries a 'small-scale model' of external reality and of its own possible actions within its head, it is able to try out various alternatives, conclude which is the best of them, react to future situations before they arise, utilize the knowledge of past events in dealing with the future, and in every way to react in a much fuller, safer and more competent manner to the emergencies which face it." The notion of a "mental model of reality" has become in the years since so widely accepted that it has grown to be almost a cliche of experimental psychology. And like other cliches its meaning is no longer called in question. From the outset Craik's

---

'This paper is based on the Lister Lecture delivered at the B.A.A.S. meeting, September
1977.

"hypothesis" begged some fundamental questions: A model of reality? What reality? Whose reality?

My dog and I live in the same house. Do we share the same "reality"? Certainly we share the same physical environment, and most aspects of that physical environment are probably as real for one of us as for the other. Maybe our realities differ only in the trivial sense that we each know a few things about the house that the other does not—the dog (having a better nose than I) knows better the smell of the carpet, I (having a better pair of eyes) know better the colour of the curtain. Now, suppose my dog chews up the gas bill which is lying on the mat by the door. Is the reality of that event the same for him as me? Something real enough has happened for us both, and the same piece of paper is involved. The dog hangs his head in contrition. Is he contrite because he has chewed up the gas bill? What does a dog know about gas bills! Gas bills are an important part of my external reality, but they are surely none of his.

If mine and the dog's realities differ in this and other more important ways they do so because we have learned to conceptualize the world on different lines. To the dog paper is paper, to me it is newspaper or lavatory paper or greaseproof paper or a letter from my friend. These ways of looking at paper are essentially human ways, conditioned of course by culture, but a culture which is a product of a specifically human nature. I and the dog are involved with different aspects of reality because, at bottom, we are biologically adapted to lead different kinds of lives.

To all biological intents and purposes the portion of reality which matters to any particular animal is that portion of which it must have a working knowledge in the interests of its own survival. Because animals differ in their life-styles they face different kinds of "emergencies" and they must therefore have different kinds of knowledge if they are to react in the full, safe, competent manner which Craik — and natural selection- recommends.

But different kinds of knowledge entail different ways of knowing. In so far as animals are biologically adapted to deal specifically with their own portions of reality, so must their nervous "calculating machines" be adapted to construct very different kinds of models. This is not to say merely that the calculating machines may be required to do different kinds of sums, but rather that they may have to work according to quite different heuristic principles. Depending on the job for which Nature has designed them the nervous systems will differ in the kind of concepts they employ, the logical calculus they use, the laws of causation they assume, and so on. They will differ in what may properly be called their "ideologies". Ideology, in the sense I use the term, means simply a framework of ideas. Ideologies provide, if you like, the "conceptual language" in terms of which questions are asked, calculations made and answers given.

Let us call these nervous calculating machines "minds". It is the thesis of this paper that a revolutionary advance in the evolution of mind occurred when, for certain social animals, a new set of heuristic principles was devised to cope with the pressing need to model a special section of reality — the reality comprised by the behaviour of other kindred animals. The trick which Nature came up with was introspection; it proved possible for an individual to develop a model of the behaviour of others by reasoning by analogy from his own case, the facts of his own case being revealed to him through "examination of the contents of consciousness".

For man and other animals which live in complex social groups reality is in larger measure a "social reality". No other class of environmental objects approaches in biological significance those living bodies which constitute for a social animal its companions, playmates, rivals, teachers, foes. It depends on the bodies of other conspecific animals not merely for its immediate sustenance in infancy and its sexual fulfilment as an adult, but in one way or another for the success (or failure) of almost every enterprise it undertakes. In these circumstances the ability to model the behaviour of others in the social group has paramount survival value.

I have argued in more detail before now that the modelling of other animals' behaviour is not only the most important but also the most difficult task to which social animals must turn their minds? In retrospect I do not think I took my own case seriously enough. The task of modelling behaviour does indeed demand formidable intellectual skill —social animals have evolved for that reason to be the most intelligent of animals — but intelligence alone is not enough. If a social animal is to become — as it must become— one of "Nature's psychologists" it must somehow come up with the appropriate ideology for doing psychology; it must develop a fitting set of concepts and a fitting logic for dealing with a unique and uniquely elusive portion of reality.

The difficulties that arise from working with an inappropriate ideology are well enough illustrated by the history of the science of experimental psychology. For upwards of a hundred years academic psychologists have been attempting, by the "objective" methods of the physical sciences, to acquire precisely the kind of knowledge of behaviour which every social animal must have in order to survive. In so far as these psychologists have been strict "behaviourists" they have gone about their task as if they were studying the behaviour of billiard balls, basing their theoretical models entirely on concepts to which they could easily give public definition. And in so far as they have been strict behaviourists they have made slow progress. They have been held up again and again by their failure to develop a sufficiently rich or relevant framework of ideas. Concepts such as "habit strength", "drive", or "reinforcement", for all their objectivity are hopelessly inadequate to the task of modelling the subtleties of real behaviour. Indeed, I venture to suggest that if a rat's knowledge of the behaviour of other rats were to be limited to everything which behaviourists have discovered about rats to date, the rat would show so little understanding of its fellows that it would bungle disastrously every social interaction it engaged in; the prospects for a man similarly constrained would be still more dismal. And yet, as professional scientists, behaviourists have always had enormous advantages over an individual animal, being able to do controlled experiments, to subject their data to sophisticated statistical analysis, and above all to share the knowledge recorded in the scientific literature. By contrast, an animal in nature has only its own experience to go on, its own memory to record it and its own brief lifetime to acquire it. "Behaviourism" as a philosophy for the natural science of psychology could not, and presumably does not, fit the bill.

Chomsky in his famous review of Skinner's Verbal Behavior[3] argued on parallel lines that it would be impossible for a child to acquire an understanding of human spoken language if all the child had at its disposal was a clever brain with which to make an unprejudiced analysis of public utterances. Chomsky's way round the problem was to propose that the child's brain is not in fact unprejudiced: the child is born with an innate knowledge of transformational grammar, and this knowledge of the grammar provides it, in my terms, with the ideology for modelling human language. Though there are snags about Chomsky's thesis, it would not, I suppose, be wholly unreasonable to suggest something similar with regard to the acquisition of a model of behaviour: the essential rules and concepts for understanding behaviour might simply be innately given to a social animal. There is, however, an alternative, and to my mind more attractive, possibility. This is to suggest that the animal has access not to "innate knowledge" but to "inside evidence" about behaviour. Nature's psychologists succeed where academic psychologists have failed because the former make free use of introspection.

Let us consider how introspection works. I shall write these paragraphs from the position of a reflective conscious human being, on the assumption that other human beings will understand me. First let me distinguish two separate meanings of what may be called "self-observation", a weak one and a strong one. In the weak sense self-observation means simply observing my own body as opposed to someone else's. It is bound to be true that my body is the example of a human body which is far the most familiar to me. Thus even if I could only observe my behaviour through "objective" eyes it is likely that I would draw on self-observation for most of my evidence about how a human being behaves (in the same way that a physicist who carried a billiard ball about in his pocket might well use that "personal" billiard ball as the paradigm of billiard balls in general).

But the importance of self-observation does not stop there. In the strong sense of the term self-observation means a special sort of observation to which I and I alone am privileged. When I reflect on my own behaviour I become aware not only of the external facts about my actions but of a conscious presence, "I", which "wills" those actions. This "I" has reasons for the things it wills. The reasons are various kinds of

"feeling"—"sensations", "emotions", "memories", "desires". " 'I' want to eat because 'I' am hungry", " 'I' intend to go to bed because 'I' am tired", "'I' refuse to move because 'I' am in pain". Moreover, experience tells me that the feelings themselves are caused by certain things which happen to my body in the outside world. " 'I' am hungry because my body has been without food", " 'I' am in pain because my foot has trodden on a thorn". It so happens (as I soon discover) that several sorts of happening may cause a particular feeling and that a particular feeling may be responsible for my willing several sorts of action. The role of a feeling in the model I develop of my own behaviour becomes, therefore, that of what psychologists have called an ' ' intervening variable' ', bridging the causal gap between a set of antecedent circumstances and a set of subsequent actions—between what happens to "me" and what "I" do.

Now, when I come to the task of modelling the behaviour of another man, I naturally assume that he operates on the same principles that I do. I assume that within him too there is a conscious "l" and that his "I" has feelings which are the reasons for "his" willing certain actions. In other words I expect the relation between what happens to his body and what he does to have the same causal structure—a structure premised on the same intervening variables—as I have discovered for myself. It is my familiarity with this causal structure and these variables which provides me with the all-important ideological framework for doing natural psychology.

Without introspection to guide me, the task of deciphering the behaviour of fellow men would be quite beyond my powers. I should be like a poor cryptographer attempting to decipher a text which was written in a totally unfamiliar language. Michael Ventris could crack the code of Linear B because he guessed in advance that the language of the text was Greek; although the alphabet was strange to him he reckoned—correctly—that he knew the syntax and vocabulary of the underlying message. Linear A remains to this day a mystery because no one knows what language it is written in. In so far as we are conscious human beings we all guess in advance the "language" of other men's behaviour.

But it may be objected that I have not really made out a case for there being any unique advantage in using introspection since non-introspective psychological scientists do in fact also allow themselves to postulate certain intervening variables such as "hunger" and "fear". And so they do. But think of how they derive them. To establish what variables are likely to prove useful to their models they must (assuming they do not cheat) make a vast and impartial survey of all the circumstances and all the actions of an animal and then subject their data to statistical factor analysis. In practice, of course, they usually do cheat by restricting their data to a few "relevant" parameters — relevance being decided on the basis of an intuitive guess. But even so their task is not an easy one. Before postulating even such an "obvious" variable as hunger the experimental psychologist must go through a formidable exercise in data collection and statistical cross-correlation (cf. Hinde)." An ordinary introspective human being has, however, no such problem in devising a "psychological" model of his own and other men's behaviour: he knows from his own internal feelings what intervening variables to go for. Indeed he knows of subtle feelings which no amount of objective data crunching is likely to reveal as useful postulates. Speaking again for myself, I know of feelings of awe, of guilt, of jealousy, of irritation, of hope, of being in love, all of which have a place in my model of how other men behave.

Before I can attribute such feelings to others I must, it seems, myself have had them — a proviso which the academic psychologist is spared. But it is generally the case, for reasons I shall come to in a moment, that in the course of their lives most people do have most of them, and often indeed it takes only a single seminal experience to add a new dimension to one's behavioural model. Let a celibate monk just once make love to a woman and he would be surprised how much better he would understand the Song of Solomon; but let him, like an academic psychologist, observe twenty couples in the park and he would not be that much wiser:

A garden inclosed is my sister, my spouse; a spring shut up, a fountain sealed. Thy plants are an orchard of pomegranates, with pleasant fruits. . . . Let my beloved come into his garden, and eat his pleasant fruits. I sleep, but my heart waketh; it is the voice of my beloved that knocketh, saying Open to me my sister, my love, my dove. My beloved put in his hand by the hole of the door, and my bowels were moved for him.

The translators of the King James Bible, who summarized these lines of the Song as: "Christ setteth forth the graces of the church; the church prayeth to be made fit for his presence" were themselves perhaps somewhat restricted in their ideological perspective.

People are I think well aware of the value of novel experiences in "broadening" their minds. I admit, pace my last example, that mind-broadening is not the usual motive which lies behind people's first experiments in making love; carnal knowledge, so called, has intrinsic attractions over and above the insight it may give into what the psalmist meant by an orchard of pomegranates. But there are times when people do apparently seek new experiences for no other reason than to help themselves "make sense", through introspection, of the behaviour of other people. The clearest cases are those where someone deliberately undergoes an unpleasant experience in order to gain insight into the associated state of mind. My mother once discovered that my young sister had swallowed twenty plumstones, whereupon she herself swallowed

l

l

'

thirty plumstones in order, she said, that she should be able to understand my sister's symptoms. My father, in the days when he was politically active, deprived himself of food for a week in order that he should know what it feels like to be a starving peasant. A colleague of mine, studying a tribe of Amazonian Indians, joined the Indians in drinking a strongly emetic and hallucinogenic drug in order that, having experienced the sickness and the visions, he should be better placed to interpret the Indian's behaviour. I could multiply examples, and so I am sure could you.

These acts of calculated self-instruction have, however, a rather artificial ring to them. They are the acts of "intellectuals", hardly to be expected of ordinary people, let alone of ordinary infra-human social animals. Yet every one of Nature's psychologists, if they are to make good use of the possibilities of introspection, must somehow or another acquire a broad base of inner experience to which they can refer. Had they but time, they might perhaps hope to pick up the requisite ideas simply by waiting passively for relevant experiences to come their way. Sooner or later, without seeking it, most animals will no doubt find that they have, say, run short of food or been beaten in a fight or had a narrow escape from danger; they may even — if they are lucky (or unlucky, depending on how you look at it) — find that they have accidentally swallowed twenty plum-stones. But what if the experience comes later rather than sooner? The costs of naivety are likely to be heavy in terms of psychological misunder-standing.

The matter is so serious that it would be surprising if it had been neglected by natural selection in the course of evolution. I believe that biological mechanisms have in fact evolved for ensuring that young animals, like it or not, rapidly receive the ideological instruction required to turn them into competent psychologists. They fall into three categories: (i) play, (ii) parental manipulation, (iii) dreaming.

The role of play in extending inner experience is so obvious as to need little elaboration. For all animals, and not just man, play involves adventures for the mind as well as for the body. If we could ask a young animal, as we can ask a child, why it is doing whatever it is doing in play, it would probably reply that it is simply "having fun": but in the course of having fun the animal is unwittingly educating itself. It is throwing itself into new kinds of interaction with the physical and social world and thereby intro-ducing its mind to a whole new range of feelings—new sensations, new emotions, new desires. Look at a child playing hide-and-seek, or look at a young monkey playing king of the castle: feelings of anxiety, of excitement, of satisfaction, of disappointment, of competitiveness, even perhaps of compassion; these and many other rarer and often unnamable ideas are being planted and tended in the youngsters' minds. One day, when the games are for real, the child or the monkey will use its introspective knowledge of such feelings to interpret and predict the behaviour of another member of its social group.

There are, however, limits to the range of feelings which animals are likely to learn about through play. They play because it pleases them to do so. How then shall they learn about the feelings associated with experiences which are in no way pleasurable? Many of the feelings most pertinent to the modelling of the behaviour of others in the social group are in one way or another disagreeable to the animal who has them—fear, anger, pain, jealousy, grief. But these are the very feelings which a young animal, left to itself, is likely to do its best to avoid. If play, on the whole, plants pleasant flowers in the garden of a child's mind, what— or who- plants the tares and weeds?

My answer may surprise you. I think that, often enough, it is the child's parents. Biologically it is in the interests of parents to increase the fitness of their offspring in whatever ways they can. Ethologists have long recognized that this is the reason why parents so often take a hand in their children's education, giving them lessons in how to do things and, of course, being active partners in their play. But there has been very little discussion of how parents might help their children by abusing them ' 'for their own good". Let me illustrate the principle with a happening I witnessed not long ago on the train to Cambridge. A woman sat opposite me in the carriage with her 4-year-old daughter. The little girl asked her mother an innocent question. The mother pretended not to notice her. The girl repeated her question, adding plaintively "Mummy, please tell me". "I'm not your mummy", said the woman, "Your mummy got off at the last station". The girl began to look anxious. "You are my mummy. I know you're my mummy." "No I'm not. I've never seen you before." And so this strange game, if you can call it such, continued until eventually the bewildered little girl broke down in tears. A wicked, heartless mother? I thought so at the time—but maybe it was an unfair judgement. That little girl was in the truest sense being taught a lesson, the lesson of what it feels like to be mystified and scared. She perhaps learned more of real importance in those few unhappy minutes than I myself have ever learned from the hundred books I have read on train journeys.

Now I believe such parental abuse of children may be much more widespread than ethologists have either noticed or perhaps cared to admit. And, following my present line of argument, I believe that its biological function may often be to educate children in the knowledge of disagreeable feelings. Children, as apprentice psychologists, need to know about being frightened, so parents frighten them; they need to know about jealousy, so parents do things to make them jealous; they need to know about pain, so parents hurt them; they need to know about feeling guilty, so parents contrive to catch them doing wrong. And so on. If you were to press me for further specific examples, I should probably continue to refer chiefly to the actions of human beings. But there is one general category of parental abuse which is well known to occur in other social animals than man. That is the "parent—offspring conflict" which occurs in relation to weaning. There are, of course, alternative theories of why mothers become progressively more hard-hearted to their children around the time of weaning, but I would suggest that at least one of the functions of the mother's behaviour is purely educational—it is in the child's best interests that it should have first-hand experience of frustration, rejection, hunger and loneliness.

The third way by which young animals may acquire their ideological grounding as psychologists is by exposing themselves to purely imaginary experiences. I mean by dreaming. Dream experience is clearly in a different class to the experience provided by play or parental manipulation; yet I would argue that as a means of introducing the animal to a range of novel feelings it is potentially as powerful. True, there may seem at first sight to be a fundamental problem here: whereas through play or parental manipulation real things happen to the infant animal and real feelings are aroused, in dreams unreal things happen and, presumably, unreal feelings are aroused. But it is a mistake to talk of "unreal" feelings. All feelings, whatever context they occur in, are internal creations of the subject's mind. Although they may be-and usually are—evoked by external happenings, it is not the external happenings as such which evoke them, but the subject's perception of and belief in those external happenings. For a feeling to occur it is a sufficient condition that the subject should have the appropriate perceptions and beliefs—that he should "think" himself to be undergoing the relevant experience. Thus for me to feel fear it is sufficient that I should think I am being chased by a crocodile: my fear will be the same whether the crocodile is an objective physical crocodile or a subjective crocodile conjured up in my imagination.

If you yourself have never dreamed of being chased by a crocodile, or if — as I hardly think likely — you doubt altogether the possibility of feelings being induced by fantasy experience, go and watch a stage hypnotist at work. Better still, go up on the stage and allow him to use you as one of his subjects: the hypnotist will, perhaps, suggest that there is a spider crawling up your neck and you will find yourself shuddering with genuine horror.

What the hypnotist does to his subjects on the stage the dreamer can do to himself as the subject of his self-generated fantasies. In the freedom of the dream he can invent extraordinary stories about what is happening to his own person and so, responding to these happenings as if to the real thing, he discovers new realms of inner experience. If I may speak from my own case, I have in my dreams placed myself in situations which have induced in my mind feelings of terror and grief and passion and pleasure of a kind and intensity which I have not known in real life. If 1 did now experience these feelings in real life I should recognize them as familiar; more important, if I were to come across someone else undergoing what I went through in my dream I should be able to guess what he was feeling and so be able to model his behaviour.

Although I have been talking now more of people than of other social animals, I have intended that most of what I have said should apply to animals as well. In people, and people alone, however, the biological mechanisms for providing ideological instruction have been supplemented in important ways by culture. All three mechanisms — play, parental manipulation and dreaming - have parallels in human cultural institutions. The play of individual animals has its counterpart in organized games and sports where youngsters, besides enjoying themselves, are encouraged to compete, co-operate, take risks, set their hearts on winning, and discover what it means to lose. Abuse by individual parents has its counterpart in "initiation rites" where adolescents are frequently subjected to bodily mutilation, to fearsome ordeals, and sometimes to forced isolation from the social group. And dreaming has its counterpart in drama and public story-telling where the actors — and their audience too — get drawn into elaborate fantasies. I am suggesting not merely an analogy but a functional homology between the cultural and the biological phenomena.
I believe it could be shown that members of a society who have, for example, been put through a brutal initiation ceremony make better introspective psychologists than others who lack the experience. At another extreme I believe that nineteenth-century readers of Dickens's serial novel The Old Curiosity Shop, who cried in the streets when they heard of the death of Little Nell, may have been better able to understand the behaviour of their neighbours when a real child died.

I do not for a moment mean to say that this is all there is to these cultural institutions, any more than a sociobiologist would say that the avoidance of inbreeding is all there is to the incest taboo. But if, as I have argued, greater insight into other people's behaviour is one of the benefits of subscribing to a cultural institution, then almost certainly it is one of the factors which keeps that institution alive.

So much for how I think that Nature's psychologists proceed. Let me turn to more purely philosophical implications of the theory. I promised at the start of this paper to say something about the evolution of consciousness.

I take it to be the case that what we mean by someone's conscious experience is the set of subjective feelings which, at any one time, are available to introspection, i.e. the sensations, emotions, volitions, etc., that I have talked of. Our criterion for judging that someone else is conscious is that we should have grounds for believing that he has subjective reasons for his actions — that he is eating an apple because he feels hungry, or that he is raising his arm because he wants to. If we had grounds for believing that a dog had similar subjective reasons for its actions we should want to say the dog was conscious too. In proposing a theory about the biological function of introspection I am therefore proposing a theory about the biological function of consciousness. And the implications of this theory are by no means trivial. If consciousness has evolved as a biological adaptation for doing introspective psychology, then the presence or absence of consciousness in animals of different species will depend on whether or not they need to be able to understand the behaviour of other animals in a social group. Wolves and chimpanzees and elephants, which all go in for complex social interactions, are probably all conscious; frogs and snails and codfish are probably not.

There may be philosophers who protest that it is nonsense to talk of a biological "function" for consciousness when, so Wittgenstein tells us, conscious experience does not even have a "place in the language game".5 But what Wittgenstein demonstrated is that there are logical problems about the communication of conscious experience — and it is not proposed by the theory that consciousness had any direct role in communication between individuals; I am not saying that social animals either can or should report their subjective feelings to each other. The advantage to an animal of being conscious lies in the purely private use it makes of conscious experience as a means of developing an ideology which helps it to model another animal's behaviour. It need make no difference at all whether the other animal is actually experiencing the feelings with which it is being credited; all that matters is that its behaviour should be understandable on the assumption that such feelings provide the reasons for its actions. Thus for all I know no man other than myself has ever experienced a feeling corresponding to my own feeling of hunger; the fact remains that the concept of hunger, derived from my own experience, helps me to understand other men's eating behaviour. Indeed, if we assume that the first animal in history to have any sort of introspective consciousness occurred as a chance variant in an otherwise unconscious population, the selective advantage which consciousness gave that animal must have been independent of consciousness in others. It follows, a fortiori, that the selective advantage of consciousness can never have depended on one animal's conscious experience being the "same" as another's.5

Maybe this sounds paradoxical. Indeed, if it does not sound a little paradoxical I should be worried. For I assume that you are as naturally inclined as any other introspective animals to project your conscious feelings onto others. The suggestion that you may be wrong to do so, or at least that it does not matter whether you are right or wrong, does I hope arouse a certain Adamite resistance in you. But allow me to elaborate the argument.

I think no one of us would object to the claim that a piece of magnetized iron lacks consciousness. Suppose now that an animal— let us call it one of "Nature's physicists" —wanted to model the behaviour of magnets. I can conceive that it might be helpful to that animal to think of the north pole of a magnet as having a desire to approach a south pole. Then, if the concept of having a desire was one which the animal knew about from its own inner experience, I should want to argue that introspective consciousness was an aid to the animal in doing physics. The fact that the animal would almost certainly be incorrect in attributing feelings of desire to magnets would be irrelevant to whether or not the attribution was heuristically helpful to it in developing a conceptual model of how magnets behave. But if this is conceivably true of doing physics, all the more is it true of doing psychology. Notwithstanding the logical possibility that every other human being around me is as unconscious as a piece of iron, my attribution of conscious feelings to them does as a matter of fact help me sort out my observations of their behaviour and develop predictive models.

Ah, you may say, but you are not really saying anything very interesting, since it can only be helpful to attribute feelings to other people— or magnets — in so far as there is something about the other person or the magnet which corresponds to what you call a feeling: the attribution of desire to magnets is heuristically valuable if, and only if, there exists in reality an electromagnetic attractive force between a north pole and a south pole, and the attribution of a feeling of hunger to a man is valuable if, and only if, his body is in reality motivated by a particular physiological state. Quite so. But the magnet does not have to know about the electromagnetic force and the man does not, in principle, have to know about the physiological state.

Magnets do not need to do physics. If they did—if their survival as magnets depended on it — perhaps they would be conscious. If volcanoes needed to do geology, and clouds needed to do meteorology, perhaps they would be conscious too.

But the survival of human beings does depend on their being able to do psychology. That is why, despite the sophistical doubts I have just expressed, I do not consider it to be even a biological possibility — let alone do I really believe — that other people are not as fully conscious of the reasons for their actions as I know that I myself am. In the case of frogs and snails and cod, however, my argument leads me to the opposite conclusion. Let me say it again: these non-social animals no more need to do psychology than magnets need to do physics—ergo they could have no use for consciousness.

Somewhere along the evolutionary path which led from fish to chimpanzees a change occurred in the nervous system which transformed an animal which simply "behaved" into an animal which at the same time informed its mind of the reasons for its behaviour. My guess is that this change involved the evolution of a new brain— a "conscious brain" parallel to the older "executive brain". In the last few years evidence has at last begun to emerge from studies of brain damage in animals and man which makes this kind of speculation meaningful.

To end my paper I want to talk about a monkey called Helen.

In 1966 Helen underwent an operation on her brain in which the visual cortex was almost completely removed. In the months immediately following the operation she acted as if she were blind. But I and Professor Weiskrantz with whom I was working were not convinced that Helen's blindness was as deep and permanent as it appeared. Could it be that her blindness lay not so much in her brain as in her mind? Was her problem that she did not think that she could see?

I set to work to persuade her to use her eyes again. Over the course of seven years I coaxed her, played with her, took her for walks in the fields — encouraged her in every way I could to realize her latent potential for vision. And slowly, haltingly, she found her way back from the dark valley into which the operation had plunged her. After seven years her recovery seemed so complete that an innocent observer would have noticed very little wrong with the way she analysed the visual world. She could, for example, run around a room full of furniture picking up currants from the floor, she could reach out and catch a passing fly.7

But I continued to have a nagging doubt about what had been achieved: my hunch was that despite her manifest ability Helen remained to the end unconscious of her own vision. She never regained what we — you and I — would call the sensations of sight. Do not misunderstand me. I am not suggesting that Helen did not eventually discover that she could after all use her eyes to obtain information about the environment. She was a clever monkey and I have little doubt that, as her training progressed, it began to dawn on her that she was indeed picking up "visual" information from somewhere—and that her eyes had something to do with it. But I do want to suggest that, even if she did come to realize that she could use her eyes to obtain visual information (information, say, about the position of a currant on the floor), she no longer knew how that information came to her: if there was a currant before her eyes she would find that she knew its position but, lacking visual sensation, she no longer saw it as being there.

It is difficult to imagine anything comparable in our own experience. But perhaps the sense we have of the position of parts of our own bodies is not dissimilar. We all accept as a fact that our brains are continuously informed of the topology of the surface of our bodies: when we want to scratch an ear we do not find ourselves scratching an eye; when we clap our hands together there is no danger that our two hands will miss each other. But, for my own part, it is not at all clear how this positional information comes to me. If, for example, I close my eyes and introspect on the feelings in my left thumb I cannot identify any sensation to which I can attribute my knowledge of the thumb's position— yet if I reach over with my other hand I shall be able to locate the thumb quite accurately. I "just know", it seems, where my thumb is. And the same goes for other parts of my body. I am inclined therefore to say that at the level of conscious awareness "position sense" is not a sense at all: what I know of the position of parts of my body is "pure perceptual knowledge" — unsubstantiated by sensation.

Now in Helen's case, I want to suggest that the information she obtained through her eyes was likewise "pure knowledge" for which she was aware of no substantive evidence in the form of visual sensations. Helen "just knew" that there was a currant in such—and—such a position on the floor.

This, you may think, is a strange kind of hypothesis — and one which is in principle untestable. Were I to admit the hypothesis to be untestable I should be reneguing on the whole argument of this paper. The implication of such an admission would be that the presence or absence of consciousness has no consequences at the level of overt behaviour. And if consciousness does not affect behaviour it cannot, of course, have evolved through natural selection—either in the way I have been arguing or any other. What, then, shall I say? If you have followed me so far you will know my answer. I believe that Helen's lack of visual consciousness would have shown up in the way she herself conceived of the visually guided behaviour of other animals—in the way she did psychology. I shall come back to this in a moment; I think you will be more ready to listen to me if I first refer to some remarkable new evidence from human beings.

In the last few years Weiskrantz and his colleagues at the National Hospital, and other neurologists in different hospitals around the world, have been extending our findings with Helen to human patients.3 They have studied cases of what is called "cortical blindness", caused by extensive destruction of the visual cortex at the back of the brain (very much the same area as was surgically removed in Helen). Patients with this kind of brain damage have been described in most earlier medical literature as being completely blind in large areas of the visual field: the patients themselves will say that they are blind, and in clinical tests, where they are asked to report whether they can see a light in the affected area of the field, their blindness is apparently confirmed. But the clinical tests —and the patients' own opinion—have proved to be deceptive. It has been shown that, while the patients may not think that they can see, they are in fact quite capable of using visual information from the blind part of the field if only they can be persuaded to "guess" what it is their eyes are looking at. Thus a patient studied by Weiskrantz, who denied that he could see anything at all in the left half of his visual field, could "guess" the position of an object in this area with considerable accuracy and could also "guess" the object's shape. Weiskrantz, searching for a word to describe this strange phenomenon, has called it "blindsight' '.

"Blindsight" is what I think Helen had. It is vision without conscious awareness: the visual information comes to the subject in the form of pure knowledge unsubstantiated by visual sensation. The human patient, not surprisingly, believes that he is merely "guessing". What, after all, is a "guess"? It is defined in Chambers's Dictionary as a "judgement or opinion without sufficient evidence or grounds". It takes consciousness to furnish our minds with the sensations which provide "evidence or grounds" for what our senses tell us; just as it takes consciousness to give our mind the subjective feelings which provide "evidence or grounds" for our eating behaviour, or our bad temper, or whatever else we do with the possibility of insight into its reasons.

So if Helen lacked such insight into her own vision, how might it have affected her ability to do psychology? I do not think that Helen's particular case is a straightforward one, since Helen was already grown up when she underwent the brain operation and she may well have retained ideas about vision from the time when she could see quite normally. I would rather discuss the hypothetical case of a monkey who has been operated on soon after birth and who therefore has never in its life been conscious of visual sensations. Such a monkey would, I believe, develop the basic capacity to use visual information in much the same way as does any monkey with an intact brain; it would become competent in using its eyes to judge depth, position, shape, to recognize objects, to find its way around. Indeed, if this monkey were to be observed in social isolation from other monkeys, it might not appear to be in any way defective. But ordinary monkeys do not live in social isolation. They interact continuously with other monkeys and their lives are largely ruled by the predictions they make of how these other monkeys will behave. Now, if a monkey is going to predict the behaviour of another, one of the least things it must realize is that the other monkey itself makes use of visual information-that the other monkey too can see. And here is the respect in which the monkey whose visual cortex was removed at birth would, I suspect, prove gravely defective. Being blind to the sensations of sight, it would be blind to the idea that another monkey can see.

Ordinary monkeys and ordinary people naturally interpret the visually guided behaviour of other animals in terms of their own conscious experience. The idea that other animals too have visual sensations provides them with a ready-made conceptual framework for understanding what it "means" for another animal to use its eyes. But the operated monkey, lacking the conscious sensations, would lack the unifying concept: it would no longer be in the privileged position of an introspective psychologist.

In the days when we were working with Helen, Weiskrantz and I used to muse about how Helen would describe her state if she could speak. If only she could have communicated with us in sign language, what profound philosophical truths might she have been ready to impart? We had only one anxiety: that Helen, dear soul, having spent so long in the University of Cambridge, might have lost her philosophical innocence. If we had signalled to her: "Tell us, Helen, about the nature of consciousness", she might have replied with the final words of Wittgenstein's Tractatus: "Whereof one cannot speak, thereof one must be silent." Silence has never formed a good basis for discussion.

Too often in this century philosophers have forbidden the rest of us to speak our minds about the functions and origins of consciousness. They have walled the subject off behind a Maginot line. The defences sometimes look impressive. But biologists, advancing through the Low Countries, should not be afraid to march around them.

Discussion
RAMACHANDRAN:

You point out that consciousness permits social interaction. I agree that my direct conscious experience of non-neutral (and emotionally coloured) states, such as pain, hunger, sex, etc., does improve my ability to interact effectively with someone experiencing similar states; especially when I assume that the other person is also conscious of these states in the same intense way that I am conscious of them. But I do not see how this argument applies to neutral states such as elementary sensations (e. g. reds, greens, etc.). How would my knowledge that the other person was consciously seeing these (rather than merely reacting to them) influence my behaviour towards him? Ifa person were consistently to report red when confronted with such and such a wavelength then I can at once begin effective communica- tion with him. It is quite irrelevant to me whether he is actually conscious of it (like I am) or not. If this is true, then why did "redness" emerge into awareness at all instead of "behaviour towards red" remaining a subconscious and neutral event like the pupillary light reflex? It seems to me that what you have given us is a theory of emotions rather than a theory of consciousness.

I see a partial answer to some of these questions in your example of the monkey Helen, who was (presumably) not conscious, although her visual behaviour could be restored; but would you like to elaborate'? Supposing I met a man whose visual performance was indistinguishable from normal (i.e. an extreme example of the kind of patient reported by Weiskrantz) but who lacked visual consciousness. Would this knowledge make any difference to my understanding him or communicating with him? If not, where does your argument stand?

HUMPHREY:

Your question about the function of "neutral states of consciousness" raises problems which, I am bound to say, I have not fully thought through. Certainly the hypothesis I've presented lends itself more readily to explaining why someone should be conscious of affective states (emotions, motives, etc.) than to explaining why they should be conscious of neutral states such as simple auditory or visual sensations. But I did not mean in my paper to sidestep the latter issue altogether, and I hope that what I say about "blindsight" does suggest where the answer lies. On pages 73-4 of my paper I do indeed discuss the question which you now put to me: "In what way would someone who lacked visual consciousness (e.g. after removal of the striate cortex) prove biologically defective?" And I answer it by suggesting that, in one respect at least, such a person would prove to be a poor psychologist, because he would find it difficult to conceive that the behaviour of another person was guided by what we call "sight" (I don't say that he could never arrive at the concept, but it might well take him a long time to catch on). A parallel of a sort is provided by the difficulty zoologists have had in accepting the existence of "alien" sensory systems, such as the electric sense in fish or the magnetic sense in birds, of which a human being can have no introspective knowledge. More pertinent still, perhaps, is the case of so—called pheromones: it now seems quite probable that human beings are, without being consciously aware of it, influenced by chemical signals from other human beings — but the idea of pheromonal communication remains strange to us because (I would argue) we cannot fit it into a conceptual framework informed by our own consciousness. Radical behaviourists did, in the early days, actually attempt to develop models of both human and animal behaviour which, borrowing nothing from human insight, made no reference to the existence of different sensory "modalities"; ordinary people, however, being dis- inclined to cut off their intuitive noses to spite their psychological faces, have always made life easier for themselves by relying on the phenomenology of their own conscious experience to generate the (genuinely) useful concepts of "sight", "hearing", "taste" and so on.

RAMACHANDRAN:

Is the distinction between ordinary consciousness and self—consciousness important to your argument?

HUMPHREY:

By ordinary consciousness or 'raw consciousness" I mean sensations, desires, etc., existing as primitive mental events. Self—consciousness or reflexive consciousness, on the other hand, involves inward observation of what is happening on the level of raw consciousness: it is thus logically dependent on the existence of raw consciousness, although it might be argued that the converse is not true, i.e. that raw consciousness is not logically dependent on the existence of reflexive consciousness. However, I know of (and can imagine) no reason to suppose that raw consciousness does as a matter of fact ever exist without reflexive consciousness: indeed, if raw consciousness were present in a subject who was unable to reflect on it, he could not (by definition) notice it, remember it, think about it or, a fortiori, tell any one else about it. Further, I am not convinced that raw consciousness as such has, or could have, any independent biological function; my own view is that raw consciousness probably evolved to provide the substrate for reflexive consciousness.

JOSEPHSON:

While we are discussing reflexive or self—consciousness, it is worth pointing out that according to some people there are two kinds of "self" involved. There is the individual self, which is the accumulation of the individual's own experiences, and the higher or transpersonal self concerned with creative insights and spiritual experience, which have the appearance of coming from a source beyond the individual and being unrelated to memory. While contact with a higher self is usually stated to be an exclusively human experience, possibly behaviour involving insight, as occurs with monkeys, indicates that they too possess this ability to a limited degree.

VESEY:

As you may know, philosophers spend a lot of their time talking about meaning. There are radically opposed views, some with quite a history to them. For instance, there is the empiricist view, held by people like John Locke, J.S. Mill, and, more recently, Bertrand Russell and A.J. Ayer. Roughly, they say that a word has meaning by being a name given to an experience. For instance, someone has a pain, gives the name "pain" to it, and then uses the same word again when he has an experience he recognizes as being similar to the one to which he first gave the name. (That is a one—sentence summary of what Mill says in Book I, Chapter 3, of his System of Logic, 1843.) This seems an attractively simple account of meaning, but there is a problem connected with it. If "pain" is a name I give to one of my experiences, and regive when I have a similar experience, what can I mean when I say that someone else is in pain? It's a bit like knowing what it means to say that it is afternoon, when one is in Houston, Texas, and then being expected to understand the remark when one is half-way to the moon. The conditions of meaningfulness have been removed. There is no zenith for the sun to be past, no horizon for it not to be past. Similarly with talk about someone else being in pain, if one accepts the empiricist account of meaning. The condition of meaningfulness, that the sensation can be recognized as similar to the one first named, no longer holds.

It seems to me that a basic presupposition of your argument is the correctness of the empiricist view of meaning. Do you have a solution to the problem l've indicated?

HUMPHREY:

Let me try to make my argument clearer with an example. Then maybe the problem you raise about meaning will be easier to resolve.

Suppose that each and every one of us owns a whistling kettle, and that it is important to be able to predict the "behaviour" of these kettles (to anticipate their whistling, etc.). The external facts I observe about my own and other people's kettles are, say, of the following kind: (i) the kettle when filled with cold water and put on the stove begins to whistle within about 5 minutes, (ii) the kettle takes less time to whistle when filled with hot water, (iii) the kettle takes more time to whistle when salt is added to the water, (iv) the kettle takes less time to whistle on top of a mountain, (v) if the kettle is filled with liquid nitrogen instead of water it whistles without being put on the stove, (vi) if the kettle is filled with treacle it doesn't whistle at all, and so on.

I suggest that, if these external facts were all I had to go on, the behaviour of the kettles might seem puzzling. I would be hard put to it to develop a theory of the relation between what is done to the kettle and what the kettle does. But suppose that, while everybody else's kettle is made of tin, my own kettle is made of Pyrex glass so that I can see into it. I look into my kettle and observe (i) that when certain things are done to the kettle the liquid inside it boils, and (ii) that when the liquid boils the kettle whistles. I am led to regard boiling as an explanatory concept, an "intervening variable" which "bridges the causal gap between a set of antecedent circumstances and a set of subsequent actions- between what happens to my kettle and what my kettle does" (cf. my paper, p. 62). Thus I now explain the behaviour of my kettle by arguing along the following lines: the kettle whistles when the liquid boils, the liquid boils when the kettle is put on the stove, therefore the kettle whistles when it is put on the stove.

But at this point something philosophically interesting has happened. While the concept of boiling has been put into my mind by a factual observation (what I actually saw when I looked into my kettle), its usefulness as an explanatory concept does not depend on the observation's having been of any particular kind; indeed, I could have observed something quite different. Suppose, for example, that when I looked into my kettle I had observed the liquid turning a red colour under just those circumstances when in fact I saw it boil, then the concept of reddening might have come to play exactly the same role in my argument as the concept of boiling: the kettle whistles when the liquid reddens, etc. Indeed as far as my new-found theory is concerned it really doesn't matter what I have actually observed (and a fortiori it doesn't matter what I choose to call what I have observed — I might as well say the liquid in the kettle is in pain).

Now, how about other people's kettles? Since they are made of tin I cannot, of course, observe the liquid inside their kettles boiling (or reddening or whatever). Can I then use the concept of boiling to help myself explain the behaviour of their kettles'? Yes. Since the usefulness of boiling as an explanatory concept is independent of any particular observation I have or could have made, the concept can play just the same role in my argument about someone else's kettle as it does in my argument about my own.

With regard to the problem of meaning, I accept that the factual propositions "The liquid in my kettle is boiling" and "The liquid in his kettle is boiling" are of different status (indeed the latter proposition is arguably, by positivist criteria, meaningless). But the explanatory propositions "My kettle is whistling because the liquid inside it is boiling" and "His kettle is whistling because the liquid inside it is boiling" are on a par.

Another example to think about: suppose that Mendel, when he was searching for a theory of inheritance, could have observed his own genes.

VESEY:

You are right: your example does make your argument clearer. It makes it clearer that it is as follows. (i) The concept of boiling is put into one's mind by what one observes on looking into kettles. Similarly, (ii) the psychological concepts one uses to explain people's behaviour—concepts like expecting, hoping, remembering, understanding, wanting, wondering—are put into one's mind by what one observes on looking into one's mind (introspecting) when one is doing these things. (iii) That one cannot look into other people's minds does not prevent one using psychological concepts to understand their behaviour.

Not only does your example make your argument clearer; it also enables me to make clear the extent and nature of my disagreement with you. I disagree with you not only about (ii) but also about (i). And the disagreement is a fundamental one, about meaning. To know the meaning of a word (="to have the concept for which the word stands") is to know how to use the word correctly. A word's being meaningful, and there being criteria of its correct use, go hand in hand. This being so, it does not make sense to talk of concepts being put into people's minds by their observing things, either inner things or outer things. Concepts are not experiences, to be put into people's minds by pointing their eyes, or their mind's eye, in the right direction. They are abilities exercised primarily, in humans, in acts of verbal communication. And the linguistic practices involved could not, even in theory, start as private practices.

JOSEPHSON:

The dilemma can be resolved by assuming that the concepts are already there in latent form in the nervous system, waiting to be triggered off by the relevant experiences. The latter do not have to be linguistic in nature.

BARLOW:

As a result of thinking about the biological role of consciousness both Nick Humphrey and I (see next paper, "Nature's Joke") have come to the same conclusion, namely that the survival value of consciousness is very much connected with its role in the social life of gregarious animals, but there is a difference between our proposals that may be important. I argue that consciousness is impossible without some kind of social interchange, so that mankind is driven to engage in social relations to preserve his consciousness. Consciousness is thus Nature's tool to make man social, just as pain can be regarded as her tool to make us avoid injury. The survival value of consciousness would result from social hominids leaving more offspring than solitary hominids. If I understand Humphrey correctly, he regards the gregarious nature of man as a prior fact, and sees consciousness as conferring an advantage in competing against other individuals within the same social group. Am I right in understanding him to say that consciousness improves social behaviour, but does not actually help to generate it, as I would claim?

I have another question relating to the use of the word "introspection", for I don't think we find out about others in this way. It is very likely true that you cannot understand certain aspects of other individuals' behaviour until you have yourself undergone the experience motivating that behaviour, and this is interesting and important. But this insight seems to come by a process of imitation rather than introspection, which I take to mean a Conscious searching of one's own mind. Sight of a pattern of muscular movements may enable one to imitate them, and I think one's feelings can imitate the emotions that generate a pattern of behaviour in another. But I don't think there is any conscious search in one's mind for them, so I would hesitate to call this process introspection.

HUMPHREY:

1. I hope Barlow will not mind if I characterize his argument as follows. Consciousness is rather like group sex: something which is a source of pleasure to the individual but which he can't achieve on his own and so is obliged to seek through interaction with others. Thus Barlow sees the biological function of consciousness—the contribution it makes to biological survival—as the provision of an incentive to being social (sociality being essential to human survival). His argument rests, as I see it, on three premises: (i) people desire to be conscious (as, for example, they desire sex); (ii) people can only be conscious through social interaction; (iii) people would not be social if they were not made to be by this "trick" which Nature plays on them. Barlow's question relates to this last point, and he is right to think that I disagree with him here. I do not believe that people remain in social groups in order to preserve their consciousness; my view is that people would, whether conscious or not, try to form social groups but that if they were not conscious they would probably fail because they would be unable to understand each other. In Barlow's view, without consciousness the social group would never get together; in my view, without consciousness the social group would fall apart. But either way, surprisingly enough, we draw the same conclusion, namely that consciousness is probably a necessary condition of being a highly social animal. And indeed we agree on a more specific prediction, namely that a dysfunction in the mechanism of consciousness (as I suggest may have occurred in Helen and Barlow suggests may occur in autistic children) is likely to show up as social maladjustment.

2. Barlow has misconstrued my argument if he thinks I'm suggesting that "we find out about others" by introspection. No, we don't "find out" about them that way; we find out about them by ordinary external observation — looking at them, listening to them, etc. What introspection does is to help us explain what we find out about them: it provides us with the explanatory concepts in terms of which we "make sense" of what we observe. This point is elaborated in my reply to Professor Vesey. But when, for example, we explain someone else's behaviour by saying "He is crying because he is in pain" we don't have to be feeling the pain ourselves (which is what Barlow seems to be implying by his remarks about "imitation").

CHAPTER 5

Nature's Joke: A Conjecture on the
Biological Role of Consciousness

H. B. BARLOW
University of Cambridge

ABSTRACT

A physiologist needs to know the function of an organ when he tries to find out how it works, and a biologist needs to know the survival value conferred on an individual by the performance of that function. This essay provides a conjectural answer to the questions "What is the function of the consciousness of man?" and "What is its survival value? ' '.

It is argued that consciousness primarily arises in the relation between one indivi- dual and another, and is not a property of a brain in isolation. One can, of course, be conscious when one is alone, but it is suggested that on these occasions one is rehears- ing future discourse with an imagined individual. This is rendered plausible by the fact that our brains are certainly adept model-makers and the character and person- ality of parents and others must be amongst the most thoroughly modelled aspects of a person's environment. Could one be conscious at all if all memories of experiences with other individuals were deleted from one's brain?

The individual values his consciousness above all else, but if it only arises in real or imagined relations with others, this will have an interesting consequence; his con- sciousness, the arena within which he makes his decisions, is not his own alone, but is influenced by and interacts with those others, real and imagined, with whom he must discourse in order to be conscious. This can be no accident, and it is suggested that Nature has constructed our brains so that, first, we seek to preserve individual con- sciousness; second, we can only achieve it in real discourse or rehearsed future dis- course; and third, important new decisions require the sanction of consciousness. These three aspects of consciousness generate a communal culture in the light of which individual decisions tend to be made. Thus the survival value of consciousness consists of the peculiar form of gregarious behaviour it generates in man; it is Nature's trick to chain him to the herd.

I became interested in the mind-body problem because I am a neuro- physiologist and try to relate subjective experiences to the physical properties of sense organs and nerve cells. Now when a physiologist wants to investigate the working of some organ he first forms a hypothesis about its function, because without such a hypothesis he is likely to waste much time studying inessential aspects of the organ; imagine, for instance, how futile it would be to investigate the lungs without knowing that the interchange of gases between the blood and air takes place in them. When thinking of consciousness my instinctive approach was to avoid studying what consciousness looked or felt like to myself, and also to avoid paying much attention to what philosophers have said about it, for this also seems mainly based on introspection. Nature does not tell us what our organs are for and is well able to make her actors think they are playing one part when they are really playing another, or serving in some quite different capacity. So instead of introspection and reading, my approach has been to observe the actors, to see what it is they refer to as consciousness, and to make a conjecture on its biological role. My conclusions depend, for whatever force they may have, first on this being (I think), an unusual approach to the problem, and second on the fact that a surprisingly simple, far-reaching and unifying concept does emerge.

In the first part of the essay I argue that consciousness is not a property of a brain in isolation, but is a property of a brain that is and has been in communication with other brains. By communication I mainly mean talk, certainly nothing mysterious or non–physical; indeed I think that much of the apparent conflict with physics in the mind–body relation disappears if one accepts that consciousness is something to do with relations between brains rather than a property of a single brain.

The second part tries to account for the prominence of these conscious interactions between brains in the conduct of the affairs of mankind. Because of its prominence one must ask the question "What is the survival value of consciousness?". I shall suggest that consciousness, and particularly the restricted nature of our conscious knowledge of our own brains, is Nature's method of making humans behave co-operatively. Our being is centred in our conscious self, but if consciousness is the relating of one brain to another, this means that one's being is centred, not in one's own brain, but in the relation between one's own brain and others. The view that consciousness is much concerned with past and present interpersonal relationships may find supporters outside the realm of neurophysiologists and biologists.

CONSCIOUSNESS A RELATION, NOT A PROPERTY

Some everyday usages of the notion of consciousness clearly reflect an appreciation that it refers to relationships. For instance, if we ask "Is he conscious?", someone will immediately try to establish contact with the person concerned to test his capacity for making such relationships. Furthermore, he is likely to apply the same test even if the person is behaving in an outwardly normal way, as might a sleepwalker or someone in an unusual, trance—like state. To be conscious is to be able to relate to others, not just to act normally.

Another fact that fits the concept well is the prominence of language in our consciousness. One is at once consciously aware of the spoken word, which seems to take precedence over almost any other sensory experience, except perhaps intense pain. And if one has something to say, this thought in one's head is certainly in the forefront of one's conscious awareness. Received and uttered speech are, of course, the most important way that two brains relate to each other.

Speech, however, is not the only way that brains relate in a way that I think qualifies for consciousness. The moment a baby first smiles at its mother seems at least as good a time to take for the birth of its consciousness as any other, and the capacity of an animal to respond personally to another or to its master might form a rather acceptable test of its consciousness: dogs and cats, yes; snails and toads, no.

So far so good, but one does not have to look far to find difficulties, for I am obviously conscious when I am completely alone. It is true that, for others to know about these conscious experiences, I must establish relationships and interact with them, but it seems quite incorrect, from our own introspective knowledge, to deny that consciousness occurs until we communicate it to others. Or is it wholly wrong? Certainly some experiences gain greatly in vividness from the telling. But I shall simply maintain that immediate conscious experience is preparation for recounting the sensory events to others, that it is a rehearsal before other brains that are embedded or modelled in the imagination.

This may sound far-fetched and evasive, but we know very well that our brains contain accurate maps and models of the physical environment, and an individual's past experience and interaction with other people must surely form the basis of models of their character and personality. It would be most surprising if we modelled people less effectively than we model the physical environment. Thus models of other brains normally exist in our heads, and one can ask "Could one have a conscious sensory experience at all if the models of all past and present acquaintances were suddenly deleted or made unavailable?". Since we learn language from others, we should have no words to represent our awareness, and it is certainly only an impoverished awareness that can occur without words. Even this residue requires, 1 would claim, an imagined person to relate to before it can become conscious. Pain, it is true, seems to require no words for its experience, but it evokes a uniquely strong urge to communicate.

Thus I conclude that subjective awareness, even of immediate sensation, is a form of imagined future discourse. Or to put it another way, the portion of the stream of sensory information of which one becomes conscious corresponds to what one is selecting for potential communication to others. It is reasonable to suppose that the brain has to make this preselection, whether or not circumstances are propitious or mandatory for actual communication, but I am insisting on the importance of one or more imagined recipients of the communication before it becomes conscious. An audience, as well as an actor, is necessary for consciousness.

Self-awareness might be thought to pose another difficulty with the view that consciousness is a relationship and not a property. If it is a relationship, what is unique about one side of it? Should not consciousness be shared between the group of brains that are interacting and relating? It is interesting that groups do sometimes claim such common consciousness, though it is certainly not usual. But in any case a relationship can have a direction, and there is no need at all for relationships to imply dispersal and sharing. Every node in a network has its own individual set of connections, and it is this set of an individual brain's directed relationships that I conceive of as its consciousness. Self-awareness would then result from a brain modelling the reaction of other brains, and incorporating the fact that the others, like itself, are nodes in an interacting network. This recognition that others are unique but like oneself implies the reciprocal, which is self—awareness: "I am unique, but similar to others." Self-awareness is a product of efficient modelling of the relations between brains.

Intentions and the making of decisions about future actions are important aspects of conscious mental activity; in fact consciousness is often thought of as the arena within which decisions are made and intentions declared. In what sense, it may be asked, is there really an audience here, people to relate to? Many decisions and intentions directly concern another person, and in these cases it is hard to believe that a brain as intelligent as the human's would fail to employ the model it has of that other person. "I shall ring him up"; "I shall kiss her"; "I shall not pay this bill": how could these intentions be declared without having in the forefront of one's mind the people most directly concerned? It must be admitted there are other decisions where the audience is not necessary; for instance "I shall go to the laboratory by way of the market place because I want to buy some apples". But my impression is that this type of decision is often barely conscious -one just finds oneself in the market buying the apples without having consciously made the decision. One can also raise the question as to why consciousness is thought of as an arena at all if there is no audience; it can hardly be the rest of one's own brain one is telling for one does not need to shout or declaim to that.

By now I hope to have established the concept of consciousness as the special quality or feeling that imbues those parts of a brain's activity that deal with the relationships of one individual with others. This requires appreciation of the model—making propensities of brains, and acknowledgement that models of other individuals, not real ones, are sometimes involved in the relationships. To strengthen the concept we must now look at the other part of the workings of a brain, those that are not accessible to consciousness.

THE UNCONSCIOUS, UNRELATING BRAIN

Consciousness has the illusion that it has access to most of the working of its own brain, but this is a foolish conceit and far from the case. There are a host of automatic actions that one can partially control consciously, such as breathing and walking, but clearly one has no introspective understanding of the intimate sequences of muscular contractions required to execute these acts. Similarly with sensory mechanisms: we see a red apple, but have no introspective access to the physiological mechanisms of receptors, retina, lateral geniculate nucleus, and primary cortex that start to label the source of excitation. These aspects of motor and sensory mechanisms are an individual brain's private business, and on the current view it is not surprising that they are inaccessible to conscious introspection.

The conceit of consciousness's claim to know its own brain's working is well brought out by our introspective ignorance of details of the models of the environment and its furniture. These enable us to walk around town, eat a meal, or drive a car, and they are extraordinarily complex, as the experts in artificial intelligence who try to imitate them have discovered. However, all we have conscious access to is the end—product, not the inner workings. A nice demonstration of our ignorance of such details is provided by the contrasted habits of the left (clutch) and right (brake and accelerator) feet when driving. The left foot has to be withdrawn slowly and skilfully when starting, but the opposite action rarely has to be done delicately and the left foot is usually depressed forcefully and suddenly. Most people are quite unaware that their feet have modelled the car's requirements, but the left foot's habit of indelicate depression is dramatically revealed if it is used for braking when driving an automatic.

What is even odder about these models is our unawareness of the process of their construction. We acquire knowledge and skill, and can consciously use the end—product, but we cannot reconstruct the steps by which they were acquired. They are formed by experience and the experience is "remembered" in the sense that it is incorporated into the model, but it is not independently available. This is certainly one important form of unconscious memory, but notice that it requires no active "repression" to explain its unavailability; this simply results from each individual memory being merged in the averaging process by which the model must be formed.

Freudian notions of the unconscious are particularly interesting, for in this case it is claimed that there are past experiences that guide and largely control an individual's behaviour, yet are repressed and thereby deliberately kept from conscious awareness. That may be so, but on the present viewpoint the working of the brain is necessarily unconscious except when it is communicated to others. Active repression is quite unnecessary, and instead some active inducement, a model brain soliciting discourse, is required to bring a thought to consciousness.

Perhaps the role played by the helpful counsellor or analyst is to solicit this discourse directly, and to provide the model for the patient to continue the discourse in his imagination. The increased range of consciousness would thereby be simply and directly explained.

Clearly one could continue to speculate along these lines. For instance, if nature normally imbues interpersonal relations with this special characteristic, perhaps she sometimes fails to do so, as she sometimes fails to provide full colour vision, or as with the occasional congenital absence of pain sensations; could this be the defect in the autistic child? At all events the main point to emphasize at this stage is that if an individual needs other brains in order to become conscious, these other brains will have a reciprocal effect on the individual: he cannot become conscious of what he cannot communicate.

Enough has been said to show that making relations its primary seat leads to a very different view of consciousness. It can no longer be thought of as something added to the physical brain, or "emerging" when the brain reaches a certain size, maturity, or complexity. Consciousness is something to do specifically with that part of a brain that deals with other brains, and this is why it is interesting. The question how the brain brings about these interactions loses interest, for there is no reason to believe that the mechanisms of the brain achieve this function differently from any others. The question whether they are bound by the laws of physics, or have an influence on them, no longer seems a specially important one. Indeed it becomes almost absurd when one appreciates that consciousness is concerned with the part of the brain that handles human relations, for then there are so many other more interesting questions to ask about it. It is like harping on the question "Is speech the physical vibration of air molecules?". Sound might be so described, but speech is so much more than sound that the answer to the question becomes unimportant; an affirmative answer leaves the interest and importance of speech unimpaired.

I hope you are convinced that one can include all the normally accepted aspects of consciousness in a view of it which makes the relating of one brain to another the primary act of consciousness. I cannot conceive how we could logically decide between this view and others which place "Je pense" or "Ich will" or "I feel" in the primary position. But just as the geocentric view of the universe lost ground because it had to be made increasingly complex to accommodate new facts, so I hope that the view that consciousness arises in interpersonal relations will gain ground because it brings out clearly and simply the cause of its own evolution: it gives a clear answer to the question about its functional role and survival value. It is this unifying role that gives point and purpose to the conjecture; but let me first illustrate with a short story.

An informed and friendly person places in your hands what he tells you is a stone, but it is soft and warm instead of being hard and cold as you expected. Surprise prompts curiosity, and you find that there is indeed a stone in your hand, but it has been heated and wrapped in a soft woollen sock. I have been saying that interpersonal relations envelop the hard physical brain in all those aspects where we talk of consciousness, and this is what makes mind appear soft and non-physical.

Now this episode is not bizarre and senseless because your friend gave you the hot stone wrapped in a sock for a purpose, namely to help you keep your hands warm. I am going now to suggest why our dubious friend, Nature, placed this warm and glowing consciousness in our heads; the answer, I am afraid, turns out to imply we are the victims of a confidence trick, but it leads to a less bizarre and senseless view of consciousness than is to be found in most philosophies.

CONSCIOUSNESS, THE INDIVIDUAL AND THE HERD

The answer I want to propose is that consciousness is Nature's trick to ensure that mankind behaves co-operatively, that he is a gregarious animal. We all regard our own consciousness as our most personal and treasured possession: not to be conscious is not to be alive, and to be unconscious is next to death. Yet if consciousness is one's own brain's discourse with other brains, there are only two ways for it to stay alive. The first is to engage in real discourse with real people for as much of the day as possible, and no one will deny that that solution promotes the gregariousness of mankind. The second, as we have seen, is to engage in rehearsal of future discourse. It is this capacity to use his models of other brains, to substitute imagined for real discourse, that seems to make man's social life so different from other gregarious creatures. How long could a purely imaginary discourse be sustained? How can one continuously check that members of the audience of one's imagined discourse are good models and correspond to their prototypes in the real world? The rehearsal and imagined discourse cannot be stored indefinitely and therefore, if I have stated the nature of consciousness correctly, an attempt must sooner or later be made to communicate it to others. This, it seem to me, is the origin of Popper's World 3, the world of human culture, books and other products of man's mind.

Thus, if people spend much time alone, it follows from the biological role of consciousness, as presented here, that culture will be produced. Perhaps cave paintings, those time-defying communications, were an early manifestation; our books, movies, and tapes the more transient modern versions. Attempting to contribute towards or to understand our cultural heritage is an activity that links one to the rest of mankind, and thus is a form of gregarious behaviour, but obviously a much more interesting one than a sheeplike desire to follow, to huddle, or to avoid being on the edge of the flock. The essence of this new trick to ensure gregarious behaviour is that one only experiences one's most personal possession, one's own consciousness, when one is sharing it with others; one only has it when one gives it away. But that is just half the story.

A convention is two-sided. I cannot say anything I like, but only what you will listen to and understand. Conventions must be established and adhered to, including those of language itself. This means that when I preserve my consciousness by entering into a real discourse with you, you have some control over what I can say, what I can become conscious of. There is nothing mystical about this and it is no more strange than a townsman not losing his way in his town because he knows it. It follows naturally from my brain's ability to model you and your ways of thought, to learn your language, to take account of your past responses, known fields of expertise, your likes, dislikes and prejudices. Because I must take these into account to converse with you, you exert some reciprocal control over my consciousness, just as the plan of the town controls the townsman's movements. And, of course, what is true of real discourse with a real person is also true of discourse with the imagined people who substitute for a real person when I am attempting to maintain consciousness by rehearsing future discourse. My family and friends, cultural archetypes, my experience of father-figures and schoolmasters, have a firm grip on what can enter my conscious mind, and because of consciousness's self-importance, they also control my decisions and behaviour.

Now let us look at it from Nature's viewpoint. Can you imagine a neater trick than this for making man a social being? She implants in you this special sensation called consciousness, telling you it is your very own possession, from birth to death; she tells you there is really nothing else that is your own, this is your very being, and it is so important that no new decision can be made, no major intention formulated, without this accompanying sensation. All this, of course, she tells you in her own script, in DNA, so that consciousness inevitably has these properties. But she also arranges that nothing whatever will give rise to that sensation of consciousness except a communication between your own brain and another brain of the species Homo sapiens, though she is not explicit about that. Without knowing why, you spend your life chasing the herd, attempting to communicate with it, and rehearsing such communication when you are unable to achieve it. Man is thus trapped by this glorious, hilarious, trick, but it is a trick with remarkable consequences. Who could have foretold that making an animal gregarious in this particular way would cause the earth to become a storehouse and museum of the products of his mind?

We had better enjoy the joke, for there is no escape; we each are chained for life to the herd and its image. But during that life we could try to ensure that the earth grows as a museum and does not become, too soon, just a dead monument to Nature's Joke.

Nature's Joke: A Conjecture on the Biological Role of Consciousnesss 91
Discussion

RAMACHANDRAN:

Let me first summarize your argument to see if I have got it right. You begin by rejecting epiphenomenalism (according to which consciousness is merely the "inner aspect" of cerebral activity) on the grounds that if this view were correct we should be conscious of everything that goes on inside our brains. Since only a tiny fraction of brain events emerge into consciousness you ask "What is it that characterizes these events and makes them different from other brain events?". Your answer is that the need to communicate certain internal states to other members of the species led to their emergence into consciousness. For instance, we shout when jabbed with a needle and maybe this is possible only because we consciously feel pain. On the other hand, it would be biologically useless to communicate (say) the pupil's response to light to another person and so this has remained an unconscious reflex.

First, your argument requires that the common denominator of all conscious brain states (as opposed to unconscious ones) is the fact that they would have survival value if communicated to members of the species. Unfortunately this does not seem to be true. I am aware of a much wider range of (say) sensations than I would ever want to communicate. For instance, I can see dozens of shades of green. If ordinary language has only one or two words for green (e.g. light green, dark green) then what exerted the selection pressure for me to become conscious of dozens of shades?

I can also see hundreds of depth planes using binocular parallax (stereopsis) and I am conscious of each of these. Why not accept the more conventional view that we are aware of depth in order to obtain visual feedback for prey catching and locomotion? What use would it be to communicate stereopsis to my neighbours? My point is that we can be (potentially) conscious of a much wider range of events than we would ever want to or need to communicate; and so language could not have exerted the selection pressure that led to the emergence of these states into Consciousness.

I would argue that events emerge into consciousness only if they are linked to the brain's decision-making mechanisms, i.e. the centre in the brain that is involved in assessing priorities of action based on certain goal criteria (what MacKay calls the "supervisory system" in this volume). Of course, this centre may be incidentally linked to language areas in man but that does not necessarily implicate language in consciousness.

Second, it wasn't clear to me whether you want to distinguish between linguistic and non-linguistic communication and between creative and non-creative use of language. Much of non-linguistic communication (popularly known as "body language") is completely unconscious, and people can unconsciously exchange a great deal of information without uttering a single word (e.g. the pupil's response to attractiveness). Conversely we often speak of "mindless babble" when people engage in unintelligent (though articulated) conversation. So perhaps you need consciousness only for particular kinds of communication.

Finally, even if we accept your argument that consciousness is intimately related to language, it doesn't follow that consciousness is causally effective in permitting or facilitating communication. If the physical world (including communicating brains) is a closed system then I don't see how consciousness can influence it. Indeed there is nothing logically impossible about brains communicating actively without consciousness ever coming into the picture. Would you agree with this?

BARLOW:

I almost agree with your summary of my argument but would like to add a bit more. I am postulating that the innate desirability of consciousness, together with the impossibility of attaining it except by communication, is an important factor in causing man to be social and gregarious. Thus the need to communicate that you refer to becomes more than a matter of finding pragmatic solutions to current problems; it becomes more like the motive for all existence. So in answer to your first question I would say that not every conscious brain state need have survival value if communicated; it is the practice of social communication that has survival value, and single communications may not. Thus I don't really find it contradictory that our sensations are more fine-grained than our communications. Surely our sensations do not contain details that we can be sure we would never wish to communicate; in seeking the right tint of paint you might ask for something "less bluey, more olive", thus using a detail of your fine»grained sensory representation in a social communication. Similarly with stereopsis you might say "nearer" or "further" in directing someone how to pour the champagne into your glass and not onto the floor.

I would also agree with your remarks about consciousness being linked to the decision- making mechanisms, but think the decisions we are most conscious of are those with social implications, whereas decisions with no social relevance are often made unconsciously.

With regard to non-verbal communication, are you sure "body-language" is entirely unconscious, to either communicant or recipient? But I don't want to evade an important issue here. Whereas I think there is little if anything in consciousness that is of purely private concern, I very much doubt if the converse is true; I Very much doubt if all the social communications we make and receive pass through our consciousness. So I agree there is a missing factor here.

Finally, why should not brains communicate without consciousness? Well, as I've just said, I think they do. In the same way I think an animal's spinal cord mediates reflex responses to noxious stimuli without experiencing pain. The subjective experience of pain puts in motion strategies for longer—term avoidance and recovery. In the same way consciousness is something we desire and seek, but can only achieve by communication; attaining this consistently calls for long-term planning.

VESEY:

You say that what philosophers have said about consciousness seems mainly based on introspection. This is not true of what Wittgenstein says about consciousness in his later works (e.g. Philosophical Investigations, Pt. I, sections 412 ff., 1953). In fact, I wonder whether you couldn't make use of a rather Wittgensteinian argument to support what you say about a connection between communication and consciousness. It is the argument propounded by Anthony Kenny in the I972/3 Gifford Lectures (A. J. P. Kenny et al., The Development of Mind, Edinburgh University Press, 1973, pp. 9I—l07). Very briefly, the argument is (i) that the behaviour of language-users is rule-governed, (ii) that if someone's behaviour is governed by a rule he must be to some degree conscious of the rule, and (iii) that it would not be possible to distinguish between correct and incorrect applications of a rule if there were not a community of language-users. Obviously it is too much to ask you for a snap decision on the validity of an argument that has quite a lot of theorizing about the philosophy of language behind it. But I'd be interested to know whether you think the sort of consideration Kenny advances is compatible with an evolutionary explanation of consciousness. (Kenny himself says that: "there seem to be profound difficulties in principle in seeing how the practice of following rules could have . . . been produced by natural selection' ' .)

BARLOW:

I did not want to insult philosophers by implying that all they have ever said about consciousness stems from introspection. However, I don't think they often go as far as they should in mistrusting common—sense introspective ideas about consciousness. My reason for this mistrust arises partly from recognition that our actions can be guided by brain events and mechanisms of which we are unconscious; you do not have to be a dogmatic Freudian to accept that this does sometimes occur. But in a more important way it stems from comparisons of the biological and introspective viewpoints on other prominent subjective experiences. Pain, to a biologist, is a signal to an animal that warns of more serious injury and is thus protective, but there is no trace of this protectiveness in the subjective, introspective, experience of pain. Similarly with love: to a biologist this is the set of emotions that guide reproduction and child rearing, but when touched by it we do not rush forward and say "Ah, how fertile and motherly you look", nor would it be well received if we did. Our nature is such that we experience conscious thoughts and feelings that make us act in certain ways, and it is these actions that serve Nature's purpose. The logic and rationality of the whole sequence is not at all evident in the part that we consciously experience, and my thesis is that this is true of consciousness itself.

I am afraid I have not attempted to tackle Wittgenstein on this matter, but I confess that 1 find Kenny's argument amazingly unconvincing. Does he really think that one must be to some degree conscious of every rule one follows? Can he give me a list of the rules he follows when choosing his footfalls on a rocky mountain path? Or when deciding that his daughter's face is the third from the left in the front row of the school photograph? Or when deciding the appropriate way to introduce his guest in a crowded room?

The suggestion that the practice of following rules could not have been produced by natural selection was, I thought, largely demolished in the discussion by Waddington and Lucas that followed Kenny's contribution to the Gifford Lectures. I do not know that much more need be added, but I am sure Kenny would agree that few human mates are selected (by either sex) without the use of language. Since language is a rule-following game, it is only too easy to see how its skilful use can be a positive factor in survival and propagation; fine words, for the human, are at least the equivalent in this respect of fine tail-feathers for the peacock. And it must be added that words and language are a great deal more useful, and confer much greater survival advantage, in other circumstances.

So in summary I do not see any reason why consciousness should not have evolved by natural selection. But to accept this you will have to look at the way it induces us to behave, not at your subjective experience of it.

RAMACHANDRAN:

In a sense your argument seems to be the exact converse of what Nick Humphrey has suggested (Chapter 4). Would you like to say anything about the relation between your views and his?

BARLOW:

I was as surprised as anyone else to find that we have been thinking about the same subject, but this surprise should be tempered by the knowledge that both of us are receptive to current ideas in sociobiology, and an extension of these ideas to philosophical problems was a natural channel for both our thoughts to flow along.

It is gratifying that we agree on the answer to the question "What does consciousness do?", namely that it promotes man's survival as a social animal. But then we seem to diverge, for I assume nothing about consciousness except that it is desirable to the individual, and only attainable in real or imagined social relations. Nick Humphrey takes a less radical view, endowing consciousness with many of the properties that we think it has by subjective introspection. Thus 1 think he assumes that, because we are conscious of some piece of information we shall make better, more rational, use of it. Now I don't see why this should be so: why should not observation of the behaviour of a conspecific creature that is hungry, love—sick or jealous lead to unconscious recognition of the cause of its behaviour? Indeed I think this does often happen, and social behaviour induced by such unconscious intuition is as rational and appropriate as that which follows conscious recognition. Nevertheless one can see that there is an important difference between the unconscious and the conscious case, for in the latter one can communicate directly about the motive state; one can say "I see you are hungry and shall give you food", whereas the unconsciously motivated response could only have been the action of giving food. Thus we see yet another link between consciousness and communication.

To put this another way, I think I neglected the aspect of consciousness that Nick Humphrey emphasizes most strongly, its input from one's own emotional state. But whereas he says this is important for understanding others, I think this understanding comes at least as effectively from unconscious intuition, and the importance of conscious awareness of one's emotional state lies in facilitating communication; this awareness adds important words to our language. And I must say I prefer my more radical view of consciousness as the prompter and initiator of man's social and intellectual life to his more conservative view that it simply facilitates these activities and makes them more effective.

# CHAPTER 6

Conscious Agency with Unsplit
and Split Brains

D. M. MACKAY

University of Keele

## ABSTRACT

The bizarre symptoms produced by section of the corpus callosum in man have led to a variety of speculations about the consciousness to be attributed to the result. Are there now two conscious minds, or even two persons, where there was one before? Is one brain hemisphere conscious, the other unconscious? Do normal (unsplit) brains embody two conscious persons all the time? Should we grant that both halves are conscious but only one is self-conscious? And so on.

In this paper I want to examine some of the presuppositions that underlie such questions, from the standpoint of information engineering. The intention is not to enter into the vexed question whether automata can be conscious, but only to take advantage of a system of concepts common to both neurology and automata theory as a scaffolding on which to feel our way around these perplexing problems. From this standpoint I shall argue that in order to attribute significant determinative power to conscious processes we have no need to rely on any breach of physical causality in the nervous system.

Our primary focus will be, not on the processing, storing and retrieval of informa- tion, nor merely on the co—ordination of sensori-motor performance, but on the evaluative aspects of conscious agency. Unless a surgical operation splits the evaluative hierarchy into two autonomously functioning wholes, there would seem to be no justification for considering the result to be two independent, conscious individuals, however elaborately absent-minded the victim might be.

In the interests of clarity, it is suggested that to speak of "hemispheres" or "brains" as conscious is to fasten on the wrong target. It is agents who may (or may not) be conscious, not brains or half-brains.

In this contribution I want to raise three questions, two of which I hope will clear the ground for the third. First, starting from the ground level of common experience, how does talk of "consciousness" arise, and what implications has it for our view of physical reality? Next, at what level of analysis, and in what categories, might we hope to find distinctive features of brain states in which a subject is conscious, as opposed to those in which he is not? Finally, under what conditions does it make sense to claim that we are confronted with two or more conscious individuals? This last question will be related particularly to the bizarre phenomena manifested in cases where the human brain has been partially split by section of the corpus callosum.

HOW DOES TALK OF CONSCIOUSNESS ARISE?

What we call "conscious experience" is for each of us the primary datum to which all our thinking must do justice —the ground on which we must build even our doubting. All our knowledge of the physical world and of other people goes back to this base; so any attempt to deny the foundational reality of conscious experience would be derisorily self-cancelling. Whatever else "exists" or "does not exist", the existence of at least one conscious agent in the world is a fact for all of us.

When trying to relate the data of conscious experience to what we believe about the physical world, I have found an imaginary visual aid' helpful in the interests of semantic hygiene. If you were asked to bear witness to the content of your conscious experience, you could in principle write down a long list of statements in a vertical column, each beginning with "I". "I see—such-and-such"; "I hear—that"; "I feel—thus—and—so"; "I remember— . . ."; ''I believe— . . ."; and the like. Call this collectively the "I-story". The I-story bears witness to data that you would be lying to deny.

Now the objective of those of us in brain research is to fill out entries in a parallel column (say to the right of the first), describing states of or processes in your central nervous system that correlate with the facts listed on the left. We may call it the "brain story". How do we get to this from our base in experience? The general answer is: through our experience as conscious agents; but to trace the logical path without jumping illegitimate gaps will take some care and patience.

As conscious agents we find ourselves having to reckon with constraints (boundary conditions) on our action and our planning of action. We cannot move freely in all directions, for example. There are "objects" in the way, and other limitations and enablements (such as "gravitational forces") to be taken into account. These constraints, many of them conditional on one another, are regular enough to be worth naming, modelling, analysing mathematically, and so forth. We attribute them to "physical reality", meaning what must be reckoned with in planning and taking action with our muscular system. Through our sensory experience our conditional readiness to reckon with the physical world is continually updated. This updating in self—matching response to the demands of sensory information we call conscious perception of the physical world?

By suitably designed exploratory and experimental interaction we can analyse and elaborate the structure of environmental contingencies and develop a scientific map of physical reality which enlarges our conditional readiness far beyond the immediate correlates of our own perceiving. The total structure of our conditional readinesses represents (embodies) what we believe about the physical world.

One relatively minute sample of physical reality we each find we have to reckon with in a special way. It is our own nervous system. This can in principle be prodded, weighed, dissected like other physical objects; but whereas changes in the rest of the physical world are known to us only by observation or report (and may be ignored by closing our eyes, ears, etc.), there is good evidence that certain physical changes in certain parts of our nervous system directly correlate with changes in our conscious experience. In many regions of that system, especially in the periphery, physical activities may normally be necessary correlates of sensory experience and the like; but (as the use of peripheral anaesthetics or the results of cortical brain damage can demonstrate) they are not at all sufficient. Only in deep central brain structures, most of which still await detailed identification, do we find physical activities so unconditionally correlated with the subject's experience that we may suppose them to be sufficient conditions of that experience. (Even then, of course, this is only a working hypothesis made plausible, but not conclusively demanded, by the data of neurology.) Although the details of the correlation between "I-story" and "brain—story" are at present obscure, the conjecture we are considering is that no change can take place in your conscious experience without some corresponding change taking place in the physical structure concerned. This in principle defines the postulated cerebral correlate of any experience: it is that physical state or process which must change if any change takes place in that experience.

The point of our visual aid is just to remind us that mental terms like seeing, feeling, thinking, hoping, believing, and being conscious all belong to the left-hand column. The corresponding places on the right, if the assumption were valid, would all be occupied by references to physical, or at any rate mechanistic, concepts such as nerve-cell firings, synaptic modifications and the like. Note that these are not translations of mental terms but only correlates, in something of the sense in which an electronic engineer's account of the physical process in a computer is a correlate of the description a mathematician might give of what the computer is doing. Neglect of this distinction causes much confusion, as when people claim that in conscious experience we perceive directly what is happening in our brains. If we actually wanted to do that, we should have to use appropriate instruments like anybody else; and as we shall now see, we might run into sufficiently profound epistemological difficulties to convince us of the difference between self—awareness and self-observation!

THE COSTS OF KNOWING

According to the working hypothesis of brain science, then, we come to know only at the cost of dedicating a relatively minute region of the physical world inside our own heads to the purpose of representing what we know or believe. Let us call this our "cognitive system". Because it is the region of our nervous system that has to represent what we know or believe, by determining the corresponding conditional constraints and enablements, our cognitive system must change significantly as what we know or believe changes; and by the same token, it is itself something not to be known by us. (In a community of persons in dialogue, as we shall see later, this dedicated region may expand to include the cognitive systems of those in dialogue with us.)

"Not to be known" here means "not to be known until afterwards". Tell me later if you like, but not now—your effort is bound to be self-stultifying. If you had a completely accurate and detailed state-description of the immediate future of my cognitive system, the changes that would be necessary to embody it in my cognitive system now must (logically must) render it out of date for me. Until afterwards, in a certain strict sense it is not just undiscoverable by me but indeterminate-for-me.

This last distinction is important.

(a) A state may be unknowable—by-A, or undiscoverable-by—A, in the weak sense that although there exists in principle a state-description which has an unconditional claim to A's assent (i.e. A would be correct to believe it and in error to disbelieve it if only he knew it) A cannot acquire this description. (Example: a detailed state-specification of the particles in the centre of the sun.) This is not a particularly interesting case, unless perhaps to a very old-fashioned variety of logical positivist.

(b) A state may, however, be indeterminate—for-A, in the much stronger sense that no fully detailed description of it exists which A would be correct to believe and in error to disbelieve.' The immediate future of my cognitive system is indeterminate-for—me (and of yours indeterminate-for-you) in this strong sense — quite regardless of the extent to which it may be affected by any physical (Heisenberg—type) ' 'uncertainty' ' .

Now of course socially, especially for scientific purposes, we share a concept of "the physical world" as if it were well defined or even (in pre-Heisenberg days) completely determinate. The point we must note, however, is that when each of us appropriates the concept for himself, he has to recognize one region of his physical world (a different one for each) to be systematically indeterminate - namely, the region that represents his present knowing. The physical world may be littered with physical relics of my past knowings and those of others, which are now fully determinate for me and everybody else. But at any given time there IS in our physical world a flickering, shifting little area of indeterminacy, different for each of us, which waits for us to determine its state by our cognitive activity.

_ In this sense, the shared social concept of a determinate physical world is strictly a make-believe, even without reference to Heisenberg uncertainty. If our brains are accepted as part of the physical world, then the true picture of that world for each of us includes an irreducible area of indeterminacy, whereof we as cognitive agents are the necessary determinants. This partly indeterminate world is the only world we know. Any Image of the physical world that showed all our cognitive systems in complete1y determinate future states would be false to reality, in the strict sense that we would be in error if we believed it to be the only one

possible. Such an image of someone else's brain may exist with an unconditional claim to my assent, if I am physically uncoupled from him; but none exists with such a claim to the assent of all. In this strict sense the social concept of a fully determinate future for the physical world is logically bankrupt. It is like a cheque which is endorsed "Not payable to anyone signing"."

LOGICAL RELATIVITY

It may be tempting to feel that if others (sufficiently detached observers) could in principle observe and predict the future of the area for which we have a "blind spot", this would show that it was not "really" indeterminate for us, but that we were just "invincibly ignorant" of its immediate future. But this would be a logical mistake. The most we could validly conclude is that the area was not indeterminate-for—others—but this was granted at the outset. The situation is relativistic, in the strong sense that no single determinate total state-description of physical reality can exist upon which all would be unconditionally correct to agree. What the detached observers know about the future of our brains is not knowledge for us. It is not that knowledge exists which we lack and cannot gain, or cannot be persuaded to accept. The truth is rather that because of our unique relationship to the subject matter, what the detached observers correctly believe would be inaccurate information for us if we had it. It would thus be a solecism to describe our lack of it as "ignorance", invincible or otherwise. This logically tantalizing conclusion is perhaps the most remarkable consequence of the assumption that conscious experience is physically embodied}

Objectors to the foregoing argument have sometimes claimed that it could be circumvented by ensuring that the future state-description was corrected to take account of the changes in the cognitive system that would be produced by embodying it there. Recondite mathematical theorems have been invoked to prove that this is in principle possible." It may be worth taking a few lines to see how this move misses the point.

Suppose that (with the help of the "Fixed Point Theorem" or whatever) a super—scientist could derive a detailed future state-description of your cognitive system that would become accurate if (but only if) you believed it. Then indeed he has found a possible description that you would be correct to believe; but unless he were allowed to interact with your situation so as to make you believe it, in the way he has assumed in making his calculations, it must remain false. Furthermore, and logically more important, what he has derived is something that you would not be in error to disbelieve! Thus whether or not in fact you were given it and believed it, his cooked-up state-description has no unconditional claim to your assent (such that you would be correct to believe it and in error to disbelieve it). On the contrary, your assent is one of the factors that will determine whether it is correct.

Even if we imagine that in a given case the super-scientist could predict that you would be given it and would assent to it, and you did, it would not follow that its logical claim on you was unconditional; for that would require him to show that you would have been in error had you rejected it. But in fact if you had disbelieved it, the assumption on which it was based would have been ipso facto false, so you would still have been right to do so! Thus although you were not mistaken to believe it, and he (having been allowed to influence you) was right to predict you would, its claim to your assent was not unconditional. You would have been mistaken to accept his prediction as inevitable-for-you.

## A DISTINCTIVE CORRELATE OF CONSCIOUSNESS?

One particularly important class of change in my nervous system is that which causes me to "lose consciousness". When this happens I become unable, then or later, to bear witness to events observable by others during the time that my system was in this abnormal state. (In passing, this common phenomenon surely throws doubt on the notion sometimes aired, that consciousness is something that arises automatically when matter is organized into structures of sufficient complexity!)

Metaphorical talk of "losing consciousness" or "regaining consciousness" might seem to lend support to speculations that "consciousness" is the name of some kind of entity, like fuel or electric charge, that can exert quasi-physical influences on the brain; but this inference would be as invalid as the conclusion that "loss of balance" in a wheel, or "loss of stability" in a servo system, must refer to the escape of intangible substances called "balance" or "stability". When I regain consciousness I become a conscious agent -1 do not acquire one. Note too that it is I who become conscious, and not my brain or CNS. Doubtless when I am conscious my CNS is organized in some correspondingly distinctive way; but it would be a solecism to claim that it is conscious.

How then may we hope to make explicit what is distinctive about the cerebral correlate of consciousness? As it happens, the requirements of industry and war for the mechanization of intelligent action have given rise to a conceptual framework that seems well adapted for the purpose. It goes by the name of information-flow analysis. Applied to a living organism it means a systematic attempt to identify needs for information in the functioning of the organism, and to discover how the necessary information is acquired?

In any organism there is a more or less rich repertoire of possible modes of action in the world, including complex "sub-routines". Information is needed to determine the running selection from this repertoire that constitutes the behaviour of the animal. Where does this information come from? In part, of course, from the receptor system; but the raw data are not enough. At a minimum, if behaviour is to be goal-directed, there must be a stage of evaluation of incoming information in the light of criteria sufficient to determine whether, and if so what, corrective action is called for. Normally the calculation of adaptive action will have to take account also of stored information, kept up to date by sensory input.

Where then do the criteria of evaluation come from? In something like a thermostat they are set by a human supervisor, but in an autonomous organism any reordering of goals and priorities is normally organized internally. Let us call the system responsible (whatever it may turn out to be) the supervisory system. Because of the complex array of alternative goals and sub-goals, norms and satisfactions pursued under different conditions by human beings in particular, the human supervisory system must be at least hierarchic, and more probably heterarchic (having feedback between levels) in structure.

At this point we face a choice. Some thinkers, of whom Sir John Eccless is a distinguished contemporary exemplar, would locate at least part of the human supervisory system outside the physical world altogether. This would mean that some of the lines of information-flow in our map would have to terminate in thin air, so to say: the "self-conscious mind" would exert non-physical influences on appropriately "open" elements of the nervous system. It is important to recognize that no scientific data rule out such speculations. Those of us who object to them do so rather on the grounds that they multiply entities beyond any demonstrated necessity. I must indeed confess to feeling the same objection to Professor Josephson's conjectures in the present symposium, even though (in his spoken presentation) he prefers to speak of consciousness as "some kind of physical substance".

But could an information-flow model without such extra-physical entities be compatible with the facts of conscious experience? Above all, could it offer any natural (rather than contrived ad hoc) distinctive correlate of the conscious as opposed to the unconscious state? It is sometimes objected that for any mechanistic theory of brain functions, conscious and unconscious states must be equally and indistinguishably reducible to "mere" configurations of nerve impulses; but this I believe is a mistake based on the wrong level of analysis — as if someone were to claim that to a telegraph engineer all messages must be indistinguishable in terms of the physical currents carrying them. Once you know the code, the opposite is true: physical data can actually provide a way of cross-checking a description of the message being sent.

EVALUATIVE SUPERVISION

The alternative view, which seems at present the more parsimonious, would be that all the lines of cause-and-effect on the information-flow map form closed loops within the brain or by way of the external physical world, and that the distinctive correlate of a conscious state should be sought first in the form of the informational activity. The conjecture I would favour° is that the direct correlate of our conscious experience is not the activation of sensory receiving centres, nor even the exercise of sensori-motor co-ordination as such, but the "meta-organizing" evaluative activity of the supervisory system. According to this conjecture, any brain centre could in principle be active without my necessarily experiencing a conscious correlate; but any change that was significant at the level of the evaluative supervisory activity in my CNS would have its necessary correlate in my experience. What is distinctive, according to this hypothesis, is the information-flow structure of the physical activity concerned, rather than any vulnerability of the physical components to extra-physical influences.

Without going into details, we may note that on this view the total physical correlate whose "informational shape" determines the content of conscious experience will constantly change with the domain of supervisory activity. In driving a car or playing a game of tennis, for example, the evaluative flow-system will have informational feedback loops reaching out not only into primary cortex but also into the outside world of action. It is well known that a blind man with a cane perceives his world as out there at the tip of his cane, not in his palm. The total physical system in which my conscious experience is embodied at a given time may thus extend far beyond the boundaries of my skin, which from this standpoint are largely incidental.

What then of the distinction between human self-consciousness and the general conscious awareness that most of us would attribute to lower animals? Without any further assumptions ad hoc, our analysis suggests a natural correlate. We can think of the human supervisory system, like that of other animals suitably equipped, as keeping up to date the internal representation of the world by an active matching response to incoming information. If, however, this response in man is organized partly at the abstract conceptual level developed for purposes of verbal communication with others, we could expect our correlated experience to link directly with Verbalized thinking: "There goes so-and-so on his bicycle"; "What's that? —Oh, it's the missing cuff-link"; and so on. If now we imagine the field of "incoming information" widened to include information about the activity of the supervisory system itself," we can by the same token expect the agent's conscious experience to include such trains of thought as: "There I go making the same mistake as before" or "Which of these do I prefer?" In short, the game of (internalized) talking to oneself about one's world, including oneself, would follow naturally on the development of the game of talking to one another, without any apparent need to suppose that the hardware of the human CNS must be open to non-physical influences of the sort postulated by Eccles[8] The re-entrant information-flow loops set up when the supervisory system became the subject of its own internal representation could, of course, be expected to introduce special possibilities of oscillatory or co-operative behaviour with qualitatively unique correlates in experience (and with the logical consequences outlined on p. 99); but none of this would seem to take us outside of the domain of normal behaviour at the physical level.

Now I have stated all this as only a conjecture, and you may well ask why we should pick on the evaluative function as crucial. My reason is that (following John MacMurray") I think of the human self primarily as an agent: one who evaluates his situation and attaches relative priorities to alternative modes of responding to it, including inactivity (passive undergoing or suffering) at one extreme and the upheaval of his whole scheme of priorities itself at the other. It is when activity becomes sufficiently stereotyped to require no evaluative supervision that we tend to become unconscious of it—even though it may employ elaborate sensori-motor co-ordination and a rich supply of stored information. Conversely, it is when we struggle with conflicting demands on our central priority-scheme that we are most acutely conscious of what we are doing and suffering.

The hypothesis is also in line with a wide range of clinical observations on the kinds of brain lesion (mostly in deep central structures) that abolish all conscious experience in coma, as opposed to those confined to cortical levels which seem merely to affect its detailed content." Moreover, it is in the central (hypothalamic and diencephalic) sub-systems of the brain that physical changes seem most closely correlated with conscious moods, desires and evaluative affect. Although in man these structures are particularly elaborately integrated with others such as the frontal lobes, damage to the latter seems only to affect the sensitivity and coherence with which priorities are evaluated and changed, rather than abolish all signs of conscious appraisal.

SPLIT BRAINS — HOW MANY CONSCIOUS INDIVIDUALS?

Dr. Ramachandran has already referred to the bizarre symptoms that result from surgical section of the human corpus callosum — the elaborate cableway of several hundred million nerve fibres that link the two brain hemispheres. Although according to Sperry" speech, verbal intelligence, calculation, established motor co-ordination, verbal reasoning and recall, personality and temperament are all preserved to a surprising degree, there is a strong dissociation of reactions to stimuli presented only to one half or the other of the split sensori-motor system. Because the speech organs are normally controlled only by one (usually the left) hemisphere, a patient may verbally report seeing only a stimulus flashed to the right of his fixation point, while at the same time correctly identifying with his left hand (controlled by the right hemisphere) an object whose name was flashed to the left (and so signalled only to the right hemisphere).

The philosophical implications of these dramatic findings are currently a matter of keen debate. In particular, how many conscious individuals are there in a split-brain patient? Sperry himself happily speaks of "two rather separate streams of conscious awareness". "Each hemisphere", he says, "has its own private sensations, perceptions, thoughts and ideas . . . its own private chain of memories and learning experiences." This claim as it stands involves the transfer of an "I-story" category to the "brain-story", and strictly makes no more sense than to claim that an unsplit brain is conscious, rather than the person whose brain it is. But the question Sperry is raising is a real one. Without committing any transgression of categories, we can still ask whether the two split hemispheres are now the brains of two conscious individual persons, who just happen to share pre-operative memories and a common body. This is neither inept nor inconceivable, and at least one philosopher, Puccetti," has argued in this way. (Puccetti actually goes further, and suggests that on these grounds each of us with unsplit brains must in reality be two persons; but to the logic of this we shall return.)

Before we jump to such conclusions, however, I believe that an information-flow analysis of the situation should sow some legitimate seeds of doubt in our minds. For obvious reasons, the surgeon splitting the corpus callosum to relieve an epileptic patient leaves intact as many oentral brain structures as possible. In particular, there is no division of the deeper central structures concerned with the most basic and dominant evaluative functions. If the normal human evaluative system is an integrated hierarchy or heterarchy (a hierarchy with inter-level feedback) then it is not at all obvious that an operation which splits at most its peripheral levels should bring into being two independently conscious individuals. Admittedly, one "split" patient has been reported to button up his trousers with one hand while unbuttoning them with the other, which suggests some independence in executive goal-setting mechanisms. But there is no evidence, and indeed it seems neurologically implausible, to suggest that more than one independent evaluative hierarchy had come into being. Bizarre though it must be, such experience of divided executive control would seem more parsimoniously bracketed with things like absent-mindedness, or the experience of "finding oneself in two minds", than with the discovery of an identical (but conflicting) twin. However vigorous the conflict at the executive level, it seems more analogous to a quarrel in a sub-committee than to the formation of an independent rival organization. The same boss, with the same ultimate criteria of evaluation, presides over both sub-agencies.

ARTIFICIAL AGENCY IN DUPLICATED STRUCTURES

At this point an example from the engineering of artificial agency may help us sort out our ideas. Suppose that a radar-guided automatic missile director were, for reasons of reliability, constructed entirely in duplicate. Each unit, we may suppose, is wired in parallel with its twin, so that both share all tasks. How many missile directors have we? The engineer's criterion is quite clear. If there is only one integrated goal-directed flow system with a single central evaluator, we have only one director. Suppose now that we begin to split the system by cutting the paralleling links. At what point would we claim to have two directors? Obviously if the splitting were 100 per cent complete, each half could thereafter function individually. But if only, say, the radar systems were split, or the information storage systems, or the "sensori-motor co-ordinating" systems that linked both of these with the missile controls, it would be quite misleading to describe the end-product as two directors. True, it would be possible, by stimulating each radar receiver separately, to have independent and even conflicting commands issued to missiles (especially if the missiles controlled by each half were different). But as long as the system had only one central evaluator defining its criteria of match and mismatch, the only directed activity would be that which was calculated to reduce mismatch and optimize match according to those criteria. The most that peripheral splitting could do would be to increase the number of degrees of freedom of the executive sub-system, at some risk to overall efficiency.

Conversely, it may be useful to remind ourselves that one and the same piece of undivided hardware can easily embody two or more autonomous artificial agents, each with its own evaluative hierarchy. A stock example would be the use of a general-purpose computer to embody two artificial chess players sufficiently independent to play against one another. Always the key question is how many independent evaluative roles can be played simultaneously. The physical singleness or multiplicity of the hardware is relevant only in so far as it bears on this question.

In the case of split brains, of course, I do not wish to argue that "hardware" considerations can safely be neglected. For all we know, our conscious experience may depend not only on having the right pattern of connections in our brains, but also on phenomena analogous to physical co-operativity15 which may (who knows?) depend on some particular combination of spatial contiguity and molecular structure obtaining only in the natural CNS. (Think, for example, of the combination of physical factors that determine whether a nuclear reactor, or a coal fire, "goes critical".) So I am very far from suggesting that we can guarantee conscious experience to any artificial agent that happens to have the appropriate evaluative structure, and I would certainly not personally attribute it to our missile director, whose "supervisory system", after all, is a human operator. The purpose of our excursion into the theory of artificial agency was rather the converse—to show how unjustified it would be to assume that peripheral splitting of the CNS, so as to produce independent sensori-motor and information-storage systems, is enough of itself to set up two independent "streams of consciousness".

HOW MANY INDEPENDENT SUPERVISORY EVALUATORS?

The key question, I am suggesting, is how many independent supervisory evaluators are able to function simultaneously. The answer for all of the human split-brain cases described so far appears to be: one. As Sperry himself admits, emotional reactions triggered by offensive signals to one hemisphere seem to be entirely unified, showing no sign of lateralization. Attempts to set up concurrent conflicting emotional sets have failed." When the non-speech hemisphere receives an offensive input, the patient as a whole blushes or giggles and bears witness to an experience of embarrassment, even though unable to say why. At the outer levels of the evaluative hierarchy that are split, it is only to be expected that independent and even conflicting sub-criteria of evaluation can be established once conflicting patterns of experience have been mediated by the two hemisphere systems; but this does not show that either has broken free of the unifying central supervisory evaluation that (according to my conjecture) identifies the conscious agent. It is significant that Sperry (loc. cit., p. 8) finds "separate parallel perform- ance on different tasks, though possible under special facilitating condi- tions, . . . not to be the general rule. . . . Attention in many tests seems to become focused in one separate hemisphere and simultaneously become repressed in the other. . . ." He points out that brainstem orienting mechanisms are undivided as well as the cerebellar controls for motor co-ordination, and warns (p. 10) that "With most of our research interest naturally concentrated on the divided aspects of brain function, it is easy to underemphasize the many components of behaviour that remain unified".

The fallacy in arguing back from split-brain data to the conclusion that two conscious persons coexist in normal human beings should now be doubly apparent. Not only are the data insufficient to prove that separate persons are embodied in split hemispheres; but even if they were sufficient (if, for example, in a science fiction world a human brain could be split so deeply that two independent conscious agents clearly had their separate supervisory evaluators in the severed halves) it would not follow logically that there must have been two before, any more than in the analogous case of our missile director. In particular we must bear in mind the evidence from ablation studies that Parkinson's Law (' 'work expands to fill the space available") seems often to apply in the neuronal domain. Any drastic change in connectivity is likely to bring about some redeployment of essential functions, possibly involving the "cannibalization" of structures whose role in the intact brain would have been quite different. This point would apply even more strongly if large-scale co—operative phenomena do play a functional part in the working of the nervous system, since the resulting dynamic patterns of activity could have the same kind of mobility and lability as the flames that flicker over a coal fire.

DIALOGUE: THE ULTIMATE TEST

Perhaps the most characteristic conscious human activity is that reciprocal interaction with others which we call dialogue. I am not now referring to the non-committal alternating monologue that sometimes passes for dialogue in our sophisticated society, but to the deep-going relationship of mutual vulnerability through which another in a special way becomes "Thou" to me and I to him. The distinction between the two seems to have an illuminating parallel at the level of information-flow analysis. As long as someone communicating with another is able to shield his own evaluative system from the address of the other, he can in principle treat the other as an object, a manipulandum, open in principle to full scientific specification like any other physical object. Once the barriers to fully reciprocal communication are down, however, a specially interesting configuration becomes possible, in which the information-flow structure that constitutes each supervisory system interpenetrates the other, and the lines of flow from each return by way of the other, so that the two become one system for purposes of causal analysis.

In this relationship, each conscious agent becomes indeterminate for the other (for the reasons explained on p. 99) as well as for himself. Each is mysterious to the other, not merely in the weak sense that the other cannot gain the necessary completely determining information, but in the strong sense that no such information exists, either for him or for his interlocutor, until after the event. There are "interaction terms", as a physicist would say, in the joint state-equation, which prevent it from having a uniquely determinate solution for either, even if the physical systems concerned were as mechanistic as pre–Heisenberg physics pictured them.

Coming back then to the two halves of a split brain, we might have strong grounds for recognizing two conscious individuals if each could be fully "Thou" to the other in dialogue; but this is just what can never be, so long as the deep central supervisory systems in these human cases are undivided. No mutual interpenetration of independent evaluators is possible, unless at the most superficial levels, where any "dialogue" that could be implemented would amount to little more than the debates with oneself that every normal individual conducts without serious danger to his individuality. Where linguistic ability was closely linked with only one hemisphere, that would of course be a further reason for doubting whether the agent embodied in the other hemisphere could be self-conscious on the lines suggested on p. 104. In this sense (though for different reasons!) I am sympathetic to the suggestion by Eccles[3] that any consciousness associated with the "minor" hemisphere might best be compared with that of a non-speaking animal.

Incidentally, it is in exchanges at the evaluative rather than simply the informative level that we see one of the most important biological advantages of human communication. It provides a vital means of resolving potential goal-conflict. To revert to an old analogy," imagine two air—conditioners operating in a common space with incompatible goal settings. What will happen? Clearly, a tug-of—war: one will run flat out heating and the other cooling until one of them breaks down, when the survivor can relax to a normal level of activity. How could this mutual attrition be avoided? One solution with some survival value would be to equip each with an arm and hand that could pull out the plug of any other it encountered (if the other did not get there first). The most viable solution, however, would be for one to use its hand to adjust the goal setting of the other into conformity with its own, so that each could settle down to carry only half a normal load.

Taking this as a parable, we can see the huge biological utility of conscious communication as a means, first and foremost, of mutually adjusting evaluative supervisory systems into compatible states. A true "meeting of minds" is a form of internalized mutual adjustment of evaluators. For such an interaction to be possible at the deepest levels between the split halves of a patient's CNS, I conjecture that more than the cerebral hemispheres would have to be separated. Without this possibility it would seem hardly justifiable to regard such unfortunates as two persons each.

CONCLUSION

The hypothesis that all our conscious perceiving, knowing, desiring and doing have specific physical correlates in the informational traffic of our CNS seems likely to accommodate all the data of human experience, both secular and sacred, without the need to postulate non-material Substances interacting with the brain. It leads directly to the conclusion that physical reality as known to each of us has a small but significant domain that waits to be determined by our cognitive agency, and has no completely determinate future state—description with an unconditional claim to our assent. It also leads naturally to a qualitative distinction

112 D. M. Mackay

between the conscious experience attributable to lower animals and the self-consciousness attributable to agents whose supervisory systems have the capacity for self-description.

I have suggested that the key element in conscious agency is evaluation —the assessment of states of affairs as desirable/ undesirable according to a (hierarchic or heterarchic) scheme of basic priorities. This implies that it is in the unity of a coherent evaluative system that a conscious individual has his unitary identity. If so, a surgical operation that divides the sensori-motor co-ordinating mechanisms but leaves intact the final levels of the evaluative hierarchy cannot be said to have created two conscious individuals, even though the resulting individual may reasonably be said to find himself "in two minds" in a distressingly wide range of special circumstances. If an operation were ever performed which left two independently viable evaluative hierarchies, the case would be different; but it may be time enough to consider this when it arises, in the light of the evidence.

REFERENCESREFERENCES

Discussion

JOSEPHSON:

You mentioned your objection to my ideas on grounds of parsimony (i.e. the inclusion within the theory of an entity, conscious experience, which has not yet been shown to be necessary to explain the data). I should like to make the point here that parsimony is one criterion, but it is not the only one. Parsimony is most relevant when we already have a complete explanation of the phenomena of interest. Neurophysiology can hardly be said to have reached this stage yet. In the history of physics it has at various times happened that explanations involving radically new assumptions have been successfully adopted, when it would have been quite possible in principle to adjust the old framework to fit the facts. An example is Einstein's general theory of relativity, based on the assumption that space is curved in the vicinity of massive bodies.

The noted physicist P. A. M. Dirac has suggested that elegance may be a strong indication of the validity of a new kind of theory. In this aspect Maharishi Mahesh Yogi's Science of Creative Intelligence, parts of which were used in my own paper in this conference, has much to recommend it, in comparison with conventional theories based on neurophysiology, as an account of the basic phenomena of intelligence.

You say in your talk that there are those who object to these new concepts because they multiply entities beyond any demonstrated necessity. Equally, there will be those who, in the spirit of Dirac, will be attracted to these ideas because of the elegance and directness of their explanatory power. The right course, it seems to me, is that both lines of attack should be followed up; the future will decide the value of each.

CHAPTER 7

Some Hypotheses Concerning the Role
of Consciousness in Nature

B. D. JOSEPHSON
Cavendish Laboratory, Cambridge

The fact of the existence of consciousness plays no part in ordinary physics (with the possible exception, discussed in the paper by Longuet-Higgins, of the process of observation in quantum mechanics). It would be wrong to conclude from this, even taking into account the enormous success which physics has had in explaining natural phenomena, that consciousness is not a parameter which needs to be included in the description of the world, for the simple reason that the physicist does not carry out experiments on systems which are conscious. In fact, the success of physics has no bearing at all on this question.

The person concerned with the study of conscious systems is the biologist, or, more specifically, the psychologist. For him, the quantitative test in the style of the physicist is not feasible; it will not be feasible until we are able to derive all of psychology in a quantitative way from neurophysiology. We are forced to conclude that the question of whether consciousness is in fact an important parameter in the scientific description of nature is an unanswered one at the present time.

From a naive point of view, consciousness obviously is a parameter having significant effects. What we can do and in fact do do when we are conscious (awake) is totally different from what we can do when we are not conscious. It seems worthwhile to study the situation and to try to frame hypotheses about the role which consciousness may play. By definition, this requires a study of subjective experience, and attempts to correlate it with other phenomena. There appears to be a problem with this, in that a subjective experience is observable by only one person, the experiencer; "objective" observation is not possible. Fortunately language comes to our rescue; in many instances the language a person uses appears to be, and would by most people be believed to be, an indication of the experience he is having. The extent to which this is true depends mainly on how adequately language has been developed to indicate the particular experience, and the practice which the person has had in describing his experience. It is true that, as a philosophically minded person might point out, we can never prove that the conscious experiences of two people are the same when they describe them in the same way. Fortunately science is not concerned with proving its assumptions, but with testing them, that is, assuming them to be true and seeing if the consequences agree with experience.

In this paper some specific hypotheses are made about the possible role of consciousness. They are based in part on common experience, but in addition I shall give arguments based on the hypotheses which connect them with systematics of intelligent behaviour. There appears, indeed, to be a distinct possibility that theories based on knowledge relating to subjective experience may provide a deeper understanding of how intelligence functions and how it comes about than theories based on neurophysiology or artificial intelligence, and this alone provides motivation for following the path taken in this paper.

Let us begin our analysis by discussing the distinction which exists between voluntary and automatic behaviour. The concept of automatic behaviour is the simpler one. Consider first an arrow being shot by an archer. It is clear that once the arrow has left the bow it is no longer under the control of the archer, and follows automatically along a trajectory determined by the laws of aerodynamics and mechanics. We have a somewhat similar situation in the case of a person running up to and jumping over a stream (the interest being in his actions, and not, as in the first example, in the free trajectory). After the person has started running he has a limited ability to control his motion or even abort the jump, but he also has the choice of not exerting such control. In this situation we may say that he carries out his action automatically. What actually occurs then is determined by laws not very accessible to us but clearly dependent on the person's competence at running and jumping, which is itself related to his past experience and learning. The person jumping appears to participate as little consciously in the details of running and jumping as does the archer in the flight of the arrow.

The last remark leads us to a consideration of the relationship between consciousness and voluntary action. The archer is conscious of a decision of it being the right moment to release the bow, and the jumper is conscious of a decision that he should now jump. And both these actions are what we would term voluntary ones.

We can deepen the discussion by linking the ideas already expressed to the concept of value. In both the cases discussed it can be seen that the voluntary actions undertaken are ones which produce a result of value to the individual concerned: hitting the target with the arrow or getting to the other side of the stream. The behaviour that comes to conscious awareness is associated with a certain degree of flexibility (the decision aspect), and this flexibility is used to maximize the value of the outcome of the current action.

If we take for granted for the moment the connection just stated between conscious awareness and flexibility, we have the basis of an explanation for what information comes to conscious awareness and what information does not. In a situation where a person has a high degree of competence, as may be the case for jumping across a stream if the stream is not too wide, there is no reasonable prospect of an increase in value of the outcome by adjusting many precise details of the run-up and jump; therefore this information need not be present to consciousness. The only information that is needed will be information such as where to land and where to take off, and this is the information that will tend to occupy consciousness.

The linking of consciousness with considerations of value suggests that consciousness may play an important role in the improvement of skills over time. There are no grounds for supposing that actions carried out automatically will improve with time; they are just like fixed computer programs in their character. But a system concerned with questions of value, as we have postulated for the conscious part of the system, may be able to recognize that certain new actions lead to consequences of higher value than would be obtained using the normal procedures in current use. We can thus assign consciousness the attribute of creativity, in that it leads to new procedures being adopted. Furthermore, in that it appears to be the case that new ideas are good ones more often than would be expected by chance, the overall effect of consciousness would seem to be a positive one of giving steady improvement over time.

To clarify a point arising in connection with the hypothesis in the last sentence, automatic behaviour displays intelligence, as well as does consciously guided, voluntary behaviour. However, according to the picture being presented, such automatic behaviour is the result mainly of past conscious experiences which had the overall effect of improving the skill of the performer (but obviously a small component of the skill is the result of innate programming and need not be attributed to consciousness).

It has been suggested above that consciousness, as well as being in effect creative, in that it is associated with changes in behaviour patterns, is also in effect intelligent, in that these changes tend to be in a positive direction. We now propose a hypothesis to account for this. The hypothesis consists in the assertion that there can be an intimate connection between conscious experience and meaning. At the moment when we understand either the situation in which we find ourselves, or alternatively the meaning of something expressed in language, we have a particular conscious experience of knowing the meaning (which after the event may dwindle to a mere memory of having had the experience). Now what do we mean by meaning here? Knowing the meaning of a situation means knowing what is in the situation for us, that is knowing what to do about the situation to ensure the best outcome from it. To give an example, when we experience the feeling of hunger we at the same time become aware that it would be good to eat some food. This will generally be followed by action to ensure the desired result. Hunger is an experience with a definite meaning to its perceiver. This example illustrates an important point, that the experience (in this particular case) has to be conscious in order to have the powerful effect that it does, which is different from merely, for example, going to a restaurant at a particular time out of habit. We may assert, in a way parallel to our discussion of the difference between voluntary and automatic behaviour, that conscious awareness of the meaning of a situation followed by an appropriate response is different from making an automatic response, since only the former is concerned with the question of the value of an action to the perceiver (in this case the value of removing hunger). One way of explaining the relation between consciousness and intelligence is to postulate that there is a basic type of subjective experience, such that as a result of something analogous to the basic laws of physics a meaningful conscious experience is automatically followed by the idea of the appropriate response (for example, hunger by the idea of obtaining something to satisfy the hunger, or danger by the idea of escaping from danger). While this idea may seem highly ad hoc, it is completely analogous to the situation we find in ordinary physics, with its various types of basic fields and their interrelating equations, such as Maxwell's equations for the electromagnetic field.

The reader will probably have realized the incompleteness in itself of the hypothesis proposed in the last section. It must be augmented by the consideration that evolution must have played an important part. Evolution must have provided an efficient perceptual system to generate conscious experiences correctly representing the environmental situation, and a planning—motor system to put the ideas generated into practice. These two additional systems are needed to allow the basic intelligence postulated to be associated with conscious experience to be effective.

I should like to acknowledge my debt to the thought of Maharishi Mahesh Yogi, whose ideas, especially those contained in his videotaped lecture course entitled The Science of Creative Intelligence, have played an important part in formulating the above.

Discussion

VESEY

In the course of talking about the freewill/determinism problem I think you said that scientists feel that a person's behaviour is determined by what happens in his head. What b ' ' ' did you mean by "behaviour"?

I would not want to dispute that the motions of a person's body are often determined,

in part at any rate, by what happens in his head. One can explain someone's arm bending, for example, by reference to nerve impulses from his brain to a muscle. But questions about a person's behaviour are rarely simply about his bodily motions. One wants to know what the chap is up to, what he is doing. And one can sensibly ask him "What are you doing?" even when one is in full possession of the facts about the motions of his body. Is he signalling a right turn or pointing out where his mother's cook used to live? The "action-description", as I'll call it, is a different sort of description from the "motion-description". I don't mean just that one can have the same motion—description but different action—descriptions, and vice versa. I mean that the action-description is authorized in a way in which the motion—description is not. The way to find out what someone is up to is to ask him, just as, when someone says he once saw Ted Heath , conducting, the way to find out whether he meant Ted Heath the band-leader or Ted Heath the Conservative politician is to ask him. In both cases it makes no sense for him to i say ''I was wondering that myself". l ' My question is this. When you talked about a person's behaviour being determined by ' what happens in his head, did you mean simply the motions of his body, or did you mean i  what he is doing? If the latter, can you please explain? Why do scientists feel that what a person does is determined by what happens in his head? is there empirical evidence to this effect? (And, if so, is the strength of their feeling proportional to the evidence?) Or do they feel it must be so? If so, why do they feel it must be so?

JOSEPHSON:

I think you have raised a very important point. When I suggested that a person's intentions are determined by what happens in his head I was basing my remark on the belief, common to physicists, that everything which is in some sense "real" has some definite location in space (and time). But perhaps this belief is not true. An argument for this, within physics, involves the theory of observation in quantum mechanics. While there exists a well—defined formula giving what happens when an observation is made, it does not seem possible to understand the formula in terms of mechanisms happening in space and time. Recent research by J. S. Bell and by H. P. Stapp, based on the so-called Einstein-Rosen—Podolsky paradox, indicate that some very counter—intuitive things are going on.

There is a point of View which transcends the distinction between whether things happen within space and time or outside space and time. This is to say that certain things have no apparent location in space and time because they happen everywhere in space and time. l This is a common View in Eastern mysticism. On this view if I intend to do some particular thing, my intention corresponds to a process happening through all space and time, but it leads to my observable behaviour at some particular time because my nervous system is, as it were, "tuned in" to that intention at that particular time.

PART III

Subjective experience

CHAPTER 8

Consciousness and Psychopathology*

M. ROTH
University of Cambridge

INTRODUCTION

At the present time, there are no indubitable criteria of "consciousness" which command general assent. The reason for this is that despite the direct experience we have of it and take for granted in other human beings, scientific understanding of it as a phenomenon is rudimentary. And precise, generally accepted definitions emerge not at the beginning of a scientific exercise but at an advanced stage of it.

In philosophical discourse regarding human consciousness and the nature of its relationship to the activity of the brain, the state of consciousness of one person is treated as comparable in essential features to the state of every other. The validity of such an assumption is questionable. There are variations in the degree and quality of consciousness and individuals differ from each other in definable ways in both respects. Degree of consciousness will be discussed at a later stage. Differences in quality arise because such characteristic manifestations of consciousness as attention, memory, the capacity for conceptual thought and logical reasoning, and the ability to perceive the world correctly are among the criteria used for the recognition of conscious states. But each of these may vary independently of the others. Thus an individual may have auditory hallucinations without insight into the illusory nature of the experience. Yet psychiatric examination will show some individuals so affected to be in a state of clear consciousness. These variations are

* Paper based on discussion comments, submitted after the conference.

important for investigations aimed at refining scientific knowledge of consciousness and have to be taken into account in the shaping of our philosophical concepts of it. In so far as answers can be obtained to these questions, they must affect the philosophical concepts of consciousness and the methods employed to refine knowledge of it.

We have neither the logic nor the language, according to Wittgenstein,' wherewith to call in question the consciousness of other human beings. But this refers to consciousness in an abstract sense. When asked to examine a man who complains of mental distress or is regarded by others as displaying behaviour out of character for him, one of the most important matters a psychiatrist has to decide is the degree to which he can be regarded as "conscious" in a specific sense of the term. His level of consciousness requires technical means for its determination and has a significant bearing upon the diagnosis and treatment of the disorder in question.

He may appear alert and awake and normal in his conduct to ordinary observation, but prove incapable of forming correct or safe judgements about his whereabouts and circumstances. If level of consciousness is lowered his memory for current events will tend to be impaired and his beliefs may be distorted or deluded.

It will be apparent that consciousness is not an all—or-none state with deep sleep or coma at one end and complete alertness at the other. There is every gradation in between. But variation is not linear all the way, in that any graph which depicts states of consciousness between these extremes must show plateaus corresponding to the states of incomplete arousal, such as clouding or delirium, in which the individual's thought and behaviour will be muddled and confused. But he cannot be awakened like the person asleep or experiencing dreams or the sleepwalker. The process of arousal has been arrested at some point intermediate between sleep and full consciousness.

Whether such an individual is capable of forming intent and carrying out complex, co—ordinated and dangerous acts is a complex and difficult question to resolve. The answer to it may be of crucial importance in the case of a man charged with murder or some other criminal act.

An elderly person with senile dementia may be able to hear noises and recognize objects in a crude and undifferentiated way, but prove incapable of retaining any current event in memory, although memory for remote events may be intact. At an advanced stage she will also prove incapable of recognizing her own relatives, be unaware that she was married and liable to stare in fear and bewilderment at the mirror image of her own face which she does not recognize. Whether or not such a person can be regarded as conscious is neither a trivial nor a purely semantic question. It poses questions about differences between certain states of mind which are inherently capable of resolution by scientific means. The answers would be bound to sharpen our knowledge of "consciousness" and the words we use to depict it.

Among the stated objectives of this conference is the ". . . study of subjective experience and . . . the relationship between subjective experience and the objective world". We are also enjoined to examine what defines the personal character or privacy of the individual's conscious experience. These are interesting questions for the psychopathologist because there are individuals who lose just these faculties conjointly. They confuse subjective experiences with perception of the external world, and through failure of this differentiation their inner thoughts and feelings are attributed to outside influences. At the same time they feel their privacy and their most intimate experiences to be encroached and intruded upon by forces beyond their control. It may be inferred from this that there is a neurological apparatus which makes possible fulfilment of this function of sifting subjective from objective perceptions.

The development of objective scientific knowledge is dependent upon our ability to differentiate between the world of phenomena independent of our perceptions and a reality distorted by or entirely determined by private inner experience. However, as Whewell pointed out more than a century ago, the ordinary individual engaged in his task of perceiving and acting upon the world is all the time testing hypotheses and so refining his knowledge in a similar manner.

In some disorders, notably in schizophrenia, the ability to differentiate between the subjective world within and the objective one beyond the boundaries of the body is undermined and in consequence the individual's picture of reality becomes distorted. It requires the knowledge held by others to diagnose the distortion and take action to correct it as far as possible by appropriate treatment.

The study of psychopathology makes it possible to describe tentatively the criteria that should be employed to characterize consciousness. These are the criteria that would have to be satisfied by any form of artificial intelligence or organism that was designed to exhibit "consciousness" in the human sense of the term.

In the sections that follow some relevant criteria will be examined and the bearing of some of the commoner forms of psychopathology for problems of consciousness considered.

MEMORY AND CONSCIOUSNESS

The ability to observe and lay down a record of events in the surrounding world, the faculty of memory, is widely regarded as a central feature of any organism or machine we would accept as having "consciousness". However, it is well known that there are disorders of the mind in which short—term memory is severely impaired so that the ability to lay down memories in, or to retrieve them from, the long-term store is severely and irretrievably damaged. Beyond a limit of about one or two minutes, such an individual will be found on direct inquiry devoid of memory for current events. But he appears alert and may be able to conduct an intelligent conversation and exercise his usual skills in dressing, eating and even in complex activities such as card-playing or the musical performance of works learned in the past.

For example, one lady suffering from such a Korsakov or amnestic syndrome had just completed a performance of Beethoven's Quartet in E Minor, Opus 59, No. 2 with three friends. Three or four minutes later, the cellist, fired with enthusiasm, said, "Let us play something else". The lady in question, who had played the violin superbly in the first performance, said "Let us play Beethoven's E minor Quartet, Opus 59, No. 2." After an embarrassed silence, one of her friends said, "But, Angela, we have just finished playing that!" One could not regard an individual with such a disorder as devoid of consciousness. The picture would, of course, be quite different if the person affected were not able to draw upon a vast memory store extending back into early life. In a progressive degenerative disease of the brain such as senile dementia, this store is also progressively encroached upon so that less and less is left. We are unable to answer the question as to whether or not there is a critical level in the extent of the loss of both short- and long-term memory which would be incompatible with "consciousness". There are no quantitative observations to guide us. But it is of interest that in advanced stages of dementia individuals are liable to sink into unconsciousness in the absence of any obvious explanation for this, such as an intercurrent infection that has caused further impairment of brain function. Death is then imminent.

Inability to remember cannot serve as a necessary criterion for the presence of consciousness for more subtle reasons. It is characteristic of human consciousness that certain experiences can be segregated or locked away in separate compartments whose contents are not accessible to ordinary methods of retrieval without special aids. In his earliest investigations with Breuer, Freud discovered that certain neurotic patients were able to recount, under hypnosis, experiences of which they had no conscious recollection. These memories had been "repressed" into the unconscious "because" they were associated with emotional conflicts beyond the capacity of the individual to resolve. Needless to say such statements tell us nothing about the mechanisms employed in such sequestration of memories. Another example is the seemingly obliterated memory of some profoundly distressing experience such as the injury received during a battle in which friends of the subject might have been killed or mutilated. Such memories can be brought into consciousness by injection of a hypnotic drug which causes partial impairment of awareness. The recovery of such memories is commonly associated with the discharge of intense emotion or ''abreaction''.

Yet another example, which illustrates the point that consciousness is not an all-or-none phenomenon of the kind implicitly assumed in some forms of philosophical discourse, is provided by certain phenomena encountered during anaesthesia. The subject, in the process of recovery, may appear to all intents and purposes unconscious. He does not respond to painful or other forms of stimulation. His eyes are closed and he is immobile. Yet the remarks made by the surgeon are heard and remembered. In some cases where they have been personal or unflattering in character they have been the subject of legal action.

It will be apparent that the ability to provide an accurate account of recent events in which the individual had been a participant cannot provide reliable information as to whether he had been conscious at the time. To direct questioning he may have no memories to report. But when special techniques are applied, he may bring to light experiences that have been recorded and sequestered as a result of repression into the unconscious in the Freudian sense. In contradistinction to this, islands of memory may remain following a state of epileptic automation in which the individual has been confused and the electrical activity recorded from the brain disturbed throughout.

CAN CONSCIOUSNESS BE REGARDED AS A CAUSAL AGENT?

It will be apparent that consciousness cannot be regarded as merely the simple aggregate of the activities that result from information transmitted from the outside world along the different modalities of perception. We know that the level of awareness of the individual cannot be judged from functional integrity of the pathways that subserve touch, proprioception, vision and hearing, or from the capacity to form a record of experiences gained alone. This is not to say that tests applied to these functions provide no information about the state of consciousness. But they are very crude indices of it.

For example, mild states of impairment are notoriously difficult to diagnose because there are no pathognomonic features. The patient will appear dull, vague and inert. His responses to external events will often be mildly inappropriate, inconsistent and undiscriminating. We know that such a mental state may be the prelude to a state of acute delirium in which the patient loses contact with his environment. And subsequently no trace of memory of the premonitory phase may survive. The human electroencephalogram will sometimes provide valuable evidence regarding the level of awareness. But it is far from being an accurate measure and the results in mild impairment of consciousness are unreliable.

Recent investigations have provided a large body of evidence in support of the view that the level of human consciousness is determined by mechanisms that are distinct from, and to some extent independent of, those which are responsible for the flow of sensory information along the specific afferent pathways to the brain. The reticular activating system whose activities have been shown to be closely related to the level of arousal2 is a complex network of neurones in the medial part of the brain stem" extending from the medulla and pons, through the mid-brain to the hypothalamus. It is distinct from the specific pathways that subserve ordinary perception in three respects. A large number of the fibres bypass the thalamus which is a relay station for ordinary afferent pathways. It is distributed to extensive areas of both cerebral hemispheres and not merely to one specific contra-lateral area. Finally, because it is a complex neuronal network any specific effects of one form of sensory stimulation are quickly abolished; different sensory stimuli have equal effects in the promotion of arousal or maintenance of level of consciousness. Damage to this part of the brain will render an experimental animal unconscious but interruption of specific pathways will not do so.

It is therefore the intensity of activity within this system and its sites of projection in the cerebral cortex that determines the level of consciousness of the individual. And this will in turn decide whether an afferent stimulus of pain, touch or hearing is or is not perceived. Thus in states of light unconsciousness it is possible to elicit electrical changes or "evoked potentials" in the cerebral cortex by sensory stimulation. But they are neither perceived nor remembered.

The answer to the question posed by the title of this section is therefore that there is a certain sense in which consciousness can be regarded as a causal agent. For its complete integrity is a necessary condition for the acquisition of information about the external world and the recording of this information in the memory store. Something more than wakefulness and reaction to stimuli is involved. The individual may appear awake and responsive to stimuli without being fully conscious.

PERCEPTION AND CONSCIOUSNESS

This leads to the relationship between perception and consciousness and the extent to which the state of the latter can be gauged from accuracy of perception.

It is widely recognized that no form of perception can be explained in terms of a passive registration of signals emanating from sense organs. Perception is an active process which creates a picture of the objective world from minimal and partially familiar cues. Hypotheses are rapidly advanced and discarded or upheld. But success in this operation of depicting external reality depends upon a consistently high level of arousal which is in turn dependent upon an adequate intensity of activity in the cerebral cortex.

We obtain some measure of insight into the role of consciousness in extracting an accurate percept from an array of signals, by observing the states in which consciousness is partially in abeyance. A good example is provided by the clouded and delirious states with impaired awareness seen in the course of chronic intoxication with alcohol and hypnotic drugs. In delirium tremens the individual is disorientated in time and place and his memory for recent events is markedly impaired. He is unable to construe happenings in the objective world correctly because, among others reasons, he suffers from falsifications of perception of hallucinations, commonly in the visual field. Large insects are crawling on his bedclothes, rats are scuttling on the floor, poisonous dust is falling from the ceiling and hideous monsters boring their way through the walls. He is surrounded by warders, spies, executioners. His relatives are at the door but are prevented from entering. There are concomitant emotional changes. The subject experiences suspicion, perplexity and intense fear which may lead him to attack his imaginary persecutors. He is living through a terrifying nightmare but he is not asleep. His perceptual world is in chaos because the errant hypotheses are being discarded and erroneous ones upheld. Perceptual gaps are filled indiscriminately with fantasy which displaces reality.

We have no more than rudimentary understanding of the neurological mechanisms underlying such processes. But normal consciousness may be conceived as the obverse of these states in which hypotheses are tested against reality and sound ones sifted from incorrect and illusory ones.

We see, therefore, that the paradigm often employed by classical philosophers of the human subject engaged in perception – usually visual perception—is a weak and unsatisfactory model for making inferences about the nature of knowledge. What we designate as human consciousness depends on the interaction of a number of disparate interdependent psychological functions: state of general arousal, perceptions of the objective and subjective world, memory, reasoning and emotion.

However, these interdependent and covarying functions display a surprising degree of autonomy. One may show marked variations without affecting the others or impairing the level of consciousness.

Illustrations have already been provided in relation to memory. The same situation holds for perception. In the case of delirium tremens, the frightening hallucinations are associated with disorientation in time and place, and inaccurate perception and judgement of the outside world; there is clouding of consciousness. But in another disorder also due to alcoholism, the subject hears hallucinatory voices in a state of clear awareness. He hears them threatening, cajoling, deriding and humiliating him. He attributes them to some remote group of tormentors intent on destroying him. But other than the hallucinations, his picture of the world is clear, detailed and accurate. Again, impairment of consciousness is prone to distort perception but not invariably so. Many patients with clouding of consciousness are merely muddled and disorientated, and the emotions of patients which have a selective effect on filtering of perceptual experiences may become grossly deranged without involving any other aspect of mental life. We call such conditions Depressive and Manic Psychoses.

MODELS OF CONSCIOUSNESS

Any attempt to create a model that replicates human consciousness must do full justice to this complexity. Such a model must be capable of making reliable and accurate observations of the outside world, and reporting on its subjective, inner activities, and must also be aware of itself as observer in both capacities. It has to replicate variation in degree of awareness or arousal as a relatively distinct function but one whose fluctuations are liable to exert a profound effect upon the accuracy of perceptions. The very characteristics that enable it to extrapolate complex percepts so effectively and accurately from minimal cues also cause it to conjure up spurious perceptions, phantoms and fantasies.

It must also make allowance for uniqueness in two respects. The first is that no individual's cerebral equipment has been shaped by genetic endowment in exactly the same way as any other, and a working consensus about its characteristics has to be achieved with the aid of comparison and communication about shared experiences. '

Finally, emotional factors have to be brought into the picture in that they have been repeatedly shown to exert an important selective effect on what is perceived. For example, a patient with a mild Korsakov syndrome may remember the visits paid during the day by his wife but nothing else. Or if he has had a stroke he may forget and ignore the paralysis of the left side of his body and even deny it completely. And the emotions of each individual are uniquely shaped by his own past experiences and vicissitudes in interplay with his genetic and constitutional endowment.

SELF—CONSCIOUSNESS

The most distinctive feature of human consciousness is the awareness that mental happenings are taking place-—or self-consciousness. This awareness is intimately dependent upon our ability to differentiate the happenings in our own inner self from those in the outside world. The phenomena of "believing", "deciding", "wishing" ' stem from this self-awareness. Uniqueness apart, they are characterized by the fact that only the individual who is experiencing these mental states has access to them. The unique privacy and freedom from control (other than through the medium of language) of these experiences have to be assumed unless we are prepared to jettison our existing scientific picture of the world. Some philosophers go further and assert that it is the states of intention, choice and belief that render the individual autonomous, make him accountable for his actions, and arouse our respect for him as a person. They are given and not susceptible to scientific investigation or analysis. We can entertain no hope of arriving at an objective account or causal explanation of the inner world of subjective experience.

Now for reasons that will emerge later it could be held that there are limits beyond which such an objective scientific account cannot proceed. But the view that the self—conscious mind is an independent entity about which the objective deterministic language of science can have nothing to say is untenable for a number of reasons.

In the first place there are conditions in which the inner self is no longer clearly differentiated from the objective world beyond it. In these disorders the thoughts in the mind of the individual are experienced as voices that emanate from some outside source. The person is hallucinated. His intentions, wishes and purposes are felt to be usurped by outside agencies. He experiences himself as the passive victim of witches, freemasons or "atomic rays". He may come to regard himself as the agent of God and act accordingly.

This summarizes some of the leading features of typical schizophrenic illness and leads to two inferences. The first is that the claim that a person's private intentions, beliefs and expectations are sacrosanct in that they cannot in any circumstances be judged subservient to the objective judgment of others is untenable. The second inference follows from the fact that such disorders often arise in minds that have been previously intact and to which the capacity to differentiate reliably between objective reality and subjective experience can be restored by specific treatments.

It follows that there must be neurological mechanisms which mediate this sifting of sense data so as to differentiate the external from the subjective world. Little is known about them at the present time. But the ignorance is of a kind which we can reasonably expect to remedy with the aid of scientific inquiry.

The view of Wittgenstein that the world of personal, private will, purpose and belief belonged to an order of phenomena that was categorically different from the world of physical phenomena is reminiscent of the distinction drawn by the psychiatrist and philosopher Jaspers between "understandable" and "causal" connections. Both believed the former to be impervious to the causal explanations and general laws that emerged from scientific study of the objective world (Wittengenstein's criticism of Freudian psychoanalytic theory was essentially similar to Japsers's3). But it is difficult to reconcile some of the findings of psychopathology with categorial distinctions of this nature.

## DEPERSONALIZATION: A MORBID STATE OF THE SELF-CONSCIOUS MIND

It may be helpful to examine another form of psychopathology in order to judge how far we are likely to succeed in explaining human consciousness in terms of the laws of physics and chemistry.

In the phenomenon of depersonalization known to psychiatrists the consciousness of the individual is divided into an observing and participating self. Here are some extracts from the self-descriptions of an intelligent and gifted observer:

When I am talking to people, especially if the conversation is difficult or important, I feel, as it were, withdrawn into myself at a great distance with difficulty in focusing my eyes or attention on the person I am talking to. At times I feel like a mind detached and nebulous without a body or a physical setting. The only thing of which I can really be aware is my own mind and the fact that it is working. Everything else shades off into unreality. It is as though I have been living automatically, reacting and behaving apparently as usual and yet with a part of me which I would call my personality not really involved. In this connection "depersonalisation" seems a very accurate word.

The feelings occasionally assume a more specific and physical form especially in the period between waking and sleeping. I feel as though mind and body were parting and expanding, and mind as it were suspended over an expanding gulf into which the body is sinking. This sensation continues indefinitely, the gulf increasing, until I switch on the light or 'get up and move about. . . . Another occasional physical sensation is that I am literally "beside myself" . . . displaced in space, almost drifting away. Here again to relieve the feeling I have to clench my fists or grasp a fold of clothing for reassurance.

Closely associated with these sensations is an inability to experience ordinary emotions. The subject not only feels one part of himself detached and viewing the other as would a passive and indifferent spectator. He complains that he moves and behaves as would an automaton or a wound-up mechanical toy. Every act, no matter how simple, and formerly carried out without reflection, seems now to require an effort of will. Eating, dressing or washing entails a special effort, and even breathing may have to be undertaken with deliberation and self-vigilance. The experience sometimes includes an actual visual hallucination of the self, usually recognized within a short interval as illusory.

It should be noted that these phenomena cannot be dismissed as irrelevant for the study of ordinary subjective experience in that they are disordered states of mind. In transient form, depersonalization commonly occurs in normal individuals.' And the central core of the phenomenon has been found to occur in a high proportion of those exposed to sudden life-threatening dangers. There is reason to believe that in such circumstances depersonalization and the sense of detachment and objectivity associated with it may enhance the chances of survival.5~5»7 When the peril has passed or ceased the mental state of the individual returns to normal. Closely similar experiences may be engendered by electrical stimulation of certain parts of the brain, namely the cortical surface of the temporal lobe. They also arise during the early stages in the development of epileptic fits that result from the discharge of a focus located in one of the temporal lobes. Since we know that experiences have a neurological substrate we may hope through scientific inquiry to acquire more knowledge about and greater control over them.

## SUBJECTIVE MENTAL STATES AND THE BRAIN

Two main questions arise which have a close bearing on certain problems of the mind—brain relationship.

The findings of neurology and psychiatry are inconsistent with the view that mind and brain are not connected. Moreover, it is clear that something more than parallelism between the two without causal connection is entailed. The investigation of abnormal mental states and related phenomena may deepen understanding of cerebral activity in its relationship to consciousness.

But is it possible that the language of physiology will become so precise and differentiated that, in terms of the activities of neurones, we will be able to describe an individual experiencing a part of himself displaced from the body in space, observing the self as a passive spectator and finding this participant and executive self distressingly unreal and unfamiliar while aware at the same time that the whole experience is morbid and illusory? Will it be possible for someone familiar with the language to recognize from such an account the feelings of depersonalization as he has experienced them in every nuance and detail during introspection? According to the materialist account of the mind-brain relationship this should be possible at some future date. The introspective accounts given by depersonalized patients, and our recognition of them through our own introspection and empathy, will then become redundant in the same way as the knowledge relating the sequence of nucleotides in DNA molecules with the amino acids they sort and assemble into proteins has rendered so much of the older language of cellular biology and genetics redundant.

For those who allow for an interaction between body and mind within the framework of a materialist account of the relationship between them, subjective experience, introspection and their disturbances should ultimately prove explicable in a causal language. They are the outcome of processes in which all the activities in one part of the brain—activities which would otherwise remain unconscious — are read off or scanned by some higher centre which undertakes a second order of perception.

According to current knowledge this centre would be located in the dominant, usually the left, cerebral hemisphere, which appears to have the main responsibility for the conceptual, linguistic and symbolic aspects of mental functioning.

In their recent book The Self and its Brain Popper and Eccles" reject this view, holding that there is no reason why such scanning activities should result in self-awareness or self-criticism.

The materialist's answer given by Armstrong° is that this scanning is identical with consciousness. There need be no residue to explain. When we have learnt to define the physiological processes, the question of all other accounts of self—consciousness will have been rendered redundant and obsolete.

Now it may be that the two accounts, the one in the language of cerebral function and the other drawn from self—conscious experience, will approximate to each other. But that ultimately physiology will entirely displace the introspective account is difficult to credit. It is impossible to conceive of a state of affairs in which the latter could be discarded as wholly superfluous. For an indefinite time ahead two separate accounts of self—conscious experience, the one complementing the other, appear indispensable.

The second question is a related one. Is it possible to envisage a state of affairs in which we will have produced a computer that will be able to experience and report upon both a subjective internal world and an objective one, and which will be able to observe itself in action as the cause of both forms of perception, empathize with others in both roles, and display insight in the presence of some forms of derangement of discrimination between subjective and objective worlds and loss of insight in others?

The answer that has to be given is that there is for the present no body of knowledge at our disposal to encourage the hope that such a feat will be possible. Nor do we possess sufficient understanding of human consciousness to apply valid and satisfactory tests so as to ascertain whether the machine we had manufactured possessed properties of a kindred nature.

CHAPTER 9

Twins, Split Brains and
Personal Identity

V. S. RAMACHANDRAN
Trinity College, Cambridge

Most of the chapters in this book are heavily inclined towards the scientific materialist view that consciousness is an emergent property of certain complex brain events. Since brains precede minds in evolution it would be hard to maintain (like the idealist philosophers did) that the existence of the physical universe depends on the existence of a conscious "observer". It is the emergence of minds that seems to require explanation. Why did certain kinds of brain activity become associated with consciousness? Is consciousness biologically useful or is it a redundant by—product of evolution? Maybe consciousness is an "epiphenomenon" like the ghostly whistling of a train —but somehow this view seems curiously inadequate. We generally think of our minds as being causally effective in all our actions, and indeed, it is difficult to think of the word "mind" having any other meaning.

One of the aims of the Cambridge symposium was to answer the question "What determines the uniqueness and privacy of an individual's conscious experience?". This question is generally considered by philosophers under the heading "personal identity". My approach to this problem will be to present a series of "thought experiments" in which the reader is invited to participate. In my view nothing more can be said about personal identity than what is contained in these examples.

Within each of us there seems to be an "I" that remains invariant in spite of continuously changing sensory impressions. If you are (say) presented a stimulus A followed by a stimulus B, there seems to be a "unifying agency" in you that relates the two sensations as having been experienced by the same person. This "I" within you also has other attributes—it claims to "will" actions' and seeks self-preservation and immortality.

Under the heading ' 'personal identity" we may include two questions:

(a) What determines the coherence and continuity of a person's consciousness — in spite of constantly changing sensory impressions?

(b) What determines the exclusive relationship of a person's mind to a particular physical brain?

Question (a) can be stated in a weak form or a strong form. The weak form of the question is "Why do I feel single in spite of changing sensory impressions?" or "Why do I not feel double even though I have two hemispheres?". In my view these questions are meaningless since there can be no circumstance in which a person can feel double—for who is there to feel the doubling? The situation is analogous to two one—eyed dogs fighting over a bone. As Descartes points out, the dogs would behave as though they saw one bone and not two! Similarly there is no sense in which a split—brain patient could feel double—even if we assume that there are two quasi—independent spheres of consciousness inside his skull.

A stronger form of the same question (a) is "What is the exact nature of the unifying agency in me that issues commands for action, etc., and relates various memories and sensations as having been experienced by the same person?". This is really an empirical rather than a philosophical question. People with frontal lobe lesions, for instance, often report losing this sense of coherence and continuity in time. The feeling that I am a particular individual with some control over my future behaviour is also associated with self-consciousness and may be the subjective correlate of what MacKay calls the brain's "supervisory system" (Chapter 6).

But now let us turn to the second question (b), which is really a special form of a problem which Jennings[2] and Eccles[3] refer to as the "Uniqueness of personal existence". Jennings asks:

What is the relation of myself, identified as it is with one particular knot in the great network that constitutes humanity, to the other knots now existing? Why should I be identified with one only? To an observer standing apart from the net, it will not appear surprising that the different knots, since they are formed of diverse combinations of strands, should have different peculiarities, different characteristics. But that the observer himself —his total possibility of experience, that without which the universe for him would be non-existent — that he himself should be tied in relations of identity to a single one of the millions of knots in the net of strands that have come down from the unbeginning past — this to the observer appears astonishing, perplexing. Through the operation of what determining causes is my self, my entire possibility of experi- encing the universe, bound to this particular one of the combination of strands, to the exclusion of some millions of others? Would I never have been, would I have lost my chance to participate in experience, would the universe never have existed for me, if this particular combination had not been made?

Eccles takes this as the starting-point of what he calls a "personalist philosophy":

This personal uniqueness and all aspects of its associated experiences are dependent upon the brain; yet it is not entirely dependent on the genetic instructions that built the brain. . . . I believe that my genetic coding is not responsible for my uniqueness as an experiencing being, as I have argued in my book Facing Reality. Of course, I have a unique genetic coding, as indeed do all of us who do not have an identical twin, but the probability of the existence of such a unique code is fantastically low: even 1 in $10^{10000°}$.

Thus the theory that the uniqueness of the code is the determinant of the unique- ness of the self results in such inconceivable improbabilities that it cannot be an explanation. Nor do my postnatal experiences and education provide a satisfactory explanation of the uniqueness of the self that I experience. It is a necessary but not sufficient condition.

We don't know how we came to be this unique self that is tied into our brain in a way we do not understand. . . . We go through life living with this mysterious exist- ence of ourselves as experiencing beings. I believe that we have to accept what I call a personalist philosophy—that central to our experienced existence is our personal uniqueness.

And in his dialogue with Sir Karl Popper, he says 3*':

I believe that there is some incredible mystery about it. What does this life mean: firstly coming—to-be, then finally ceasing-to-be? We find ourselves here in this wonderful rich and vivid conscious experience and it goes on through life, but is that the end? . . . Is this present life all to finish in death or can we have hope that there will be further meaning to be discovered? I don't want to define anything there. I think there is complete oblivion about the future, but we came from oblivion. Is it that this life of ours is simply an episode of consciousness between two oblivions? . . . Our coming-to—be is as mysterious as our ceasing—to—be at death. Can we therefore not derive hope because our ignorance about our origin matches our ignorance about our destiny?

It is obvious that (a) my genetic uniqueness, (b) the uniqueness of the experiences I have had, and (c) the particular physical matter that now constitutes my brain are all necessary conditions for my existence as I am now; but what Eccles seems to be asking is whether these conditions alone are sufficient to explain my personal uniqueness. Maybe questions such as these are meaningless since they seem to imply a metaphysical origin for my existence and unlike Eccles I do not believe in soul—like "agents" inhabiting brains. But I do share his view that each person's conscious existence in this world is an extraordinary mystery.

THE MIND—BODY PROBLEM AND "PERSONAL IDENTITY"

The history of science is full of examples of large conceptual gaps which were bridged by sudden flashes of insight. One recalls Maxwell's equations and the large gap which once existed between "life" and "non-life" before the advent of molecular biology. But when confronted with the mind—body problem one has the uneasy feeling that somehow we are dealing with a different kind of "gap". As Konrad Lorenz[4] points out:

The "hiatus" between soul and body . . . is indeed unbridgeable, albeit perhaps "only for us". . . . I do not believe that this is a limitation imposed just by the present state of our knowledge, or that even an utopian advance of this knowledge would bring us closer to a solution. . . . It is not a matter of a horizontal split between sub- jective experience and physiological events, nor a matter of dividing the higher from the lower, the more complex from the more elementary, but a kind of vertical dividing line through our whole nature.

The mind—body problem and the problem of "personal identity" are really two sides of a coin — and there is an interesting thought experiment which illustrates this.

Imagine that you are a "super—scientist" with complete access to all laws of physics and brain-function. Supposing you were using microelectrodes to record from nerve-cells in the brain of a person A looking at (say) a red flower.

As soon as he sees the red flower you find that certain cells begin to respond vigorously (in those areas of the brain which are known to be involved in colour perception). You then come up with what is essentially a complete and detailed state—description of his brain when he is confronted with such—and—such wavelength.

Now the important thing to note is that this description you have produced is not in principle different from a description of (say) what happens when a computer solves differential equations. Each is an intellectually satisfying account of a sequence of events in the external world; and we can describe these events at both "soft—ware" and "hard-ware'' levels.

Now you can repeat the experiment in any number of people (B, C, D, E, etc.) and you would come up with the same description each time. There may be minor differences in detail due to statistical fluctuations in each individual's brain state) but the information—flow diagram specifying "perception of redness" would be the same for everyone.

If brain-science is sufficiently advanced you may even be able to record from cells in your own brain when you are looking at a red flower. The description you would then come up with would be identical to descriptions you produced earlier for the brain states of other people looking at a red flower. If you compare the diagrams you have produced for A, B, C, D and E with the one you produced for your own brain state, you will not discern any difference.

But now we have a curious discrepancy. You have no reason to doubt that the descriptions of other people's brain states are complete. But when you examine the description of your own brain's response to the red flower you will notice that it seems to leave something out —namely, the actual conscious perception of "redness". From an "objective" point of View your brain has the same logical status as other brains. You have studied 11 brains of which yours is one. And yet you find there seems to be something fundamentally incomplete about your description of one of these n brains (i.e. your own) but not of any of the others. The description of your own brain is identical to A, B, C, D and E but incomplete. Hence there seems to be an asymmetry in nature between the observer's brain and the brains of those whom he observes. Or, to put it differently, there is no one—to—one correspondence between an objective description of the world and what you experience — since your perception of redness just is not contained in that description.

This leads us to what might be called a definition of consciousness. Consciousness is that "property" which makes a detailed state-description of the observer's own brain seem incomplete (in some philosophical sense) when contrasted with the descriptions of the brains of other people whom he observes —even if these descriptions are identical to his own in every other respect.

In everyday life, of course, we conveniently forget the special philosophical status of the observer. This has been called the fallacy of "objectivation" by Erwin Schrodinger?

Without being aware of it and without being rigorously systematic about it, we exclude the Subject of Cognizance from the domain of nature that we endeavour to understand. We step with our own person back into the part of an onlooker who does not belong to the world, which by this very procedure becomes an objective world. This device is veiled by the following two circumstances. First, my own body (to which my mental activity is so very directly and intimately linked) forms part of the object (the real world around me) that I construct out of my sensations, perceptions and memories. Secondly, the bodies of other people form part of this objective world. Now I have very good reasons for believing that these other bodies are also linked up with, or are, as it were, the seats of spheres of consciousness. I can have no reasonable doubt about the existence of some kind of actualness of these foreign spheres of consciousness, yet I have absolutely no direct subjective access to any of them. Hence I am inclined to take them as something objective, as forming part of the real world around me. Moreover, since there is no distinction between myself and others, but on the contrary full symmetry for all intents and purposes, I conclude that I myself also form part of this real material world around me. I so to speak put my own sentient self (which had constructed this world as a mental product) back into it—with the pandemonium of disastrous logical consequences that flow from the aforesaid chain of faulty conclusions.

Elsewhere he says:

So we are faced with the following remarkable situation. While the stuff from which our world picture is built is yielded exclusively from the sense organs as organs of the mind, so that every man's world picture is and always remains a construct of his mind and cannot be proved to have any other existence, yet the conscious mind itself remains a stranger within that construct, it has no living space in it, you can spot it nowhere in space.

The questions raised by Schrödinger in these eloquent passages lead us inevitably to the "why" of personal existence. Why did one tiny corner of the Minkowski space—time diagram become suddenly "illuminated", as it were, by my conscious awareness?

For millions of years the universe must have been a "play before empty benches". Then, quite suddenly, I was born in a little corner of the world. In a sense I created the world the moment I was born and my mind gave it substance and form. I am told that the world existed before my birth and I infer that it will continue after my death. Yet in what meaningful sense may the world be said to continue when the cognitive agent who perceives the world has ceased to exist?

These are rather self—centred ideas. Contrast them with the more intellectually satisfying (but equally pessimistic) view —the so-called

"objective" world—view of a detached external observer. From his vantage—point my brain is just one of many thousands of brains and obviously has no special philosophical status —no "privileged access" to the world. The "I", in his view, is a mere evolutionary novelty and so my coming to be and passing away have no special significance.

WHAT DOES THE UNIQUENESS OF MY EXISTENCE DEFEND ON?

As a brain scientist I am puzzled by the following facts about my existence:

1. In my lifetime I have had a great variety of experiences — sensations, emotions, thoughts, etc. But one thing that all have in common is that they are all my experiences.

This mind of mine is always experienced as one in spite of the diversity of sense-data. I also feel convinced that this mind did not exist before the biological birth of my body, i.e. before a particular egg of my mother had been fertilized by a particular sperm of my father. Also, I experience "gaps" in my memory when my brain activity is arrested using an anaesthetic and conclude that I exist only as a result of the activity of a particular brain. That brain is in turn located in a particular body that people have named "R". Moreover, when people refer to "R" I realize that they mean me and not my body.

2. One of the most mysterious questions I can ask about myself is the following: "How did my conscious agency (the 'I' within me) come to occupy this particular body which people call 'R'? In other words, why was I born in a particular place at a particular time? Why am I me rather than someone else? Why was I not born, say, a thousand years ago in Egypt or Rome?"

This question may strike the reader as being mystical or even meaningless but its exact significance will become clearer when he reads the "paradoxes" I shall soon describe. You may be tempted to brush aside the question by saying to yourself: "I am me rather than someone else simply because no one else has the same unique brain organization that I have. I am me, by definition. . . ." But this leads to further questions:

3. Does my existence (i.e. the existence of the conscious "agency" that "inhabits" the present body and experiences pleasure, pain and

emotions) depend on the particular physical matter that now constitutes my brain? In other words, if you were to replace all the carbon, H2, 02, Na+, Kt and other atoms in my brain with identical atoms picked out from the environment at random, then would I continue to exist? Would the same conscious agency then experience pain and pleasure, that experiences pain and pleasure now?

4. Does my existence depend on the particular environment I was raised in? To answer this question, let us do a "thought experiment". If I asked you "Would you mind particularly, if I tortured you ten years from now?". You would answer "Yes". If I went on to ask "If you spent those ten years in Africa, would you mind being tortured after that?", you would answer "Yes". Your answer would also have been "Yes" if I had asked "Would you mind being tortured after you have spent ten years in Finland? ' '. Hence one is convinced that environmental programming is irrelevant to personal existence although it determines the content of personal awareness. Even if you were regressed back to early childhood and asked "Would you mind being tortured 20 years from now" you would answer "Yes . . . irrespective of where I am brought up.

"PERSONAL IDENTlTY" — ONTOLOGICAL AND SEMANTIC QUESTIONS

There are really three different questions about personal identity that people are often confused about. It is important to keep these questions separate, since although their meanings overlap a great deal, failure to distinguish between them can lead to all sorts of verbal quibbles.

Question 1

First, there is the empirical question of what gives unity and coherence to my "mind". This is really a problem for brain physiologists. What is often referred to as the mysterious "unity of mind" may simply reflect some particular kind of neural organization that integrates different sensory impressions and issues commands for action based on certain "goal criteria". Our "minds" also construct symbolic representations of the outside world and we can even enact various roles in this symbolic world before doing so in the real world. What we call "self-awareness" must have emerged in evolution when one's own body became a part of this symbolic representation.

Question 2

Second, there is the philosophical question of "asymmetry" between the observing self and other agents. Supposing you are sitting in a red room and hundreds of other agents exactly identical to you are sitting in rooms which are coloured differently from yours. Then from your subjective point of view there is only "redness"; but in the objective description there are hundreds of people and hundreds of colours. Your perception of redness has no special place in that description. So there is no one-to-one correspondence between the "objective description" of the world and your own subjective experience.

In each one of the thought experiments I am going to describe I shall begin with an asymmetry between the self (A) and others (0). Now we can do certain things to A. We can remove all the information in A's brain and insert new memories, programmes, etc. (as in Bernard Williams's° example) or we could replace all the atoms in his brain either one by one (gradually) or suddenly. You would then come up with what might be a new asymmetry between the apparently new agent A' and 0. Now the key question is how does the old asymmetry A—> 0 relate to the new asymmetry A'—> 0? This is the only real or ontological question we can ask about personal identity and almost all other questions which philosophers have dealt with in the past (for example, if an agent A' claims to be Napoleon based on his memories and based on his resemblance to Napoleon — would we want to regard him as Napoleon?) really boil down to this question. In the rest of this essay when I ask is A' existentially continuous with A, what I am really asking is whether the original asymmetry between A and 0 continues as the asymmetry between A' and 0 after certain transformations have been applied to the physical world (including A); or whether a new asymmetry has been created. All other questions about P.I. are trivial in the sense that they are of no fundamental philosophical importance. For instance, to take the extreme case, it would be trivial to ask whether A' is the same as A after A had undergone plastic surgery—since it is obvious that nothing would have happened to the original A—>O asymmetry. Of course, external Os often use the face as a criterion for identity, but their choice or criterion has no bearing on what I shall call ontological or existential identity. So, like Popper,' I would argue that Strawson's ideas are somewhat irrelevant to the true philosophical question underlying personal identity.

The "paradoxes" I shall state are really a summary of everything that can ever be asked about what I have called the ontological personal identity question—the question of how the A—>O asymmetry persists after certain perturbations have been applied to the physical world (including A's brain).

Question 3

Finally we come to a third question about personal identity—namely the question of how you would identify a person A1 as being different from other agents and as being the same as an agent A whom you have seen in the past. This question (which some philosophers have been interested in) is really quite different from the ontological question (2) although it often masquerades as (2). I shall call this the empirical identity question since it ought to interest only policemen and detectives. (It is not different in principle from asking how one goes about distinguishing chickenpox from measles.)

Take the question of criminal responsibility. A has just committed murder and is to be punished. Supposing I were gradually to replace all the atoms in his brain over a half—hour period to create A'. Now should I punish A' since he talks and behaves like A originally did (he would even remember the murder!) and since he satisfies all the conventional criteria that an external observer would use for determining the identity of A? I think we would feel justified in punishing A' only if we felt sure that he was existentially the same as A — i.e. only if we are sure of his ontological identity in terms of whether the A —> O asymmetry has really continued as A'—> O asymmetry.

Hoping to tackle some of these questions I invented a series of paradoxes involving twins and split~brain patients. As I pointed out earlier, these paradoxes are really a way of separating the question of ontological identity from empirical identity. It turns out that the paradoxes cannot be resolved at all; suggesting that the really interesting questions about personal (ontological) identity can never be answered. This does not mean that no answer exists but simply that we cannot ever hope to answer them — even in principle.

Of course, in everyday life we assume that what I have called "empirical identity" corresponds closely with ontological identity—but this assumption can never be proved; it is merely a belief that we accept for convenience.

THE TWIN PARADOX

Experiment I

1. Does my existence, the existence of the 1 within me (the cognitive agent that experiences joy, pain and pleasure) depend on the particular atoms that now constitute my brain? The answer is clearly no; from the following argument. I experience continuity of "self" right from childhood. However, there is continuous metabolic turnover of the atoms in my brain. Every few months or so the atoms in my brain must undergo almost complete replacement as I excrete wastes and eat new food. Yet I do not experience a "jerky" existence.

This argument also makes sense from another point of view. It is obviously the way in which nerve cells are connected together that determines my conscious awareness. In other words, it is the processing of information in my brain that leads to awareness, and the actual atoms that "carry" the information are quite irrelevant.* Otherwise my existence today would depend on the particular apple pie or Christmas cake I ate last week!

Having accepted that an individual's awareness would continue even if the particular atoms in his brain are replaced let us go on to do a "thought experiment". Supposing a "super-scientist" were to create a being that is exactly identical to you down to every fine detail. Ignore, for the moment, any limitation that might be imposed by Heisenbergian uncertainty. Imagine that this identical twin' is now seated in the next room (that is identical to your own) and keep in mind that there is nothing logically impossible about this whole situation. (Although practically the experiment would be very difficult to perform.)

---

* We shall examine this conclusion more critically later (p. 153), but let us assume for the
moment that it is true and see what it leads to.

If I were to ask you ' 'Shall I torture the chap in the other room? Would you mind terribly if I did so?" you might answer "I don't mind your torturing him because I won't feel the pain" (although, of course, you might feel some concern for his well-being).

If I were then to ask you: "I am afraid I have to kill you now. After killing you, would you mind if I tortured your twin? Do you mind particularly what I do to him after you are gone?" Although you might have some ethical concern for his well—being, you would probably answer: "I don't particularly mind what you do to him —though obviously I wish him well. . . ."

And now we come to a "paradox". I kill you instantly and grind you up or cremate you. I then bring the twin brother and make him sit down in your chair. This is logically exactly equivalent to replacing all the atoms in your brain with new atoms.* It must follow that if I now torture your brother you will experience pain and if I make him happy you will be happy. In short, you will survive death and will "continue" in your brother! So you should be just as much concerned about your brother's future welfare as you are about your own!

The curious implication of this is that so long as you are alive nothing happens to you (i.e. you don't feel pain) when 1 pinch your twin. However, the moment I destroy your brain and bring the twin to your room, "you" will feel pain when "he" is pinched!

2. What would happen if I were to destroy you and replace you with two identical agents instead of one? Would you then continue in each of them? If not, what or who decides which one you should continue in?

If information=ontological identity then you ought to continue in both the twins. This would be a fortunate state of affairs if you are now motivated by two conflicting goals—say two careers you would equally like to pursue or two women whom you would equally like to marry. You could then destroy yourself after having asked each of your twin brothers to pursue one of these two goals. Perhaps "you" would then be simultaneously satisfying both your desires!

'The fact that the replacement is done suddenly rather than gradually is irrelevant to the
question of ontological identity. See Appendix I.

3. We shall now go on to consider a more instructive version of the "twin paradox' '.

Is the paradox applicable to a situation where your twin had been brought up in an environment that was different from yours?

Assume that you actually have such a twin living now in (say) Paris and assume that he has been much more fortunate in life than you have been. Maybe he is more famous and has more money than you do; in which case you may envy him and may wish that you were him.

Perhaps when you were both still very much alike (in early childhood — say, when you were both 5 years old) your parents were divorced. Perhaps you were at that stage adopted by your mother and brought up under less favourable circumstances in Cambridge than your brother who was adopted by your father in Paris. Since you both began with the same genetic potential it was the environment alone that made all the difference. Your fate was decided at that critical moment when your parents were divorced. "How wonderful it would have been", you may feel, "if I had been adopted by my father; and my brother by my mother."

Assume that a time machine existed now. I regress both of you back to the stage when you were relatively undifferentiated. Assume, for the sake of argument, that the two of you were completely identical down to every fine detail at the time when the divorce took place. (This assumption is not critical to my argument but it simplifies the logic considerably.)

I now exchange two 5-year-olds so that you (the reader) are subsequently brought up by your father. I then bring both of you back to the world of the present. What would be the situation now? Would you now be your brother or would you still be you anyway in spite of the exchange? Would you now exist as your brother and experience all those fortunate circumstances which once belonged exclusively to him?

There are three ways of approaching this problem:

(a) There is an obvious linguistic sense in which you would not and indeed cannot exist as your brother. You could take the stand that since "I am by definition the person who is here and now, it is meaningless to even ask whether I would exist as my brother".

(b) However, hidden behind this linguistic riddle is a more fundamental philosophical problem.

Many of us often regret some (retrospectively) foolish decision or other which we made in the past. For instance, you may dislike your present career as a philosopher. At the age of 14 you may have decided on philosophy instead of (say) medicine and perhaps you very much regret having made the wrong choice. Given a chance to live your life again you would obviously choose to do medicine.

Yet if the linguistic argument presented in (a) is strictly correct you would not even have existed if you had chosen medicine. In fact the argument would imply that it is not even legitimate for you to regret your past (or wish that you had chosen a different career) since the only alternative would be non-existence!

This linguistic argument (a) must surely be false since it seems flatly to contradict common sense. There clearly is a sense in which you would have existed even if you had chosen medicine —and in fact all your daily actions are based on that fundamental assumption. As I pointed out on page 146, it seems likely that the particular environment you were brought up in is irrelevant to your existence (ontological identity) although it determines the content of your awareness. (One way of looking at this would be to suggest that the "I" within you is analogous to the program of a computer. The existence of these programs would clearly not depend on the particular inputs that the computer was called upon to handle.)

It seems to me, therefore, that no linguistic resolution of the twin paradox is possible, and this takes us back to where we started. If you had exchanged places with your twin brother early in life would you now exist in your brother's body in Paris—i.e. would you experience those sights, events and sensations which once belonged exclusively to him?

Let me explain this a little further since it is central to my whole argument. Supposing you (A) have lived in Cambridge all your life, and your twin (B) has been brought up in Paris. Assume that at least one complete cycle of metabolic turnover of brain atoms has occurred since you were separated (in early childhood). You ate English food (composed of E atoms) and your brother ate French food (F atoms). So your brain now is made of E atoms (in England) and your existence is "tied" to these atoms while your brother's existence is tied to F atoms.

Now supposing A and B had exchanged places early in life. Then B would have eaten E atoms and A would have eaten F atoms; but the final physical state of affairs in the world would be exactly the same as it would have been if the exchange had not taken place. In spite of the exchange there would now be one brain in England made of E atoms and one in Paris made of F atoms.

Since the physical world now is the same as (not merely indistinguishable from) what the physical world would have been if no exchange had occurred, it must follow that your existence would also remain unaffected. Hence you would now be in Cambridge even if you had exchanged places with your twin!"'

BIOLOGICAL CONTINUITY AND PERSONAL IDENTITY

Until now we have been assuming that existential continuity is unaffected by replacement of brain atoms; and this resulted in a series of "paradoxes". Perhaps the paradoxes prove that this central assumption is wrong. Maybe if the atoms of your brain are replaced (whether gradually or suddenly) you will cease to exist and a new person (identical to you but not the same as you) may begin his existence. The fact that you experience continuity in spite of metabolic turnover is no guarantee that you are not existentially a new conscious agent. Supposing I were now to suddenly replace the atoms in your head and supposing a new agent is thereby created. There is no way in which either I (the experimenter) or you (the new agent) could know that you were a new agent since you would experience an uninterrupted continuity of memories with the "old" agent. The question of whether you have ontologically a new existence can never be answered. But if it is really the case that every complete replacement of brain atoms leads to a new existence—then several bizarre consequences follow. For instance, one implication would be that it would be quite unnecessary to plan your life more than a year or two ahead since metabolic replacement of your brain would have occurred by then and you would, therefore, in effect be planning the future of someone else's life! Of course, that person would resemble you a great deal and have the same memories, etc., but he would be existentially new (just as an identical twin is existentially different from you)—and you would, therefore, really be planning the life of a future identical twin!

'I would like to emphasize that these paradoxes make sense only if the reader thinks of himself as one of the twins. From the point of view of a detached external observer the situation is completely symmetrical.

Here, and elsewhere, I have been asking the question "Is this agent A' existentially the same as the agent A who existed a few minutes ago?". But what exactly do we mean by existential "sameness"? Is this a pseudo-question arising from misuse of language?

To answer this, let us remind ourselves of the definition of consciousness which we considered earlier—in terms of the asymmetry between the observer and other agents. All my thought experiments involve changing the world in some way to find out whether this asymmetry persists. The question is "How does the new asymmetry which arises after applying a certain transformation (e.g. replacing brain atoms, replacing with an exactly identical twin, or replacing memories) relate to the asymmetry we began with before applying the transformation?". Does your "self" or consciousness as defined by this asymmetry continue after I have applied the transformation?

This is the only precise meaning we can give to the question of whether two agents are existentially the same or not. To a detached external observer the phrase "existentially continuous" is quite meaningless since from his point of view there was no asymmetry even to begin with. The criteria he would use for deciding whether A would continue as A' are largely arbitrary. For instance, if A is suddenly replaced by a replica (A1) made of new atoms, he may choose to call A' a new "person"; but if A's atoms are gradually replaced one by one to produce A' he may decide to continue to use the same name label A. His choice has no bearing, however, on what we have earlier referred to as ontological identity.

Further, the continuity of experienced consciousness is neither necessary nor sufficient logically to guarantee existential continuity. It is not necessary because you experience discontinuity on waking up from sleep (or anaesthesia) and in such situations there would obviously be no grounds for assuming that you were existentially new each time you woke up. And it is not sufficient either, because, although I can replace your brain atoms gradually (one by one) in such a way that your experienced consciousness is unaffected, there is no way in which either you or I could be sure that you were ontologically the same as the original agent.

SPLIT-BRAIN PATIENTS

The normal human brain consists of two nirror-image halves (the cerebral hemispheres) which are connected together by a band of nerve fibres called the corpus callosum. Almost all memories and skills that are acquired during an individual's lifetime are laid down simultaneously in the two hemispheres; and there is evidence to indicate that one of the functions of the corpus callosum may be to permit such duplication of memories.

In most people one cerebral hemisphere appears to be specialized for speech and so it would not be strictly accurate to speak of all memories being duplicated. However, in some rare individuals, both hemispheres appear capable of speech production; and in such persons one hemisphere is almost literally a mirror-image of the other.

During the last decade or so the brains of several human patients have been surgically divided into two by cutting the corpus callosum. The procedure was originally used to prevent the spread of epilepsy from one hemisphere to the other. When these "split-brain" patients8' were subjected to a battery of psychological tests, it was found that they often behaved as though they were inhabited by two "minds" or spheres of consciousness. In one well-known example (quoted by Sperry8) the two hands (controlled by different hemispheres) even tried to perform mutually incompatible actions. For instance, while one hand was trying to button the jacket worn by the patient the other hand simultaneously attempted to unbutton it!

Almost everyone (except a few theologians) now accepts Sperry's views that surgical bisection of the brain actually creates two "minds" or conscious agents where only one existed before.

Of course, it is true that these patients often look normal and even behave normally except when special tests are used to reveal the presence of two minds. There are two explanations for this. First: it is possible that the "dominant" hemisphere is dominant not only for speech but for initiating motor commands as well (just as a dominant spouse can sometimes completely suppress the individuality of his more submissive partner).

Second: it must be borne in mind that the brain of a split-brain patient has not been divided at the output level — the thalamic, bulbar, and spinal motor output centres have not been split. It is possible that when these centres receive conflicting "commands" from the two hemispheres, some simple strategy is adopted to resolve the conflict. One such strategy would be simply to obey the first command and ignore the second. These strategies may be embodied in the circuitry of the motor output system and may not need instructions from higher centres.

In spite of these strategies (which may help the patient avoid conflicts), there is at least one sense in which he really has two minds. If I pinch his left hand only his right hemisphere feels the pain and if I pinch his right hand, pain is felt only by his left hemisphere. So if we were to consider sensations and the reactions to sensations as being of prime importance, then we are really dealing with two independent spheres of consciousness here, although the person's motor response appears to issue from one mind because of limited "channel capacity" (or even actively adopted strategies) at the output level.

Experiment 2. Split—brains and personal identity

Our interest in split—brain patients arises from the fact that they can be used to construct bizarre thought experiments—a possibility that has already been recognized by Derek Parflt.9

Assume, for the sake of argument, that you (the reader) have speech centres in both your hemispheres. This assumption is not unreasonable, since, as I pointed out earlier, such cases are actually known to exist.

Now supposing it has become necessary to remove one of your hemispheres for some surgical reason (e.g. to relieve pressure from a tumour). You ought not to mind this since removing one hemisphere alone would not affect your existential continuity (in the other hemisphere). It would also make no difference to you whether I removed the right hemisphere or the left; since you would be confident that you would continue existentially in either hemisphere.

But now, let us assume that in order to simplify the surgery it has become necessary to cut your corpus callosum before removing one hemisphere. There is no reason why you should object to this minor change in surgical procedure: since the eventual outcome of the operation would remain unchanged you ought not to mind having your corpus callosum divided before hemispherectomy.

I then divide the corpus callosum, thereby instantaneously creating two minds. Assume that the "mind" associated with the right hemisphere is you (the reader). I whisper to you "I am going to destroy you now since I shall be removing the right hemisphere. But, of course, you don't have to worry since there is a spare 'you' located in the same body." (This message is delivered to the right hemisphere alone by whispering it through the left ear.)

Now you might realize at an intellectual level that there was another "spare" hemisphere available in which you would continue to exist even if the right hemisphere is destroyed. Yet in spite of realizing this you would probably object violently to being destroyed. In what sense, you might wonder, is the mind of that "other" chap sufficient to replace your own? We have a situation here that is analogous to the "twin paradox" except that both the conscious agents in this case can legitimately claim direct existential continuity with one conscious agent who existed just a few minutes ago. This ought to increase the confidence of each of the two conscious agents that he will "continue" in the other hemisphere if he is destroyed!

CAN THESE PARADOXES BE RESOLVED?

Supposing an agent exactly identical to you is sitting in the room next door. Your two minds are ontologically different at least in one sense, i.e. in the sense that I can do things to you (such as cause pain) while at the same time sparing the other person. Since the two of you simultaneously coexist in space, you are numerically different and there would be no grounds for "confusing" one for the other. Our "paradoxes" arise only when one agent A is destroyed and replaced by a replica A' made of new atoms. Also, when we are considering the ontological continuity of A, it is irrelevant whether the replacement is done suddenly (as with a twin) or gradually (as in metabolic replacement).

If we accept the position that ontological identity=information and that the "carrier" of the information is irrelevant then A should continue as A'.

But this would have several curious implications. One implication would be that if you are replaced by two or three agents (A', B', C') who are completely identical to you, you ought to continue completely in each of them. Another implication would be that if A' only partially resembles A, containing (say) only 90 per cent of the information that A contains (e. g. I could replace you with another man instead of an identical replica), then A's ontological identity ought to continue at least partially in A'. (We cannot be sure of this but it seems reasonable.) So, when you die, you ought to "continue' ' at least partially in all other people!

But supposing the "carrier" of the information in your brain—i.e. your brain—atoms—are also necessary determinants of your ontological identity then you ought not to even survive another two or three years since metabolic replacement of brain atoms would have occurred by then. And, again, the fact that you would appear (to other people) to have survived, or that you experience an uninterrupted continuity of memories right from childhood, is irrelevant to the question of your ontological identity.

So, after considering all these thought experiments we are, in a sense, back where we started. It looks as though questions about "empirical" identity are philosophically trivial and questions about ontological identity can never be answered!

But can we learn anything at all from the examples we have been considering? Some of our "paradoxes" seem to imply that we all go through life making certain assumptions about the nature of our existence. We sometimes accuse others of holding "super—natural" beliefs about souls and life after death without realizing that our own life is sustained by beliefs that are even more superstitious. For instance, we assume that we shall survive metabolic replacement of brain atoms and that we shall continue to remain more or less the same person in the near future. Sometimes we regret our past and assume that if we had lived elsewhere our lives would have been more fortunate. All these assumptions seem reasonable enough but if you examine them carefully (as we have done in this chapter) you will notice that they are all mere beliefs and that none of them can actually be proved.

Our revels now are ended. These our actors,
As I foretold you, were all spirits and
Are melted into air, into thin air;

. . . We are such stuff
As dreams are made of, and our little life is rounded with a sleep.

CONCLUSION

According to Wittgenstein "The results of philosophy are the uncovering of one or another piece of plain nonsense.' '

Wittgenstein's remark seems particularly appropriate to some of the problems we have been dealing with in this chapter. Perhaps the best that philosophers can hope to do is to state more concisely the nonsense that has already been uncovered by other philosophers; and in a sense that is what we seem to have achieved for the problem of personal identity.

I began by making a distinction between what I called "empirical" and "ontological" identity. I pointed out that the empirical identity question is philosophically trivial and then went on to explore all possible ramifications of the ontological identity question by inventing a series of "paradoxes". These paradoxes encompass all the questions that men have ever asked about souls, transmigration and immortality, including metaphysical questions such as "What am I?".

It may turn out that the ontological identity question can ultimately never be answered. But at least we have succeeded in understanding the question as clearly as possible and that is the best that one can hope to do in philosophy. Also, our analysis seems to have taken us slightly further than Hume, who believed:

The whole of this doctrine leads us to a conclusion, which is of great importance in the present affair, viz. that all the nice and subtle questions concerning personal identity can never possibly be decided, and are to be regarded rather as grammatical than as philosophical difficulties . . . we have no just standard by which we can decide and dispute concerning the time, when they acquire or lose a title to the name of identity. All the disputes concerning the identity of connected objects are merely verbal. . . .

APPENDIX I: SUDDEN V. GRADUAL REPLACEMENT OF BRAIN ATOMS

The paradox I have considered rests on the assumption that sudden and gradual replacements of brain atoms are logically equivalent. This assumption seems permissible to me since the final result of the replacement is exactly the same whichever procedure is used (i.e. there now exists a completely identical brain that is composed of new atoms).

The objection that sudden and gradual replacements are not equivalent might arise from the common tendency to confuse material objects with functions. If consciousness were a lump of "something" attached to the brain then it might become dislodged if the brain were replaced suddenly but might "stick" to the brain if the replacement were done gradually. But if consciousness is a function (as it almost certainly is) the rapidity of replacement becomes irrelevant (i.e. the physical momentum of sudden replacement would not give it a "jolt" as it would if consciousness were something like a bit of matter attached physically to the brain).

It is true, of course, that if the replacement is done gradually then continuity of function would be preserved (like planks being replaced one by one in a bridge so that it is never allowed to collapse), while sudden replacement would interrupt, albeit briefly, the continuity of function.

But temporary interruption of function is not detrimental to the preservation of existence: complete cessation of brain activity (as during deep anaesthesia), followed by recovery of activity, interrupts the continuity of the stream of consciousness but the person who wakes up after the discontinuity (i.e. after anaesthesia) is exllstentially the same as the original person; unless one adopts the supernatural position that the original "soul" departs and is replaced by a new one.

In the case of planks being replaced in a bridge it is largely a linguistic problem whether we choose to call the "new" bridge (arising from the replacement) the same as the original one or merely identical to it. We may choose to define anything that results from gradual replacement as the same as the original object and that which results from sudden replacement as a new but identical object; and this nomenclature is entirely arbitrary.

But the question of whether my existence would continue if my brain atoms were replaced cannot be reduced to this kind of linguistic analysis. An outside observer may choose to call me the same person if the replacement had been gradual but not if the replacement had been sudden. But the question is not what I should be called but whether I would ontologically continue to exist or not——and this cannot be answered by merely considering what criteria people generally use in such situations.

APPENDIX II: THE MIND—BODY PROBLEM (p. 142)

Not everyone would find it necessary to believe in a "split" of the kind described by Lorenz or implied in my thought experiment. Grover Maxwell has argued, for instance, that since the ontological or intrinsic (as opposed to descriptive or structural) properties of the world are fundamentally unknowable to science, the possibility is open that some of these properties are just the ones that are exemplified in the events that constitute our own private experience. In that case, mental events would be merely one kind—perhaps a rather special kind—of physical event (G. Maxwell, in Consciousness and the Brain, ed. G. G. Globus et al., Plenum Press, 1976). However, the assumption of such a split is not necessary for creating the "paradoxes" that I have described.

Discussion

Mackay:
You argue that if my physically indistinguishable twin sat down in my chair, that would be "exactly equivalent" to replacing all the atoms in my brain, so that "in a sense" I would be sitting in that chair. But (1) for this to be at all plausible, my twin and I would have to have had identical experiences (including meeting the same people at the same time and in the same geometrical relationships) at every point throughout our two lives. This is impossible in principle, unless you could have us in duplicate worlds with duplicate people – i.e. unless you had already solved the problem of producing identical twins!

(2) Something that makes me permanently distinct from any other cognitive agent is that in principle I can be "Thou" to him and he to me. Even theoretically identical twins would become distinguishable in principle the moment they were able to engage in dialogue. One of them, for example, would have to listen while the other talked, and so on. (3) If, as I would argue, the identity of a conscious agent is associated with the interpersonal roles he can play, then no other conscious agent who exists simultaneously with me in the same world can be confused with me, since in principle we have the capacity to be "Thou" to one another. To take two conscious role players, however indistinguishable, and exchange them is different from taking one conscious role player and exchanging his bodily atoms, precisely because the first requires the prior existence of two conscious role players, and the second only one. The question "Which was which?" is simply resolved by tracing the roles, active and/or passive, played by each up to and through the point of exchange.

RAMACHANDRAN:

My question is: "What would happen if an agent exactly identical to me were to be created?" The fact that this is impossible in practice is irrelevant to my argument.

Your second question is an extremely interesting one. My reply would be that in the kind of situation I have been considering (i.e. in a Laplacian world) the two agents would not be able to engage in "dialogue" even if they were allowed to confront each other. Dialogue requires exchange of information and since both our agents contain exactly the same information no such exchange can occur. Each agent would say and do exactly the same thing as his twin. Let anyone who believes in his "free will" imagine this situation!

Supposing I now create a replica of you (Donald MacKay) and allow the two of you to confront each other in a completely featureless room—so that your brain states are completely identical right up to the point of confrontation. You would then try to be "Thou" to him by starting a conversation but to your surprise you would discover that he always simultaneously utters the same words as you. You might then even go on to explain to him that you wanted to prove to the audience here at the conference that you could be "Thou" to him—but you would be unsuccessful and the audience (who are watching you through closed-circuit TV) would notice that the other Donald MacKay was simultaneously making equally futile attempts to be "Thou" to you!

Fortunately, there is a trick you could employ to break the "deadlock". You could use a radioactive device designed to (say) emit a signal either towards you or towards him. You could then decide in advance that only the person towards whom the signal was emitted:

should start speaking. This would at least help you start a dialogue but even then the conversation would soon become quite boring since your brains contain identical information.

Furthermore, in all my thought experiments the exchange of bodies is done before the two agents have had a chance to initiate a dialogue. So although your question is an interesting one it does not create problems for the examples I have been considering.

CHAPTER 10

Mind—Matter Interaction in the
Psychokinetic Experience

SUZANNE PADFIELD
West Wickham, Cambridgeshire

For many of you, the topic about which I am going to make some comments, namely psychokinesis or the apparent ability to move or alter matter by paranormal means, will seem startling, remote and implausible. May I stress here and now that the remarks I shall make are aimed fundamentally at bringing us merely to the starting-point of discussion, not at attempting to provide a scientific answer under the guidelines of science at present in existence.

Just as we cannot measure an electric current with a ruler, neither can we register a step in the evolution of mankind with an attitude or frame of reference proper to a much cruder and outworn part of the history of man's mental understanding. We have to adjust our attitudes, alter our frame of reference and angle of approach. The startling and intensely uncomfortable nature of psychokinetic phenomena provides the reasons which make the study and discussion of them a very good starting-place.

For the purpose of aiding scientific understanding, I have in the past demonstrated some psychokinetic effects under controlled conditions. Most of these experiments took place at the Paraphysical Laboratory, Downton, Wiltshire, with Dr. Benson Herbert, and the principal experiment involved a piece of apparatus known as a light mobile system. This apparatus consisted of a single strand of polyester fibre, the polymer known as polyethylene terephthalate, 25 cm long and 16 microns in diameter (chosen for its high tensile strength and low electrical and thermal conductivity because of its low moisture content). A straw beam 8 cm long was attached to one end of the fibre by means of sealing wax and the other end attached by the same means to a cork which fitted tightly into the neck of a large glass bottle in which the straw became suspended horizontally. The straw was balanced by two differently coloured pieces of plasticine, one at either end, and the sides of the bottle marked with vertical lines, enabling the angle of rotation of the straw to be observed accurately.

The system was placed on a vibration-free surface in a room free from disturbances and left for 24 hours, being monitored during that time to ensure that none of the known factors which might produce an effect on the straw beam were in operation. At the selected time for the experiment (no detectable movement of the beam, i.e. less than half a degree movement, having been recorded for 24 hours) I would enter the room quietly and stand 5 or 6 ft away from the system. I always wore a visor to reduce effects of heat radiation from my face and electrostatic charge from my hair.' I would then commence to "direct" the beam a certain number of degrees towards or away from me, either by free choice, having stated the number of degrees of rotation and direction beforehand, or at the command of the experimenter who was also present in the room but only near enough to the system to allow accurate observation, usually 10-12 ft. The experiment was successful about 70 per cent of the time and the straw beam would rotate the required number of degrees and remain still until a further direction or degree of rotation was chosen. A series of up to fifteen runs of psychokinetic influences could be accomplished during one experimental period with successful deflections of the beam from 5 to 90 degrees, fatigue usually deciding when the period would end.

This particular experiment was carried out almost weekly for a period of nine years, and various refinements were made at different times. For example, when the subject and the experimenter entered the room a period was allowed during which any effect upon the mobile due to a change in temperature or humidity caused by the addition of two human bodies to the room would, if it was going to occur, have been observed. In fact we found that the addition of two people to the room caused no detectable effect on the system, no movement of the beam being observed prior to the start of the experiment. When a large number of observers wished to be present this did pose a great problem as the disturbance caused by temperature and humidity change and by general bustling around, no matter how strict one tried to be, usually caused an oscillation in the system or alignment of the beam. We overcame this problem by giving observers visual access via the window of the room containing the light mobile system. Later, at the Stanford Research Institute in California in 1976, I was able to produce deflexions of the beam successfully using only the monitor of a video camera for visual contact, the system itself being in an adjoining room with no one present. I have also been able on a large number of occasions to demonstrate "psychometry", which is the ability to tell the past of an object, or events in the history of an object, merely by handling it and with no other information available?

There are similarities and differences in my subjective experience of what is taking place when I do both psychometry and psychokinesis, which I believe are indicative of a new attitude and framework which may provide a useful tool for future understanding of the nature of matter and of consciousness.

In the case of psychometry I am aware of a feeling of scanning the past events of the object I am holding and am aware of sequences or memory tracks, some of which become actual events and others which existed only as possible events. Both are explored and the actual events are singled out and emphasized in the same way as one might retrieve a memory trace.

In the case of psychokinesis I am also aware of a sequence of possible events, as it were, in stages which I feel myself to be exploring. It differs from psychometry in that I am aware of the possibility of future events which are open to me and I am able to choose one of them, which becomes the actuality. In the case of the light mobile system it is the new position it will occupy.

In both cases there is the subjective experience of exploring possibilities rather in the way one might remember what one did yesterday and the things one might or might not have done in retrospect.

I must emphasize here that I feel myself to be a part of these processes and these events and not in any way separate from them. I am a part of the events, of the sequences, not merely observing them.

This subjective experience of knowing all the possibilities and being a part of each one as it occurs is not mine alone. It occurs frequently in the revelations of mystical literature, "I am That",3 and traditionally the second stage of spiritual development embodies just this experience, the notion that the perceiver is not separate but is a part of the process. In 800

B.C., Patanjali, the founder of the school of Raja Yoga, wrote in his Yoga Sutras:" "There is identity of relation between memory and effect-producing cause, even when separated by species, time and place." Obviously the word "species" may cause some misunderstandings. The text and commentary 1 use as my source'* discusses the problem of correct translation at some length. If I do as Alice Bailey suggests to her readers and apply my own concept of what Patanjali was saying in the light of my own experience and the rest of his teachings, I would insert "differences of form" rather than and in place of "species". But it is the phrase "identity of relation" that is the key. The dynamics is determined by the identity of structures, one with another, rather than spatial positions. Let me explain further.

Consider the brain as an atomic organization or society. The process of thought may be considered as the reciprocal activity taking place between different atomic organizations. The memory trace is the effect, within the brain as an atomic organization, of its interaction and degree of identification with other atomic organizations. By identification, 1 mean that what we know as a memory trace occurs when the sequence of atomic codings within the brain becomes identical with those of other atomic organizations. Patanjali says: "The past and the present exist in reality, the form assumed in the time concept of the present is the result of developed characteristics and holds latent seeds of future qualities."5 Here again I believe the translation has suffered and the context of the Sutra strongly suggests replacing the obscure word "qualities" by the more precise words "states", which makes absolute sense in the light of my own experience.

Those "developed characteristics" of which Patanjali speaks are the sequences of atomic codings. They are, I believe, what I encounter and interact with when I do psychometry. "The latent seeds of future 'states' " embodies the notion of a choice of one among many possible atomic rearrangements of an encountered atomic organization, via the interaction with the brain.

When two codings match, you have a memory: at the instant they match you have the possibility, w'a the interaction, of creative thought, imagination to one degree, the macroscopic alteration of form to a larger degree. Obviously people will look for and try to make some comparison with more conventional psychology. They might expect some change in the nervous system in the case of psychometry which allows images of past events to be re—created. What I am saying is that there is no re-creation of images, but that what is taken to be the re-creation of an image is in fact a newly created event arising out of the identification or matching of codings within the brain and any encountered organisation (or object). People term the new event a "memory" because of degrees of similarity. But I am saying that what we term "memories" are in fact new events never precisely identical (as those who have to deal with eye-witness accounts in court will testify).

Concerning the similarity between the past and the present: in the case of psychometry people would normally expect that the experiences that an object had undergone must have changed it in such a way that the psychometrist could re—create the image of those things. The normal way of thinking would be that the object could "pick up traces" which would stimulate analogous memory traces in the mind of the psychometrist, therefore implying some kind of passive memory store. This is not so. Of course each organization is structurally altered atomically by its encounters with other organizations. But each encounter (including its encounter with a psychometrist) is actually a new event and the image the psychometrist perceives is that of a new event. One asks why those experiences tend to bear a startling similarity to events someone recognizes and verifies the psychometrized object to have been involved in. To answer this question I can only say briefly that the elements which go to make up the total object or organization are connected in a similarity space in which distances are defined by degree of similarity and where time and space do not automatically appear at all. In such a space a natural form of connectivity is a sequence of elements or patterns of elements in which neighbours differ only minimally. I further have to postulate that cerebral tissue has evolved the special function of rapidly producing structures which match the coding of some parts of such a sequence and thus get connected to the others. These will seem to be in the past of the object. This form of connectivity is unfamiliar in current physics. On the other hand, there is a growing interest in discrete or combinatorial approaches to physical foundations and in such approaches sequences in similarity space appear naturally at a more primitive level than space and time. (See, for example, Bastin and Noyes, "Possible physical interpretations of the combinatorial hierarchy", to be published in proceedings of the July 1978 Tutzing conference on "Quantum Theory and the Structures of Time and Space' ' .)

People might interpret what I have been saying about psychokinesis as my having suggested that the mind of the subject would have the power of exploring a great range of possibilities which are open to a given object or physical situation, so that by choosing and working on one of them the "mind" could make it come about. This is a step in the right direction of thinking from my point of view, but the emphasis is wrong. If you look back at my description of influencing the suspended mobile you will see that this summary gives too much autonomy to the mind. Possibilities are in fact explored, but which one is to be chosen is dictated to a large extent by the possibilities themselves.

I am well aware that I have encroached upon the subject of physical particles and what they can do, and that I have postulated that, in some way, the structures formed from them may have the power, via identity of structure, of matching with and leading what we call our consciousness forwards or backwards along sequences determined by the interaction of the microscopic organizations both within and outside the brain. I am also aware that this idea is foreign to current physics, where this kind of connectivity has not been noticed. Formerly physicists would simply have said that no such thing could happen; now they are not so unequivocally certain about the matter. Some writers are seriously investigating the freedom allowed for psychokinesis by current quantum theory. Even they, however, have only demonstrated that a good deal of freedom exists. They have said nothing positive about the way their information organizes the details.

My function in this situation is only to present you with my experience and to invite you to examine my argument that particular forms of connectivity are dictated by these experiences.

Editorial note:

The above has been printed with only minor changes from the author's original manuscript. A restatement of the theoretical ideas in more conventional terminology based on editorial discussions with the author, may be of some value, provided it is borne in mind that the concepts may not be capable of exact translation. The basic ideas are: (1) laying down of memory is not the laying down of a precise copy of an image. but the creation of a structural change which encodes the event; corre-spondingly the recall process is an active one and does not in general re-create the original event exactly; (2) objects are capable of laying down memories of events they have experienced, by a similar mechanism to that of personal memory; (3) the structure of an object may encode a future possibility as much as it may encode a past event; (4) by generalizing the sense of identity, a psychic may perceive images connected to an object in the same way that we normally perceive images related to our nervous systems: (5) psychometry is explained as a perception created in this manner from the coding of a past event in an object; (6) psychokinesis is explained as a process of first creating an image of a future possibility for the object out of a structure within the object which encodes that possibility, and then interacting with that struc- ture so as to trigger off a causal chain leading to that possibility being realized.

It is an interesting question whether the author's theories, if valid, would reduce the paranormal phenomena she describes to normal ones. The explanations she gives would be quite conventional if one were to accept the idea that an external object could by some mechanism function as part of a person's nervous system, a possibility which it is difficult to deny on purely logical grounds. While Dr. Ramachandran in his paper asks why one person's experiences should be linked to one particular nervous system, Ms. Padfield argues that the principle just stated can on occasion be violated; while again Mrs. Noakes (see Josephson's Afterword to the Conference) suggests that physics as currently interpreted gives an incomplete description of physical reality, and that subtler aspects of a person's identity exist which are not necessarily confined to his usual physical body. These papers all point towards the idea that personal identity and the relation between objective and subjective reality are questions of crucial importance to science. [B. D. J.]

APPENDIX

During the conference, several questions about experimental
procedure, and in particular about the separation of psychokinetic effects
from movements due to familiar causes, were put. The experimenter is
very aware of the complexity of the problem of isolating effects, and in
my opinion the safest course by far is to pursue what I will call a
pragmatic approach which does not presuppose that one has a complete
knowledge of relevant effects. If one observes the suspended beam for
long enough, one can make an estimate to any desired measure of
accuracy of the probability of a given effect taking place as a result of
uncontrolled effects whatever these may be. Then, provided only that one
is satisfied that the introduction of the subject has not altered any of the
ambient conditions significantly, one can give an upper bound to the
probability of the subject's effects being fortuitous.
In the experiments at Downtown, this pragmatic approach was
consistently used. The beam was observed every half—hour or so continuously for 24 hours before an experimental session, and all excursions of the beam greater than a fixed angle were recorded. What happened was that during the 24 hours no excursion greater than one or two degrees was observed; usually there was no excursion at all. During the experimental sessions I was able to produce excursions of the beam through angles of say 45 degrees, at will, every few seconds for as long as I was asked.

In circumstances like these I would hardly bother with probability calculations, but they would be there in principle for sticklers on experimental protocol.

It is very unfortunate that I seem to have given a misleading impression in my talk, when I spoke of the 24 hours' observation period. Some of my questioners evidently thought that I said that only one experiment (i.e. excursion of the beam) was recorded in one 24-hour period, and some of the questions are misdirected in consequence. What I meant to say was that there would be a 24-hour observation period before each experimental session, the session including an indefinite number of excursions of the beam.

Discussion

MACKAY:

The "information rate" (number of bits of information per day) claimed for the alleged communication channel here is so low that our normal instincts for the dangers of correlated disturbances can be unreliable. For example, an extremely minute correlation between the process by which the "commands" were selected and the pattern of earth tremors, etc., that might physically influence the beam could give rise to a spurious appearance of information transmission at these low rates. How did the scores vary according to the method of selection?

PADFIELD:

I would agree with Professor MacKay's criticism if the experimental procedure were as he supposes, and I hope my remarks above (see Appendix) have cleared up the misunder- standing. I hope, too, that my description of our "pragmatic approach" assures him that our reliance on "normal instincts" had a proper basis.

A lot of attention was given to the selection of commands. At an early stage in the course of experiments, trials were made in which the instructions were selected by a suitable random process using random number tables to dictate the timing, and direction of the excursion of the beam which was to be aimed for by the experimenter. It was found that the degree of success of the subject was not related at all to the method of selection.

MacKay mentions earth tremors. In some experiments a simple seismograph was kept running. There was never any correlation between earth tremors and other effects of any sort.

BARLOW:

Obviously you cannot give details of all the control observations and other precautions you took when doing these experiments, but I wonder if you could give us a few particulars in order to show us how easy, or difficult, it was to come to the conclusions you have come to?

First, what is the natural period of your device when you are not trying to influence it in any way? Second, I think you said it took a day to recover fully from a perturbation, but I wonder if you can specify rather more precisely the time constant of the decay of oscillations following an imposed perturbation? Third, I wonder if you can specify the range, and perhaps the standard deviation, of positions observed if the reading was taken at, say, daily intervals following a long period without any deliberately imposed perturbations? Fourth, what was the amplitude of the perturbations you thought you achieved, how long did it take you to achieve them, and how did you decide what direction of perturbation you would attempt on any given day?

The system may not be simple enough to give straightforward answers to these questions, but even very approximate answers would indicate to us rather more clearly the nature of the task of deciding whether your device is influenced in the way you believe that it is.

It is not always easy to decide whether a signal has emerged from the noise, even when the noise behaves well and observations can be repeated every few seconds. It must be a fearsome task if the noise is less regular than expected and if you can only make one observation per day.

PADFIELD:

Firstly, the system is certainly far beyond the point of critical damping, in the direction of very low Q. I don't know exactly how far, but the sensation one gets is always of a beam which drifts, certainly not one which oscillates. The fibre which was used for most of the experiments (after a great many had been tested) was a single strand of a polyester- polyethylene terephthalate with a diameter of 16 microns. It has a very low torsional elastic constant. In fact the elastic constant plays a very small part in the thinking about the experiment. For reasons which are not understood, these beams seem to have a natural alignment to which they settle down (quite apart from the activity of the subject) and you have to turn the torsion head several times before the torsional force is great enough to overcome this tendency to alignment.

The second question is answered by my answers to MacKay, particularly in relation to the confusion over the 24-hour period, and by my statements about the restoring torsional force and the damping. There must presumably be a time constant which characterizes the exponential relaxation after an excursion of the beam, but I have no idea what it might be. When 1 cause the beam to rotate it moves through a finite angle and stays there.

In view of my foregoing comments on oscillations questions 3 and 4 seem to boil down to the question "How far do you customarily move the beam?". The answer is that it usually moves anything from 5 degrees to 90 degrees and that it is pretty much under my control how far it goes.

VESEY :

To the best of my knowledge I have never moved anything psychokinetically, so I was very interested in your account of the experience of doing so. From what you said there would appear to be one respect in which the experience of moving something psychokinetically is like the ordinary experience of, say, moving one's arm. Lotze once said that he felt "thoroughly at home" in his voluntary bodily movements, as distinct from certain other bodily activities. (I suppose he was thinking of things like digestive processes—things we would not ordinarily say were "done" by the agent, although he certainly does something else, namely eating and drinking, to bring them about.) Now, you said that you had the experience of "not being separate" from the psychokinetically induced movement. I took you to mean that it felt rather like making an ordinary bodily movement. But you went on to say something which seemed to me to conflict with that. You said — didn't you? — that you had to visualize the desired movement in order to bring it about. I would have thought that to the extent to which you had to do that it would seem to you that you were separate from the movement. It would make it more like "willing" dice to fall in a certain, visualized, way. I wonder if you could say a little more about what you meant by not feeling separate from the movement in the psychokinetic case.

PADFIELD:

I was really concerned to make the point that to get the effect observed, you had to feel a part of the whole system and process including the mobile, as distinct from as it were giving it instructions through what MacKay calls a communication channel. (Indeed, this second process means nothing to me experientially beyond enunciating the words of the instructions.) I actually gave much more detailed instructions than merely to visualize the movement, for you had to get sufficiently a part of the detail of the system to intervene between two states. The misunderstanding may be due to a use of the word "visualize" which carries a sense of seeing as a process where a message is carried from a thing to a mind. On the other hand, "visualize" also carries a sense of reproducing something of what has been seen, which seems to militate against separation.

# CHAPTER 11

## Phenomenal Space

M. J. MORGAN

University of Durham

One of the attributes of mental events, such as thoughts and sensations, that has been most persistently described as distinguishing them from physical things is that mental events do not have an obvious location in physical space. If we are trying to catch a cricket ball, a physicist could tell us the trajectory of the ball, and where it is at a given instant. But if called upon to say where our perception of the ball is he would obviously be much more puzzled about what was required as an answer. One course of action that might occur to him is to get the observer to point to, or otherwise indicate, the position in which he sees the ball to be. In this manner a trajectory of the perceived ball might be plotted out. Such a trajectory might differ, and indeed usually would differ, from the trajectory of the actual cricket ball itself, because of such factors as the speed of the visual response. Suppose the physicist, having established the perceived trajectory, were to examine the point in space which the perceived ball occupied at a particular time. He might do this, if he were very innocent, in the hope of seeing what a perceived ball looks like. Of course, he would find nothing there. No matter how hard he looks in the space occupied by cricket balls and the like he will not find perceived or phenomenal cricket balls. This is the sort of consideration that has led to the notion that perceptions do not occur in physical space at all, but rather in a purely mental or "phenomenal" space. If this claim is true it is clearly a very powerful reason for maintaining a mind—matter dualism. I therefore think it important to point out that the concept of a phenomenal space is mistaken, or at best confused, and this is what I shall try to argue in the following pages.

Let us first of all analyse applications of the concept of phenomenal space a bit further. Consider a well-known illusion such as the "Pulfrich Pendulum". The observer looks at an object swinging from left to right on a length of string at right angles to his line of sight. If he places a neutral density filter in front of one of his eyes, carefully keeping both eyes open, the observer now sees a very striking effect: the pendulum, instead of moving in a plane at right angles to the line of sight, now seems to move in an ellipse, constantly changing its apparent distance from the observer. The orbit is clockwise in depth (as if viewed from above) with the filter over the left eye and anticlockwise with the filter over the right eye. The details of this illusion are not important for the present discussion; what I wish to draw attention to is the fact that in this case, as in other illusions, it is possible to apply conflicting spatial descriptions to the object. We say that the physical object (the bob of the pendulum) is moving in a straight line, whereas the perceived object is moving in an ellipse. In the case of the physical object we are accustomed to saying that it moves "in" space. But if this description is applied, what shall we say of the movement of the perceived object? Is the ellipse also "in" space? If so, what space is it "in": the same space as that of the physical bob, or some special space reserved for perceptions? '

A widely canvassed answer to this question is that there is indeed a separate space for perceived objects, a "phenomenal", "subjective" or "mental" space, quite distinct from the space in which the physical object moves. A phenomenal object, phenomenally moving in this phenomenal space, can be meaningfully described, on this theory that I am outlining, as moving in ellipses, straight lines or whatever. This is not supposed to be a merely idle analogy, a sloppy use of the same word "space" to cover two concepts that share nothing whatsoever; on the contrary, as we shall see, it is often thought that phenomenal space shares sufficient properties in common with physical space for it to be meaningfully described as having a geometry — although, as we shall also see, it has been supposed that these geometries do not have to be identical.

It would be unfortunate to give the impression that the concept of a phenomenal space has arisen only out of perceptual illusion. Consider another example, in which Shepard and Cooper showed people drawings representing complex three-dimensional shapes with several limbs and angles. The observers were given the task of judging whether two such shapes, presented together, were the same or not. One of the shapes could be rotated relative to the other, or both rotated and mirror-imaged: in the first case the observer was meant to say that it was the same shape, in the second case that it was different. The finding was that the greater the angle through which the shape was rotated, the longer the observer took to decide whether it was "the same" or not. When asked how they did the task, observers straightforwardly replied that they mentally rotated one of the shapes until it coincided with the other version, to see if they matched. It seems that the greater the angle through which they had to carry out this mental rotation the longer they took over it. Obviously, the observer is not rotating the physical object on the paper. As in the case of the Pulfrich Pendulum, it is tempting to say that what is really moving is an image or phenomenal object, and that it is rotating in a phenomenal space.

Gregory's theory of the geometric illusions provides another illustration of the way in which the concept of phenomenal space might be used, although Gregory has not emphasized this aspect explicitly. In this theory certain features of line drawings are thought to trigger constancy scaling mechanisms normally involved in three-dimensional representations of objects in space. Lines that are indicated by primitive perspective features as being further away from the observer are expanded, and those indicated as nearer are relatively contracted. One interpretation of what is going on here is that out of the line drawing a representation of an object in a 3-D phenomenal space is being constructed, and that the observer is making judgements of lengths of lines in the phenomenal figure. However, we must be cautious here, because Gregory stresses that the illusions may be seen even when no depth is perceived in the figure. In such cases the illusion is treated as a judgement of line length determined by an unconscious process of "primary scaling". Even this, however, is treated as a scaling operation, which seems to demand a spatial representation of some sort.

To conclude this brief introduction to uses of phenomenal space, I give the following quotation, which may be more aptly considered as a blunder than as a reasoned statement, but which nevertheless illustrates a certain popular conception:

One important hypothesis suggests that the brain contains a model of the outside world. We are so familiar with this model that we think it is the outside world, but what we are really aware of is an imitation world, a tool which we manipulate in the way that suits us best and so find out how to manipulate the real world which it is supposed to represent. . . . When we cross the road and avoid traffic we are really dodging the moving buses and cars in the mind.

In other words, there are two sets of moving buses and cars; one set in the physical world, which are dodged by our real bodies, and another set moving in a purely phenomenal space, which are dodged by a phenomenal version of our bodies. It is fortunate indeed that these two dramas are utterly distinct, for if it were not so a collision with a phenomenal bus might injure our physical body, with disastrous results. Collisions between real buses and phenomenal bodies are probably less to be feared on the whole, although bus drivers might think differently on this point. Luckily these speculations about cross-modal traffic accidents need not detain us, for the supporters of phenomenal space insist that it is utterly distinct from physical space and that there is no possibility of interaction between the two. Indeed, this is what dualism is all about.

The first point that needs to be stressed about the doctrine of the "two spaces" is that it differs in a very important way from representational theory as we normally apply it to perceived qualities such as colours, smells and the pitch of sounds. We do not say that there are two distinct sets of colours, one physical and the other phenomenal. On the contrary, we say that phenomenal colours are the only kinds of colours there are: they represent, not a further set of colours, but differing wavelengths of light. Colour science made little progress until it was realized that colour mixture, for example, was a property of colours rather than lights. There was much confusion concerning the interpretation of Newton's experiments until the Young—Helmholtz theory finally became established and cleared up this logical point. Similarly, we believe that there are only phenomenal smells. (The philosopher Bradley was mistaken in ascribing to physiologists the belief that when we smell rotting fish we are aware of the stinking state of our nervous system.) There is no need to multiply examples. However, the doctrine of the two spaces is not like this, because the same word "space" is still used to describe both the phenomenal and the physical referent, and it is considered that the two spaces share a number of important features. Physiologists could not agree, I take it, that there is only a phenomenal space, in the same way that there are only phenomenal colours. If they held that there was only a phenomenal space, it would be hard to understand why they should attempt to explain various aspects of our perception (such as binocular vision of depth) by drawing spatial diagrams, and by appealing to the fact that light travels in straight lines. If "straight lines" and the like were purely phenomenal, explanations couched in these terms would not be physiological explanations at all. They would resemble a "colour theory" such as the one of Goethe, who insisted that explanations of colour should remain within the realm of colour. There may be something to be said for a rigorous phenomenology of this kind—I do not want to enter into this argument at present, but merely to point out that the scientific treatment of perception has rejected pure phenomenology in the case of colours and smells, but has not succeeded in doing so when it comes to the spatial aspects of perception.

Exactly the same point can be made about the treatment of time and duration in psychology. We speak of physical events such as eclipses having a duration, but we also talk as if it were meaningful to apply exactly the same word to experiences. It is recognized that "subjective" or phenomenal time will run more or less slowly on different occasions in comparison to a physical clock, but it remains "time" none the less. William James discussed Helmholtz' 5 treatment of this problem:

If asked why we perceive the light of the sun, or the sound of an explosion, we reply "Because certain outer forces, either light waves or air waves, smite upon the brain, awakening therein changes, to which the conscious perceptions, light and sound, respond." But we hasten to add that neither light nor sound copy or mirror the ether or air-waves; they represent them only symbolically. The only case, says Helmholtz, in which such copying occurs, and in which "our perceptions can truly correspond with outer reality, is that of the time-succession of phenomena. Simultaneity, succes- sion, and the regular return of simultaneity or succession, can obtain as well in sensa- tions as in outer events. Events, like our perceptions of them, take place in time. . . ." (W. James, Principles of Psychology, Vol. 1, pp. 627-8.)

Helmholtz calls this the "only case", but his use of the word "outer" gives the game away for space as well. This is a spatial term that can be applied to physical events and to phenomena. I perceive a tree as outside my head and my headache as inside; a physiologist would say that the brain events corresponding to both these phenomena are inside the brain. We may have arguments on this point but we use the same words. I think it was Winston Churchill who described the Americans and the British as "two nations separated by a common language". What Kant called the inevitable ambiguity of terms like "outer" and "inner" has very much this divisive effect in the philosophy of perception. In the Critique of Pure Reason Kant blames scepticism of the Berkeley variety entirely on this ambiguity.

Reference to Kant brings me to a second general remark on the concept of phenomenal space. This is that Kant, although he may be held in part responsible for the intrusion of phenomenal space into psychology (through figures such as Lotze and Hering), nevertheless did not himself believe in "two spaces". The whole point of his argument is that space is purely phenomenal. He believed it to be a grave error to postulate a physical space separate from the one involved in our perception. For this he had two main motives, which —if we follow critical arguments by P. F. Strawson and others — do not fit very comfortably together. First, there is the point I have already discussed: that we apply a full—blown representational argument to colours and smells, but draw back on the brink of applying a similar rigour to space and time. Without really saying why, Kant found this privileged position of space and time inelegant, and urged that the representational argument be pushed to its obvious conclusion. Arguing in this manner Kant poses as an empiricist who wants to make empiricism more rigorous, just as Berkeley had poured scorn on Locke's distinction between primary and secondary qualities. Like Berkeley, Kant criticizes the distinction between primary and secondary qualities as ' 'merely empirical".

This is Kant as a wolf in the empiricist fold, devouring the sheep on the pretext that a few less animals will make the flock stronger. Or, to change the metaphor, he is throwing out the baby to keep the bathwater clean. This inspired benevolence has not on the whole had a very good reception from scientists at their bench, who are distressingly tolerant of possible logical flaws in their outlook provided they carry on getting good results, and who tend to react with gross ingratitude to offers of assistance from the like of Kant and Berkeley. But Kant had a much more powerful reason for his theory than a desire to make empiricism self-consistent, and here his thinking has had much greater influence, if no more ultimate success. This was his conception of geometry, an account of which will take us to the heart of the present problem.

When we are taught Euclidean geometry in school (at least, as it used to be taught) we are offered one or more "proofs" of Pythagoras' theorem, which purports to be a metrical statement about triangles, namely, that the square of the length of the hypotenuse of a right-angled triangle is equal to the sum of the squares on the other two sides. If we now go out into the field and carefully construct a right—angled triangle, we shall find, within the limits of experimental error, that the Pythagorean metric adequately summarizes the results of our real measurements. This agreement between theory and observation is amazing. Kant, at any rate, found it simply staggering. On the face of it, unaided reason has enabled us to predict the results of real measurement, carried out on physical objects. Kant called reasoning of this kind a priori synthetic: a priori because it does not seem to depend on previous experience, and synthetic because it is about facts, not about tautologies. Kant's philosophy of space and time is concerned with the problem of how such reasoning is possible.

Without going into the details, which are forbiddingly complex, his answer was once again that space and time are phenomenal: but now much more fundamentally so than the level on which colours and smells are phenomenal. We can imagine experiencing a world in which colours and smell are absent; but (according to Kant) there is no form of experience intelligible to ourselves that does not involve space and time. There are thus certain truths which could be stated to hold for experience even before we have the relevant experience: they are the minimum conditions for any experience whatsoever. This has some analogies to the Socratic method of showing that we have innate ideas of geometry, but with an important difference; for Kant it is not merely a matter of demonstrating that we are, as a fact, born with certain innate ideas of space —the important point is that without just these ideas we should, as experiencing beings, never have been born at all.

It may seem a long way from these general considerations to Euclidean geometry in particular, and indeed it is too far, for Kant never shows beyond the utmost generalities how a derivation of Euclidean geometry might be managed according to his principles. It is now generally held that his effort was doomed in any case, because other geometries have been discovered that are at least as self-consistent logically as the Euclidean variety (see below).

Nevertheless, a revisionist version of Kant's claim has been claimed in recent years which depends upon the "two—spaces" concept. This is not really a Kantian notion at all, for to distinguish between a phenomenal space and a physical one is to abandon the basis of Kant's philosophy entirely. However, Strawson has suggested that while Kant is clearly wrong in claiming that we know Euclidean geometry to be true a priori for physical space, he may none the less be correct in saying that is true a prion' of a purely phenomenal space:

If we can make sense of this notion of a phenomenal interpretation for Euclidean geometry then perhaps Kant's theory of pure intuition can be seen, at least up to a point, as a perfectly reasonable philosophical account of it. To bring out the status of the propositions of such a geometry, it is best to take an example. Consider the pro- position that not more than one straight line can be drawn between any two points. The natural way to satisfy ourselves of the truth of this axiom of phenomenal geometry is to consider an actual or an imagined figure. When we do this, it becomes evident that we cannot, either in imagination or on paper, give ourselves a picture such that we are prepared to say of it both that it shows two distinct straight lines and that it shows both these lines as drawn through the same two points. (P. F. Strawson, The Bounds of Sense, pp. 282-3.)

This is a very strong claim indeed for the existence of a purely phenomenal space. Strawson is claiming not merely that we might be able to make sense of such a concept, but that we could describe phenomenal space as having a geometry all of its own. On this theory, the very strong intuitive appeal of Euclidean geometry is to be explained by its being the geometry that describes such things as phenomenal straight lines and circles. The theory makes no claims about the proper geometry of physical space. It would not be worrying, on this account, if physical triangles were found to have angle sums greater than two right angles: phenomenal triangles could still be Euclidean. This is a "two spaces" doctrine with a vengeance.

I have tried to explain in detail elsewhere' why I do not think that this idea of a phenomenal geometry will work. The main problem can be best brought out by considering first of all how theories of physical space and geometry have progressed during the last few hundred years. We have already seen that a remarkable feature of Euclidean geometry is that it appears to make metrical assertions about figures, as in the case of Pythagoras' theorem. Obviously the origin of the metrical aspects must be buried somewhere in the axioms, and in fact Pythagoras' theorem depends upon Euclid's 5th Axiom, the famous "parallel axiom": "Given three straight lines p, q, r one of which p intersects the other two, then if the sum of the interior angles of intersection on the same side of p is less than two right angles, then if r and q are produced indefinitely they will meet on that side of p. " It is the presence of the two right angles in this axiom that carries the metrical burden of the whole geometry. It is possible to do without the parallel axiom, but only if we are prepared to substitute a statement that would normally be a theorem, such as "There exists a single triangle with the angle sum of two right angles."

In 1773 the Jesuit priest Saccheri tried replacing the angle sum of two right angles by various alternatives, and attempted to show that the resulting geometries were not self-consistent. He failed, and the idea at length became established that several self-consistent geometries exist. This was not finally established with certainty until Klein found a method of mapping each theorem of a non—Euclidean geometry on to a corresponding theorem in Euclidean geometry. If this can be done, the non—Euclidean geometry must be at least as self-consistent as the Euclidean geometry. Klein's method was roughly as follows. Euclidean geometry makes reference to a number of "primitive elements" such as "straight line" and "intersection". In a formal statement of the geometry these elements can be replaced by symbols. The same can be done for the non-Euclidean geometry using a different set of symbols. Such axiomatic systems are said to be "uninterpreted" in the sense that the primitive elements have not yet been co—ordinated with points, lines and so on. Let us now interpret the elements of a non—Euclidean geometry by relating them to the interpreted elements of Euclidean geometry, such as "straight line", "line on the surface of a sphere" and so on. We shall now have a set of axioms that could be said to be true or false in Euclidean geometry. The question is whether a suitable interpretation can be found such that all these interpreted axioms would be true in Euclidean geometry. Klein showed that such an interpretation was possible, and in fact several are now known to exist.

For example, suppose that there is a figure T in the non-Euclidean geometry, composed of three L's, the angle sum of which is greater than two right angles. If we co-ordinate T with a Euclidean "triangle" and L with a Euclidean "straight line" we have an inconsistency. But if we adopt the mapping T; "spherical triangle" and L E "shortest distance between two points on the surface of a sphere" there is no inconsistency. If, given this mapping, no further inconsistency turns up we may conclude that our new geometry is as self-consistent as the Euclidean.

Another approach is to interpret the primitive elements arithmetically, and to show that an interpretation is possible at least as consistent as arithmetic itself. This was done by David Hilbert.

We can now take the argument a stage further. Even the interpreted elements of a geometry, such as "straight line", are not yet interpreted in a physical sense. Before we can apply a geometry to physical measurement we have to decide on interpretations like: "straight line" E "path of light ray in empty space". Until we have done this we have no right to speak of one geometry as more or less true than any other. If we took a Euclidean "straight line" to mean the shortest distance between two points on the surface of the earth, we should find as a matter of fact that it is not true, in the sense that triangles would have angle sums greater than two right angles. Whether it is true of the paths of light rays in empty space is a matter of experimental investigation. Einstein put this very clearly as follows:

For example, Euclidean geometry considered as a mathematical system, is a mere play with empty concepts (straight lines, planes, points, etc., are mere "fancies"). If, however, one adds that the straight line be replaced by a rigid rod, geometry is transformed into a physical theory. A theorem, like that of Pythagoras, then joins a reference to reality.

I hope it can be seen without much further elaboration that the idea of a purely phenomenal geometry is highly suspect.' Suppose I ask you to imagine a triangle and then I wish to find out whether the sum of the angles of this phenomenal figure is 180 degrees. Since this figure is purely phenomenal there is no measurement 1 can carry out on it. My only recourse is to ask you, the imager, to measure it for me. You may even feel that you can do this, and reply "two right angles". But what has been established here? If I agree with your answer, all that has been established is that we use the words "triangle", "straight lines" and "two right angles" in the same way. Consider what we should say if someone obstinately proclaimed that their phenomenal triangles had angle sums greater than two right angles; or, worse, that he could imagine five different lines in the same plane passing through a point and none of them intersecting a sixth line in that plane. His statement could be true of our curves but not of our lines. How do we know that he is talking about our lines rather than our curves? Concluding that our eccentric has a non-Euclidean phenomenal geometry would be just like saying that the geometry of a sausage is non—Euclidean. If we decide that lines drawn at right angles to the long axis of the sausage are straight, then all "straight lines" on the sausage are parallel and its geometry is indeed non-Euclidean. But Euclidean descriptions of sausages are not beyond our power if we adopt a different definition.

It may be objected that eccentrics of the kind just described do not crop up in reality, and that we all know perfectly well what a phenomenal straight line is. Indeed we do, and this is precisely because we have been taught to use the word by other people. It can hardly be maintained even by the most convinced nativist that we are born with an association between a phenomenal straight line and the verbal utterance "straight line". We are taught it by being shown physical straight lines. A mother cannot say to her child: "Imagine a straight line. Now this is what we adults call a straight line." Nor is it any improvement to say: "Imagine the shortest distance between two points: this is called a straight line." For the term "shortest" has not yet been defined, nor can it be without escaping if only momentarily from the phenomenal domain.

Thus any phenomenal geometry must be entirely parasitic upon physical geometry. 1 do not see how it can make sense to speak of different geometries for physical and phenomenal triangles. Even if there were such things as phenomenal triangles we could only learn to call them triangles by being presented with real triangles. We have a set of rules determining which kinds of line drawings shall be called triangles. They must, for example, be drawn with straight edges, not curves. If we show these to people they are simply not allowed to say: "This gives rise to a phenomenal square." For what then should a square be called?

Equally intractable difficulties arise when one considers congruence operations, and what they might mean in a purely phenomenal domain. Superficially it looks as if the congruence relation "the same length as" can be replaced in phenomenal geometry by "looking the same length as". I think I have said enough to indicate the broad lines on which this can be attacked, but the argument is lengthy and I would refer the reader who thinks this point important to a previous article.'

If we abandon the notion of a phenomenal geometry, as I think we must, what other properties could phenomenal space have to qualify as being space—like? If we ask what are the properties possessed by physical space and try to find counterparts in the phenomenal domain we shall soon see that phenomenal space is extremely impoverished. Of course, arguments still rage about whether physical space is meaningfully described as having properties at all. "Relativists" continue to attempt the replacement of all space-properties by the properties of fields and relations between bodies. Nevertheless, both relative and absolute theories of physical space will have to cope, as Hinckfuss points out, with properties like the following:

Electrical, optical, and electromagnetic properties of space
(i) Empty space is a poor conductor.
(ii) The magnetic permeability of empty space is $471 \times 10^{-7}$ henrys per metre.
(iii) The permittivity of empty space is $8.55 \times 10^{-12}$ farads per metre.
(iv) The speed of light in empty space is $2.9978 \times 10^{8}$ metres per second.
(v) Empty space is transparent.

I cannot say with confidence whether or not empty phenomenal space is transparent, still less what its permeability might be. One is reminded of a remark by Meyerson: "The real appears to us a fact — a datum. Now, reason would like to consider it as necessary. Hence the extravagant attempts to reduce it to space, namely to nothingness. . . ." If this applies to physical space, how much more does it apply to phenomenal space, the chief distinguishing feature of which seems to be that it has no properties whatsoever.

It may seem somewhat perverse to deny the existence of a phenomenal space so thoroughly. I am aware that it is in a sense setting up a straw man to argue that phenomenal space has no metric and no properties corresponding to the physical space. After all, no one has ever really supposed that we can measure phenomenal distances in feet or microns. There are weaker versions of the representational hypothesis than these I have discussed, and they can very probably be phrased in an entirely acceptable form. Of course, I do not wish to deny that we have a representation of spatial relations; for example, in the sense that as I look out of my window I see a line of trees as further away from me than the lawn. What this does not mean, however, is that I somehow measure the distance between phenomenal trees and lawns in a phenomenal space.

The perceived space between trees and lawn, I am suggesting, should not be thought of as having any existence prior to the judgement of the distance. And this judgement refers to the distance between the real trees and lawn, not to some special and separate phenomenal distance between them. Our perceived space is constituted by perceptual judgements of angle and distance, rather than existing prior to these perceptions and permitting them to be made?

I began this essay with some examples and would like to conclude with them also. Shepard and Cooper's subjects describe themselves as carrying out their task by "mental rotation' '. The proper description of what they were doing, I suggest, is that they were imagining a rotation—not that they were rotating an image. Shepard and Cooper are actually very careful to say that "mental rotation" does not imply the rotation of anything. The advantage of the description "imagining a rotation" is that it does not imply that any actual rotation occurred, thus obviating the need for a space in which something might rotate; the phrase "rotating an image", on the other hand, implies that something was actually rotated, with all the difficulties this brings in its wake. In the Pulfrich Pendulum the target is seen as rotating in an ellipse; this does not mean that the observer is inspecting a phenomenal target that is actually moving elliptically. There are possible advantages here in Gregory's "hypothesis" terminology for perceptions. The ellipse is the observer's hypothesis or judgement of the pendulum's path; it is not a geometric description of some special phenomenal kind of trajectory. Hypotheses are not themselves elliptical or straight any more than judgements are, and they are about physical events in physical space, not about a separate phenomenal domain.

Discussion

VESEY:

Kant held that spatial properties are entirely phenomenal. Like you I think he was wrong. But my reason for thinking him wrong is one which brings me into conflict with you on something else. My reason is a very general one about the conditions of meaningful discourse. I hold that for the remark "X looks ø", where X is some object and ø is some property, to be meaningful, "X is ø" must be meaningful. There must be an accepted practice with "X is ø' which we cotton on to when we are learning to talk, and by refer- ence to which it can be settled whether or not we are using the term "ø" correctly. Only then can it be allowed that we know what we are talking about when we say "X looks ø".

Some of the things you said — for instance, about "straight" and a brick—led me to think you would agree with me about this. But then, if I'm not mistaken, you went on to contrast spatial properties with properties like colour and smell—as though a word like "blue" could have an entirely phenomenal sense. Do you really think this?

D. E. Broadbent once wrote (Behaviour, London, 196]):

When a man sees blue, his experience is intensely real to him, but the essence of it cannot be communicated. All he can do is to say a word which labels that experience, so that he can tell other people whether or not some fresh situation gives him this same quality of awareness. No man can tell whether another is really feeling the same as he does himself when he looks at a colour.

Personally I think this is bad philosophy, rather than good psychology. What do you think?

MORGAN:

We seem to be in close agreement about "phenomenal" space. My purpose was to argue that, for example, there was no sense in which a line could be said to "look" straight unless there were agreed ways of showing that a line is straight. I think perhaps I wanted to go a little further than this, and to suggest that when we say a line "looks" straight we are actually more or less covertly carrying out exactly the same operations as we should on a real line. I'd be interested to know how you see this claim as relating to your logical argument.

Concerning colour and other "secondary qualities" we probably also agree; although I hesitate to dismiss Broadbent's views as "bad amateur philosophy", since he is only expressing a view that has been held by the overwhelming majority of scientists since the seventeenth century. Confusion may have arisen in my chapter because I was trying to explain Kant's point of view (not mine) that while colour and space are both phenomenal, they differ in that colour is subjective (a property entirely of the perceiver) while space is, in the weird terminology peculiar to Kant, a priori objective, in that it refers necessarily to objects "outside" the observer. Kant seems to have swallowed the received doctrine that colour words are names for wavelengths, or for mysterious inner experiences aroused in some unknowable manner by wavelengths. My own View is that a colour word is a name for a property shared by a particular class of objects in the world; the sensations and emotions aroused in us by an instance of such a property have to do with our experience of the whole object class. Even so-called "colour-blind" people, who have great difficulty in distinguishing between objects on the basis of wavelength information alone, frequently use colour words more or less correctly. This is presumably because they have learned the limits of class membership without too much reliance on wavelength information—in which case it seems to me mistaken to call them "colour—blind".

I believe it was Iris Murdoch who justified a painter's saying to a novitiate "you _don't understand Red". Perhaps physiologists find colours mysterious and incommunicable because they have not done the right sort of work to understand them, and have scant respect for the patient endeavours of the phenomenologists, who have shown that colours have different "weights", "distances" and so on. I don't know how to remedy this real lack of communication.

VESEY:

On your first point —there are different senses of "looks": I may say that a coin looks elliptical if seen at an acute angle, meaning by this that if I were to trace its outline on a transparent screen between the coin and my eye, at right angles to the line of vision, the tracing would be elliptical. In this sense of "looks", the look is determined by the laws of perspective. Again, I may say that the lines in the Muller—Lyer figure look unequal. In that case the look has nothing to do with perspective. What I mean is that on looking at the figure I would judge the lines to be unequal if I didn't know better. Perhaps your remark about our "more or less covertly carrying out exactly the same operations as we should on a real line" can be interpreted to cover both these senses of "looks". My own point IS that, in both of them, "looks" makes sense only because "is" would make sense.

On your second point, about what I called "bad philosophy" being scientific orthodoxy, perhaps I should make my own position a bit clearer. There is the everyday world in which there are flashes of lightning, colours and so on. And there is the scientific world in which there are discharges of electricity, wavelengths and so on. For certain purposes (explanation, prediction, control) the scientific world has priority. For certain other purposes (including knowledge of the scientific world!) the everyday world has priority. What I regard as philosophically bad is to suppose that the question "Which world is prior?" has some meaning tout court, that is, without any specification of purpose. For the same reason it is bad philosophy to ask "Which is the real one, the everyday world or the scientific world?" without specifying one's criteria of reality. Once the unqualified "Which is real?" question has been allowed, the scientist feels professionally committed to giving it the answer "The scientific world", and so to saying that the everyday world is unreal. Then, when questions about sensible colours are raised, he has to say that, since they are not needed for explanatory purposes in the scientific world, they exist only in the other one, the one he has had to dismiss as "unreal". Or, as he is inclined to put it, colours are "only subjective". The fact that philosophers and scientists have been saying this sort of thing for thousands of years does not make it either good philosophy or good science. But I suspect, from what you say, that we are in agreement on this.

# Afterword to the Conference: The Prospects for Consciousness Research

8. D. JOSEPHSON
Cavendish Laboratory, Cambridge

In the conference recorded in these Proceedings a group of scientists discussed various topics concerned with conscious experience and the relationship of conscious experience to the physical world. What exactly are we doing when we discuss conscious experience, and what are the future prospects for research in this area? These are questions I should like to explore in this concluding chapter.

To some extent scientific inquiry is just an extension of an activity which occurs naturally in everyday life. Both involve learning about the world in order to be able to make successful predictions about it, and in order to be able to carry out such actions as will have desirable outcomes. One of the principal differences between them is that the processes of science are more self-conscious and in a certain sense more public. In science hypotheses are deliberately stated, and steps are taken to test them, using both experimental methods and intellectual analysis. By these criteria, the papers presented in this conference are scientific in character, in comparison with general conversation on the same topics by a randomly chosen group of people. However, conscious experience is a field of inquiry in which application of the usual methods and techniques of science is particularly difficult. Two difficulties in particular are worth discussing in some detail. One of these is that conscious experience is personal in nature, and hence not open to public inspection, and the other is the problem of suitably describing conscious experience.

The first difficulty is one which in practice we have even with conventional scientific experiments. While it is true that a certain proportion of experiments are mechanized by using chart recordings or photography, in other cases readings are taken directly by the individual experimenter, and in such cases the general scientific community does not have direct access to the original experiment itself. Acceptance of such results is dependent upon the reputation of the experimenter and the repeatability of the results by other people. In such a situation we can never have an exact repetition of an experiment, but only a close approximation to it, and in, for example, psychological experiments it is only possible to reproduce a situation which is qualitatively the same. This does not prevent general conclusions from being drawn (a good illustration being the case of linguistics; it is possible to infer grammatical rules from a collection of utterances in which no two speakers are talking about the same thing). In the light of these remarks, the drawing of general conclusions about conscious experiences may not be an impossible task.

If we want to consider conclusions of a reasonably precise nature about conscious experience, we must consider the question of what kind of description of conscious experience is to be used. There seem to be three main possibilities. Firstly, there are the verbal descriptions given by the conscious subject himself. Secondly, there is the possibility of measuring physiological or behavioural correlates to conscious experience. And finally, there is the more remote but also more exciting possibility that some future formulation of physics may describe inner experiences as well as the external world, and hence provide its own way of quantifying subjective experience.

I shall deal fairly briefly with the first two possibilities. With the first possibility we are dealing with the use of language to describe experiences. The fact that language can be used at all must imply similarities between the experiences of different people, and furthermore, since words are used to distinguish between different possibilities, must imply that there are particular differences between conscious experiences that we can be trained to notice and attach linguistic labels to. To this extent language constitutes a kind of measuring instrument, though an imprecise one. It may be important to note that, since the meanings of the words are tied to the conscious experiences, for one person to understand another's description fully it may be necessary for him to have had a similar kind of experience, a problem to which Charles Tart has drawn attention.' On the other hand, it may be possible to understand strange experiences on the basis of a mathematical model, in the same way that we can understand curved spaces or multi-dirnensional spaces beyond our own experience mathematically.

Let us turn now to the second possibility. It may be possible to find physiological correlates to conscious experience, for example the EEG. If this could be done, the result would be to add precision to the verbal description. Similarly, there might be definite behavioural changes associated with a conscious experience, a familiar example being the effects of alcohol intoxication. Another example would be the more subtle changes following experience of the meditative state, involving for example improvements related to attention, discrimination and value judgements.

So far we have been concerned with the study of conscious experience at a purely phenomenological level. What is the possibility that a more quantitative, mathematical theory might be feasible? One way in which this might come about is through an extension of existing physical theory. The existing basic theory of physics, quantum mechanics, while in one sense a good description of nature is, from a different viewpoint, highly inadequate. I am referring not to the well-known difficulty that its predictions are statistical in character, but to the fact that it is not entirely
clear how it is to be applied to the real world. The situation is that while we understand how the theory can be confirmed, in terms of controlled experiment, there is no well-defined prescription for how predictions can be made in a general uncontrolled situation, where the knowledge of the state of the system does not necessarily take the form required to apply the quantum theory, i.e. measurement of a physical quantity. Is it the situation that inner observation as well as observation of the external world should count as a quantum—mechanical measurement, and if so, what is it a measurement of? The fact that quantum theory is a theory of what can be deduced from observation, as much as it is one of what exists, seems to force such matters upon our attention; if we exclude such matters we cannot legitimately regard quantum mechanics as a comprehensive theory.

I should like, finally, to consider in this light the talk given in this conference by Mrs. E. M. Noakes,* to which I shall give more emphasis

* The Editors reluctantly agreed to a request 'by a number of participants that Mrs. Noakes's paper should not appear in these proceedings, on the grounds that its methodology lay outside the paradigm of science as perceived by these participants.

than normal on account of it not being available for these Proceedings. In her talk, based on a particular mystical tradition, she described various entities which are supposed to have important effects in the life of an individual human being. An example given was the so-called "astral body", i.e. the collection of the feelings and emotions of the individual concerned. Now while the average scientist might recoil at even the mention of a term such as "astral body", it must be admitted that feelings and emotions form a relatively unchanging pan of an individual's make-up, and that furthermore they do have well-defined effects on the world publicly observable (through the agency of the individual's nervous system, presumably). Following along this line of thought, we can argue that awareness of feelings and emotions, or other inner experiences, constitutes an observation of the world, which may later have publicly observable effects. If we were to try to say instead that all the physics is to be described in terms of the properties of neurones and synapses, we should run into the practical difficulty that we are not able to observe the details required in order to make prediction (perhaps not even in principle, in a living being) in the way that we can observe feelings and emotions. This argument again suggests the necessity of including inner experience within the subject matter of physics.

I cannot detail here the remaining subtler entities and processes to which Mrs. Noakes made reference in her talk. I can only conclude by making the point that while mystical experience is not at the moment considered by the majority of scientists to be a matter worthy of scientific attention, this is to some extent purely an arbitrary decision. The desire to probe more deeply into the interaction between man and the world in which he exists will ultimately lead to the systematic study of the mystical experience and to its incorporation into science.

List of Participants

(Contributors of papers are indicated by an asterisk)

*Professor H. B. Barlow, The Physiological Laboratory, University of Cambridge, Cambridge CB2 3EG.
Dr. E. W. Bastin, Pond Meadow, West Wickham, Cambs. .
Professor J. W. S. Cassels, Department of Pure Mathematics and Mathematical Statistics, 16 Mill Lane, Cambridge.
Professor Sir Alan Cottrell, The Master's Lodge, Jesus College, Cambridge.
Professor O. R. Frisch, Trinity College, Cambridge CB2 1TQ. _
*Professor R. L. Gregory, Brain and Perception Laboratory, Medical School, University of Bristol, Bristol BS8 1TD.
Dr. J. R. Henderson, Cavendish Laboratory, Madingley Road, Cambridge CB3 OHE.
*Dr. N. K. Humphrey, Sub-Department of Animal Behaviour, University of Cambridge, Madingley, Cambridge CB3 8AA. .
*Professor B. D. Josephson, Cavendish Laboratory, Madingley Road, Cambridge CB3 OHE.
Dr. A. J. Leggett, Department of Physics, Sussex University, Brighton BN1 9QG. _
*Professor H. C. Longuet-Higgins, Department of Experimental Psychology, Sussex University, Brighton BN1 9QG.
*Professor D. M. MacKay, Department of Communication and Neuroscience, University of Keele, Keele, Staffordshire ST5 SBG.
*Dr. M. J. Morgan, South Road, University of Durham, Durham DH1 3LE.
*Mrs. E. M. Noakes, Sidmouth House, Sidmouth, Devon EX10 8ST.

198 List of Participants

*Mrs. S. Padfield, Pond Meadow, West Wickham, Cambs.

*Dr. V. S. Ramachandran, Trinity College, Cambridge CB2 ITQ.

*Professor Sir Martin Roth, Department of Psychiatry, University of Cambridge, Cambridge CB2 IEL.

*Professor G. Vesey, Department of Philosophy, Open University, Milton Keynes MK7 6AA.

Dr. P. Whittle, Psychological Laboratory, Downing Street, Cambridge.

Prof. O. L. Zangwill, Department of Experimental Psychology, Downing Street, Cambridge.

Name Index

Subject Index

The letter D after a page number denotes a discussion comment.